\documentclass[12pt]{article}
\usepackage{amssymb} 
\usepackage{hhline}
\usepackage[scale={.75,.75}]{geometry}
\usepackage[T1]{fontenc}
\usepackage{bbm} 
\usepackage{fleqn} 
\usepackage{mathrsfs} 
\usepackage{cite} 
\usepackage{longtable}
\usepackage{caption}
\usepackage{cite}
\usepackage{nicefrac}

\usepackage{color}
\usepackage{amsmath}
\usepackage{mathrsfs}
\usepackage{verbatim}
\usepackage{graphicx}
\usepackage{upgreek}
\DeclareFontFamily{OT1}{pzc}{}
\DeclareFontShape{OT1}{pzc}{m}{it}{<-> s * [1.200] pzcmi7t}{}
\DeclareMathAlphabet{\mathpzc}{OT1}{pzc}{m}{it}	

\usepackage{lscape}
\usepackage{psfrag}
\usepackage{hyperref}

\newcommand{\q}{\textbf{q}}

\newcommand{\be}{\begin{equation}}
\newcommand{\ee}{\end{equation}}
\newcommand{\bea}{\begin{array}}
\newcommand{\eea}{\end{array}}

\newcommand{\bes}{\begin{split}}
\newcommand{\ees}{\end{split}}

\newcommand{\tp}{t}
\newcommand{\tm}{y}

\newcommand{\Slash}[1]{\displaystyle{\not}{#1}}
\newcommand{\bath}[1]{#1}
\newcommand{\fermionmass}[1]{\mathfrak{#1}}
\newcommand{\C}{\mathcal{C}}
\numberwithin{equation}{section}
\numberwithin{table}{section}
\allowdisplaybreaks

\begin{document}

\title{ 
{\normalsize     
}
%
\textbf{On the Role of Quasiparticles and thermal Masses in Nonequilibrium Processes in a Plasma\\[8mm]}}
%
\author{Marco Drewes\footnote{Electronic mail address: \href{mailto:marco.drewes@epfl.ch}{marco.drewes@epfl.ch}}\\ 
\textit{Institut de Th\'eorie des Ph\'enom\`enes Physiques,}\\
\textit{\'Ecole Polytechnique F\'ed\'erale de Lausanne (EPFL),}\\
\textit{CH-1015 Lausanne, Switzerland}
{\normalsize\it }
}
\date{}
\maketitle

\thispagestyle{empty}


\begin{abstract}
\noindent
Boltzmann equations and their matrix valued generalisations are commonly used to describe nonequilibrium phenomena in cosmology. On the other hand, it is known that in gauge theories at high temperature processes involving many quanta, which naively are of higher order in the coupling, contribute to the relaxation rate at leading order. How does this accord with the use of single particle distribution functions in the kinetic equations? When can these effects be parametrised in an effective description in terms of quasiparticles? And what is the kinematic role of their thermal masses? We address these questions in the framework of nonequilibrium quantum field theory and develop an intuitive picture in which contributions from higher order processes are parametrised by the widths of resonances in the plasma. In the narrow width limit we recover the quasiparticle picture, with the additional processes giving rise to off-shell parts of quasiparticle propagators that appear to violate energy conservation. In this regime we give analytic expressions for the scalar and fermion nonequilibrium propagators in a medium. We compare the efficiency of decays and scatterings involving real quasiparticles, computed from analytic expressions for the relaxation rates via trilinear scalar and Yukawa interactions for all modes, to off-shell contributions and find that the latter can be significant even for moderate widths. Our results apply to various processes including thermal production of particles from a plasma, dissipation of fields in a medium and particle propagation in dense matter. We discuss cosmological implications, in particular for the maximal temperature achieved during reheating by perturbative inflaton decay.
\end{abstract}

\newpage
{\footnotesize \tableofcontents}

\newpage\index{}

\section{Introduction}
Many important features of the observable universe can be understood as the result of out of equilibrium processes during the early stages of its history, when it was filled with a hot primordial plasma. 
This includes the decoupling of the cosmic microwave background, the creation of light elements, dark matter production, baryogenesis, and, in inflationary cosmology, the production of particles altogether during reheating. 
In many cases the relevant energy scales by far exceed those that can be realised in any human made experiments and provide an excellent laboratory to test particle physics models beyond the Standard Model (SM). 
Nonequilibrium dynamics also play a crucial role in the understanding of signals from heavy ion colliders. Thus, a quantitative understanding of nonequilibrium processes is crucial for cosmology as well as particle physics.

Boltzmann equations and their matrix valued generalisations can accurately describe many nonequilibrium phenomena. They are known to suffer from uncertainties when the coherence lengths are large, but are usually believed to be accurate in absence of such effects. 
On the other hand, it is known that in gauge theories at high temperature resummations are necessary because processes involving many quanta, which naively are of higher order in the coupling, contribute to the relaxation rate at leading order.
How does this accord with the use of single particle distribution functions in the kinetic equations? 
Medium effects are often included in effective kinetic equations by the use of thermal masses for quasiparticles. What are the limits of the validity of this procedure? To which degree can the quasiparticles be treated as particles? Do the thermal masses act as kinetic masses?

All these problems are related to the assumption that the system should be described in terms of the properties of individual particles as asymptotic states. The definition of these is, however, ambiguous in a dense plasma. We show that the above questions can be answered in a consistent and intuitive way when swapping the single particle phase space distribution functions as dynamical variables for correlation functions of quantum fields. This formalism in addition allows for a full quantum treatment of coherent oscillations and quantum memory effects. 

In this work, we study the relaxation of scalar and fermionic quantum fields in a large thermal bath. 
Depending on the initial state of the out-of-equilibrium fields, they either gain energy from or dissipate it into the bath.
This resembles a large number of interesting phenomena including thermal production of particles from a plasma, propagation in dense matter and cosmological freezeout processes.
It is, apart from reporting a number of new results summarised below, one of the objectives of this article to explain and promote the formalism employed here to a wider audience.  It provides powerful tools to treat nonequilibrium quantum systems in terms of quantities that have a clear physical interpretation, without semiclassical (on-shell) approximations or a gradient expansion.
We aim to make the connection between nonequilibrium quantum field theory and effective kinetic equations transparent also to readers without background knowledge in the former. Therefore we try to avoid technicalities and restrain to language commonly used in particle physics.

\newpage
\begin{center}
\textbf{Outline of this Article}
\end{center}
In section \ref{QuantumFieldsAndParticles} we introduce our notation and review the formalism we use, including the physical interpretation of the various quantities.

Section \ref{scalarsectioN} is devoted to the relaxation of scalar fields in a thermal bath.
In section \ref{trilinearsection} we study in detail the kinematics of the relaxation via a trilinear interaction with a bath of other scalars. In the first part we recall the interpretation of the known result in the quasiparticle approximation.
In the second part we derive a formula that includes corrections beyond this approximation, using resummed perturbation theory. 
These appear as off-shell contributions and give rise to apparent violation of energy conservation in the quasiparticle picture. In section \ref{ScalarwithYukawa} we consider the case that the scalar is coupled to a bath of fermions with gauge interactions via a Yukawa coupling. 
In section \ref{quarticsection} we discuss the kinematic differences between relaxation via 3-vertices and 4-vertices, using the simple example of a quartic interaction. 

In section \ref{comparisonsection} we compare the contributions to the relaxation rate from decays and scatterings of real quasiparticles to off-shell contributions.  

Section \ref{sectionaboutfermion} is devoted to the relaxation of a fermion with Yukawa coupling. In \ref{fermioncorrelators} we give exact expressions for the nonequilibrium two-point functions of a fermion in a thermal bath. In \ref{fermionviayukawa} we compute an analytic expression for the rate of relaxation via a Yukawa coupling in the quasiparticle regime. Details of the calculation are given in appendix \ref{YukawaSelfEnergy}.

In section \ref{comparisontoBE} we compare the time evolution of the energy density in the Boltzmann and field theoretical approaches.

In section \ref{InflatonSection} we apply our results to reheating of the universe via perturbative inflaton decay. We discuss the possibility of an upper bound on the temperature due to closure of the phase space for decays by large thermal masses.

Further implications and possible extensions are discussed in section \ref{discussion}.
In appendix \ref{KBEessentials} we review some basic ingredients of nonequilibrium field theory used in the analysis, and in appendix \ref{YukawaSelfEnergy} we derive an analytic expression for the relaxation rate of a fermion with Yukawa coupling in a thermal plasma that has, to the best of our knowledge, not been reported in the literature.\\

The main new results reported in this article are the discussion of the nonequilibrium propagators in sections \ref{relaxratesection} and \ref{comparisontoBE}, the explicit computation of higher order contributions in section \ref{OffShellSection} and the related discussion in section \ref{analsection}, the numerical comparison between leading order and resummed results in section \ref{comparisonsection}, the expression for the nonequilibrium propagator for Dirac fermions in section \ref{sectionaboutfermion}, the analytic formula for the fermion relaxation rate in section \ref{fermionviayukawa}, the comparison of the energy density in the Boltzmann and quantum field theoretical approaches in section \ref{comparisontoBE} and the application to cosmic reheating in section \ref{InflatonSection}.
We embed these into detailed discussion that aims to draw an intuitive physical picture that is coherent without previous knowledge in nonequilibrium quantum field theory.


\section{Quantum Fields and Particles}\label{QuantumFieldsAndParticles}

\subsection{Motivation}\label{motivation}
\paragraph{Quantum Fields and kinetic Equations}
Many nonequilibrium processes can be treated in a canonical way by
means of Boltzmann equations (cf. \cite{kt90}) with sufficient accuracy. 
These equations are semiclassical in the sense that they describe a system of classical particles which propagate freely (on-shell) between isolated interactions, the cross sections for which are computed from S-matrix elements in vacuum. They have been used very successfully to study nonequilibrium processes in a dilute, weakly coupled gas. 
However, in a dense plasma and in the presence of strong interactions the validity of these approximations is questionable. 
Corrections due to thermal effects 
have been discussed in the framework of Boltzmann equations \cite{Covi:1997dr,Giudice:2003jh,Kiessig:2010pr,Anisimov:2010gy}.
Recently much progress has been made to derive consistent quantum kinetic equations from quantum field theory \cite{Sigl:1992fn,Buchmuller:2000nd,Aarts:2001qa,Prokopec:2003pj,Konstandin:2004gy,Asaka:2005pn,Asaka:2006rw,Lindner:2005kv,Garbrecht:2008cb,FillionGourdeau:2007zz,DeSimone:2007rw,Garny:2009rv,Cirigliano:2009yt,Herranen:2010mh,Beneke:2010dz,Boyanovsky:2004dj,Anisimov:2008dz,Anisimov:2010aq,ABDM3}.
Most of this work aims to include quantum interferences, coherent oscillations and non-Markovian effects. 

Here we focus on another aspect.
It is well known that the properties of particles are modified if they propagate in a medium. This e.g. includes Debye screening of charges and medium related modifications in the dispersion relations, which sometimes can be parameterised by thermal masses. Following early work \cite{Wolfenstein:1977ue,Klimov:1981ka,Weldon:1982aq,Weldon:1982bn,Weldon:1989ys} on the field theory side, cosmological implications were soon after discussed in the context of Affleck-Dine baryogenesis in \cite{Linde:1985gh}. 
Recently the topic has received increasing interest in the context of thermal leptogenesis \cite{Giudice:2003jh,Kiessig:2010pr,Anisimov:2010gy} and in models where dark matter is produced resonantly due to a change in neutrinos dispersion relations \cite{Laine:2008pg}.

An interesting observation has been made in \cite{Kolb:2003ke}. The authors suggest that an upper bound on the temperature of the universe during reheating by perturbative inflaton decay can be inferred when the would-be decay products acquire large thermal masses. These increase with the plasma temperature and eventually block the phase space for further decays.
Such bound would have far reaching consequences as the temperature in the early universe plays a crucial role for the abundance of thermal relics (in particular the gravitino problem \cite{Pagels:1981ke}), thermal leptogenesis \cite{Fukugita:1986hr}, the fate of moduli \cite{Yokoyama:2006wt} and the decompaticfication of extra dimensions \cite{Buchmuller:2004xr}. 
Indeed the mechanism would only be relevant during reheating, but whenever a significant amount of entropy is released by the dissipation of a component that redshifts like non-relativistic matter (e.g. a quickly oscillating field or non-thermal relic) at temperatures much larger than the corresponding mass.

The validity of arguments based on modifications of the phase space due to thermal masses was challenged in \cite{Yokoyama:2005dv}, where the author pointed out that medium related corrections to the widths of the resonances in the final state may have a significant effect.
The widths of decay products are known to be relevant in other contexts. 
In \cite{Beneke:2010dz} it was found that they are essential for the damping of flavour oscillations. A first principles calculation of the lepton asymmetry generated in the decay of heavy Majorana neutrinos in \cite{Anisimov:2010aq,ABDM3} shows large deviations from the result obtained from Boltzmann equations unless the finite widths of leptons and Higgs fields are taken into account. However, there the role of the width is related to the loss quantum coherence and memory, which leads to a behaviour that is local in time and allows to understand the emergence of classical behaviour in the quantum system. The argument brought forward in \cite{Yokoyama:2005dv}, in contrast, refers to kinematics.

\paragraph{The Particle Concept and its Ambiguities}
Any conclusion based on kinematic arguments about the mass, width and energy of (quasi)particles pre-assumes that the system can be well described in terms of the properties of single (quasi)particles that it is composed of. 
In quantum field theory this translates into the statement that integrals over spectral densities are strongly dominated by pole contributions. 
In typical collider experiments the language of stable or instable particles is suitable to describe the behaviour of quantum fields at asymptotic times because far away from the point of an interaction, elementary excitations of the fields propagate like free particles. 
It is, however, well-known that there exist various physical systems for which a description as a collection of (real) particles is not suitable. The most obvious example is a coherent classical (e.g. electromagnetic) field. It has furthermore been known for long \cite{ES1939} that the particle concept is ambiguous in a time dependent background, which may be provided by an external field or by a gravitational background \cite{BD}.
In nonequilibrium systems it is also not clear that a description in terms of (quasi)particles is suitable in the presence of strong interactions because the definition of asymptotic states is ambiguous due to the omnipresent background plasma. 
For instance, the experimental results from heavy ion collisions have revealed that the QCD plasma near the temperature of hadronisation cannot be well-described by an effective quasiparticle model \cite{Adams:2005dq}. 
 
A consistent description of nonequilibrium quantum fields without reference to particle numbers or asymptotic states is always possible in terms of quantum mechanical correlation functions. Therefore correlation functions and expectation values of operators form the suitable language to treat nonequilibrium quantum fields in the early universe. Our analysis is based on these techniques, mainly using the formalism reviewed in \cite{Berges:2004yj} and applied to a similar problem in \cite{Anisimov:2008dz}. We will use the notation and several results of that work, which are summarised in section \ref{relaxratesection} and appendix \ref{KBEessentials}. However, we believe that most of the following arguments can be understood without prior knowledge of nonequilibrium quantum field theory.

\subsection{Elements of Nonequilibrium Quantum Field Theory}\label{relaxratesection}
In this section we recall the ingredients of nonequilibrium quantum field theory used in this work, using the example of a real scalar field $\phi$, and introduce our notation. Some additional formulae are summarised in appendix \ref{KBEessentials}.
For a detailed introduction we refer the interested reader to the references \cite{Chou:1984es,LeB,Berges:2004yj}. 

We are interested in systems in which few nonequilibrium degrees of freedom are weakly coupled to a large thermal bath. This situation is realised in many interesting physical systems, including freezeout or thermal production of particles in the early universe and the propagation of energetic particles in a plasma. Our results can be used for a full quantum treatment of those.

As the authors of many earlier works, we assume that the fields in the bath thermalise fast on the time scale associated with the dynamics of the out of equilibrium degrees of freedom. Then the background plasma can be thought of as a heat bath with a well-defined temperature $T$ at each moment in time. This allows to use expressions known from thermal (equilibrium) quantum field theory for all correlation functions of the fields that make up the bath.

\paragraph{The relevant Correlation Functions} 
A thermodynamical system is represented by a statistical ensemble.
In quantum field theory, this does not correspond to a pure quantum state, but 
is described by a density matrix $\varrho$. 
The expectation value for an operator
$\mathcal{A}$ is given by 
\begin{equation}
\langle\mathcal{A}\rangle={\rm Tr}\left(\varrho\mathcal{A}\right)\label{expvalue}
,\end{equation}
where we have adopted the usual normalisation ${\rm Tr}\varrho=1$.
The averaging $\langle\ldots\rangle$ defined in (\ref{expvalue}) includes statistical (ensemble) as well as quantum averages.
Direct computation of the time evolution of $\varrho$ is difficult in practice and only possible in few cases.
Generically the von Neumann (or quantum Liouville) equation which governs it can only be solved perturbatively for a reduced density matrix with an effective Hamiltonian. In most practical applications to date an number of additional assumptions (including an on-shell limit and gradient expansion) are made that lead to matrix valued Boltzmann equations which take account of coherent oscillations\footnote{See \cite{Gagnon:2010kt} for an example of a calculation which avoids most assumptions.}.

Instead of directly looking at $\varrho$, it is equivalent to study the time evolution of all correlation functions of the fields. This can be seen in loose analogy to the Bogoliubov-Born-Green-Kirkwood-Yvon hierarchy in classical statistical mechanics. However, the description in terms of correlation functions avoids all ambiguities related to the definition of asymptotic states and particle numbers since they are always well-defined. They allow to compute the expectation values for all observables at all times.
Though in principle knowledge of the infinite tower of $n$-point functions is required to describe the infinitely many degrees of freedom of the density matrix, it is in practice often sufficient to study the time evolution of the mean field or one-point function $\langle\phi(x)\rangle$ and two independent connected two-point functions. This is the case when the observables of interest can be expressed in terms of field bilinears in a controllable approximation. A common choice is given by the connected Wightman functions
\begin{eqnarray}
\Delta^>(x_1,x_2) &=& \langle \phi(x_1)\phi(x_2)\rangle_{c}, \label{forw}\\ 
\Delta^<(x_1,x_2) &=& \langle \phi(x_2)\phi(x_1)\rangle_{c} . \label{back}
\end{eqnarray}
Here the subscript $_{c}$ indicates {\it connected}.
Alternatively one can consider any linear combination of these, e.g. time ordered and anti time ordered propagators. A particularly convenient choice is given by 
\begin{eqnarray}
\Delta^{-}(x_{1},x_{2})&=&i\langle [\phi(x_{1}),\phi(x_{2})]\rangle_{c}=i\left(\Delta^>(x_1,x_2)-\Delta^<(x_1,x_2)\right), \label{DeltaminusDef}\\
\Delta^{+}(x_{1},x_{2})&=&\frac{1}{2}\langle\{\phi(x_{1}),\phi(x_{2})\}\rangle_{c}=\frac{1}{2}\left(\Delta^>(x_1,x_2)+\Delta^<(x_1,x_2)\right).\label{DeltaplusDef}
\end{eqnarray}
Here $[ , ]$ and $\{ , \}$ denote commutator and anticommutator, respectively.
In general the functions $\Delta^{\gtrless}$ and $\Delta^{\pm}$ depend separately on the two four-vectors $x_{1}=(t_{1},\textbf{x}_{1})$ and $x_{2}=(t_{2},\textbf{x}_{2})$. Here we consider spacially homogeneous fields which depend on $t_{1}$, $t_{2}$ and the relative spacial coordinate $\textbf{x}_{1}-\textbf{x}_{2}$ only. Furthermore we are interested in the behaviour of fields that are weakly coupled to a large thermal bath with many constituent fields, to which we collectively refer as $\mathcal{X}_{i}$. The interactions that keep the bath in equilibrium are much stronger than the coupling to $\phi$.
This is absolutely crucial because backreaction-effects are suppressed and one can compute self energies from $\mathcal{X}_{i}$ propagators only. Contributions from diagrams involving $\phi$-propagators in loops are double-suppressed by the weaker coupling \textit{and} the much smaller number of diagrams (as there are many more $\mathcal{X}_{i}$-fields).
It can be shown that in this case $\Delta^{-}$ is time translation invariant and depends on properties of the bath only \cite{Anisimov:2008dz}.

The two point functions $\Delta^{\pm}$ have an intuitive physical interpretation. This can most easily be seen by looking at their Wigner transforms $\Delta^{-}_{\textbf{q}}(t_{1}-t_{2})$ and $\Delta^{+}_{\textbf{q}}(t_{1},t_{2})$, i.e. the spacial Fourier transforms in $\textbf{x}_{1}-\textbf{x}_{2}$. 
$\Delta^{-}_{\textbf{q}}(t_{1}-t_{2})$ is proportional to the spectral density $\rho_{\textbf{q}}(\omega)$ associated with the field operator $\phi$, 
\begin{eqnarray}
\Delta^-_{\q}(t_{1}-t_{2})=i\int^{\infty}_{-\infty}\frac{d\omega}{2\pi} 
e^{-i\omega (t_{1}-t_{2})}\rho_{\q}(\omega)\ 
\label{dmin}
.\end{eqnarray}
Thus $\Delta^{-}$, which is also known as \textit{spectral function}, characterises the spectrum of excitations in the plasma.
To illuminate the physical interpretation of $\Delta^{+}$ we define the energy momentum tensor for $\phi$ as $T_{\mu\nu}^{\phi}=\partial_{\mu}\phi\partial_{\nu}\phi-\eta_{\mu\nu}\mathcal{L}|_{\mathcal{X}_{i}=0}$, where $\mathcal{X}_{i}$ are all fields other than $\phi$. 
The contribution to the $\phi$-energy $\epsilon^{\phi}=\langle T_{00}^{\phi}\rangle$ from the mode $\textbf{q}$ can be written as
\begin{align}\label{Energie}
\epsilon^{\phi}_{\q}(t) &= \frac{1}{2}\left(\partial_{t_1}\partial_{t_2} 
+ \omega_{\q}^2 \right)\left(\Delta^+_{\q}(t_1,t_2)+\langle\phi_{\q}(t_{1})\rangle\langle\phi_{\q}(t_{2})\rangle\right)\big|_{t_1=t_2=t} 
.\end{align}
The $\langle\phi\rangle\langle\phi\rangle$-term in the brackets is due to the mean field while the $\Delta^{+}$-term is the contribution from fluctuations that can be interpreted as particles. Thus, the statistical propagator can be viewed as a measure for the occupation number of the mode $\textbf{q}$ that remains well-defined even in a dense plasma. It does not depend on any reference to asymptotic states or particle numbers. 

The functions $\Delta^{\pm}$ have to be found by solving the coupled integro differential equations given in (\ref{abzug2}), (\ref{addi2}), known as Kadanoff-Baym equations \cite{kb62}.
The solutions for scalars have been found in  \cite{Anisimov:2008dz}. The spectral function is given by (\ref{dmin}) with
\begin{eqnarray}
\rho_{\q}(\omega)&=&\left(\frac{i}{\omega^{2}-\omega_{\q}^{2}-\Pi^{A}_{\q}(\omega)-i\omega\epsilon}-\frac{i}{\omega^{2}-\omega_{\q}^{2}-\Pi^{R}_{\q}(\omega)+i\omega\epsilon}\right).\label{spectralfunction}
\end{eqnarray}
Here $\Pi^{R}$ and $\Pi^{A}$ are the retarded and advanced self energies.
The expression (\ref{spectralfunction}) is identical to the well known result from thermal field theory, but was found as an exact solution for the nonequilibrium equation of motion (\ref{abzug2}).
The solution for the statistical propagator can be expressed in terms of $\Delta^{-}$, 
\begin{eqnarray}\label{solution}
\Delta^{+}_{\q}(t_{1},t_{2}) &=& 
\Delta^{+}_{\q,\text{in}}
\dot{\Delta}^{-}_{\q}(t_{1})\dot{\Delta}^{-}_{\q}(t_{2})
+\ddot{\Delta}^{+}_{\q,\text{in}}
\Delta^{-}_{\q}(t_{1})\Delta^{-}_{\q}(t_{2})\nonumber\\
&+&\dot{\Delta}^{+}_{\q;\text{in}}
\left(\dot{\Delta}^{-}_{\q}(t_{1})\Delta^{-}_{\q}(t_{2})
+\Delta^{-}_{\q}(t_{1})\dot{\Delta}^{-}_{\q}(t_{2})\right)\nonumber\\
&+& \int_{0}^{t_{1}}dt'\int_{0}^{t_{2}}dt''
\Delta^{-}_{\q}(t_{1}-t')\Pi^{+}_{\q}(t'-t'')\Delta^{-}_{\q}(t''-t_{2}) 
.\end{eqnarray}
Here the initial conditions have been parameterised as
\begin{eqnarray}
\Delta^{+}_{\q,\text{in}} &=& 
\Delta^{+}_{\q}(t_{1},t_{2})|_{t_{1}=t_{2}=0} ,\label{in1G}\\
\dot{\Delta}^{+}_{\q,\text{in}} &=& 
\partial_{t_{1}}\Delta^{+}_{\q}(t_{1},t_{2})|_{t_{1}=t_{2}=0} 
= \partial_{t_{2}}\Delta^{+}_{\q}(t_{1},t_{2})|_{t_{1}=t_{2}=0}
, \label{in2G}\\
\ddot{\Delta}^{+}_{\q,\text{in}} &=& 
\partial_{t_{1}}\partial_{t_{2}}
\Delta^{+}_{\q}(t_{1},t_{2})|_{t_{1}=t_{2}=0} \label{in3G}
\end{eqnarray}
and the self energy $\Pi^{+}$ is defined in appendix \ref{KBEessentials}, its Fourier transform can be related to the retarded self energy $\Pi^{R}$ via (\ref{PiKMS}) and (\ref{ipa}). 
The equations (\ref{in1G})-(\ref{in3G}) define initial conditions for an ensemble, which at initial time is described by a density matrix that can be written as a direct product $\varrho_{\phi}\otimes\varrho^{eq}_{\rm bath}$. 
$\varrho^{eq}_{\rm bath}$ is the equilibrium density matrix of the bath, which can be fully characterised by the temperature (and in more general cases all chemical potentials). $\varrho_{\phi}$ characterises the initial state of $\phi$. We assume that $\phi$ initially has Gaussian correlations only, in which case $\varrho_{\phi}$ is fully characterised by the five parameters for each mode, which can be chosen as (\ref{in1G})-(\ref{in3G}) and the initial values $\phi_{\q,\text{in}}=\langle\phi_{\q}(t)\rangle|_{t=0}$ and $\dot{\phi}_{\q,\text{in}}=\partial_{t}\langle\phi_{\q}(t)\rangle|_{t=0} $ for the mean field. For $\ddot{\Delta}^{+}_{\q,\text{in}}\Delta^{+}_{\q,\text{in}}-(\dot{\Delta}^{+}_{\q,\text{in}})^{2}=\frac{1}{4}$, $\varrho_{\phi}$ corresponds to a pure state \cite{Berges:2004yj}.
The \textit{memory integral} in the last line of (\ref{solution}) contains all non-Markovian effects.
The time evolution of the mean field mode $\textbf{q}$ is given by
\begin{eqnarray}\label{fieldvalue}
\langle\phi_{\textbf{q}}(\tp)\rangle=i\left(\dot{\phi}_{\q,\text{in}}+\phi_{\q,\text{in}}\partial_{t}\right)\int\frac{d\omega}{2\pi}\rho_{\textbf{q}}(\omega)e^{-i\omega t}
,\end{eqnarray}
where we assumed $\langle\phi\rangle=0$ in the ground state.
Equation (\ref{fieldvalue}) is valid if the initial deviations $\phi_{\q,\text{in}}$ and $\dot{\phi}_{\q,\text{in}}$ are not too large. 
It can be found from the finite temperature effective action \cite{Yokoyama:2004pf}\footnote{In \cite{Yokoyama:2004pf} contributions to the effective action from four point functions where included, leading to an equation of motion that goes beyond the Langevin type, but equilibrium propagators where used for $\phi$ in loops, restricting the analysis to the regime where the relevant $\phi$ modes are close to equilibrium.} which leads to an effective Langevin equation for $\phi$ \cite{Boyanovsky:2004dj}, that in the case of consideration here is equivalent to the Kadanoff Baym equations \cite{Anisimov:2008dz,Greiner:1998vd}.
Using the fact that $\Pi^{A}=(\Pi^{R})^{*}$ for a real scalar field, (\ref{spectralfunction}) can be rewritten as
\begin{eqnarray}\label{spectralfunction2}
\rho_{\q}(\omega)={-2{\rm Im}\Pi^R_{\q}(\omega)+2\omega\epsilon\over 
(\omega^2-\omega_{\q}^2-{\rm Re}\Pi^R_{\q}(\omega))^2+({\rm Im}\Pi^R_{\q}(\omega)+\omega\epsilon)^2} . 
\end{eqnarray}
This expression fulfils the well-known sum rule
\begin{equation}\label{sumrule}
\int d\omega\rho_{\textbf{q}}(\omega)=2\pi
.\end{equation}
The poles of (\ref{spectralfunction2}) determine the spectrum of propagating field excitations. Their positions in the complex $\omega$-plane depend on the self energy $\Pi^{R}$, which depends on temperature\footnote{We have not made this dependence explicit here and elsewhere because $T$ here is not a dynamical variable, but an external parameter. All self energies, and consequently spectral functions, rates, dispersion relations etc. that appear in this work are temperature dependent unless stated differently.}.
The self energy can be written as a sum of a temperature independent part, which coincides with the expression in vacuum, and a temperature dependent correction due to the medium. The former includes the same divergences known from
 the vacuum theory and has to be renormalised. The temperature dependent piece does not contain additional divergences as the medium does not affect the physics at short distances \cite{LeB}. The renormalisation conditions can be imposed at $T=0$, and the renormalised spectral function, expressed in terms of renormalised quantities, has the same shape as (\ref{spectralfunction}) \cite{Anisimov:2008dz}. We are not concerned with the details of the renormalisation here, but it is important that, as the renormalisation conditions are imposed at a particular temperature (typically $T=0$), it is not possible to absorb any shifts of the poles at some other temperature by renormalisation. 

The correlation functions (\ref{dmin}) and (\ref{solution}) are exact solutions of the quantum equations of motion (\ref{abzug2}) and (\ref{addi2}), without semiclassical or Markovian approximations. Here we are mainly interested in kinematic aspects, but they may also be used to study other quantum effects. If, for instance, there are several fields that carry an additional index (e.g. flavour), the definitions (\ref{DeltaminusDef}) and (\ref{DeltaplusDef}) are matrix valued and include correlations between fields of different flavour. Then the equations of motion (\ref{abzug2}) and (\ref{addi2}) are matrix equations and the self energies (which in general are non-diagonal in flavour space) can give rise flavour oscillations. The formalism presented here allows a full quantum description of these.
 
\paragraph{The Quasiparticle Approximation} Let $\hat{\Omega}_{\textbf{q}}$ be a pole of $\rho_{\textbf{q}}(\omega)$ with $\Omega_{\textbf{q}}={\rm Re}\hat{\Omega}_{\textbf{q}}$ and $\Gamma_{\textbf{q}}=2{\rm Im}\hat{\Omega}_{\textbf{q}}$. If the inequality
\begin{equation}\label{quasiparticle}
\Gamma_{\textbf{q}}\ll\Omega_{\textbf{q}}
\end{equation}
is fulfilled, the spectral density features a sharp peak at $\omega=\Omega_{\textbf{q}}$ that can be interpreted as a quasiparticle with well-defined energy and a dispersion relation given by $\Omega_{\textbf{q}}$. 
This interpretation requires, in addition to the condition (\ref{quasiparticle}), that the energies of all quasiparticles with the same quantum numbers are sufficiently well-separated ($\delta\Omega > \Gamma$) to be viewed as individual resonances\footnote{In \cite{Jakovac:2011vw} the spectral density has been used to define an effective number of degrees of freedom that takes the classical value if all resonances are separated and allows to interpolate into the regime of degenerate masses.}.
Then one can approximate
\begin{equation}\label{GammaApprox}
\Gamma_{\textbf{q}}\simeq-\frac{{\rm Im}\Pi^{R}_{\textbf{q}}(\Omega_{\textbf{q}})}{\Omega_{\textbf{q}}}
\end{equation}
and interpret $\Gamma_{\textbf{q}}$ as the width of the quasiparticle.
As the denominator of (\ref{spectralfunction2}) can be a complicated function of $T$ and the wave vector $\textbf{q}$, there can be poles in addition to those that remain in the limit $T\rightarrow 0$. They can be interpreted as collective phenomena or plasma waves. In a hot QED plasma of photons and electrons, for instance, there are propagating low momentum modes of longitudinal photons known as plasmons, whose dispersion relation differs from that of their transverse counterparts. There is also an additional fermionic excitation with negative helicity over chirality ratio \cite{Klimov:1981ka}, sometimes referred to as plasmino or hole.
As the following discussion will show, the origin of the poles (dressed particle or collective) is irrelevant for the kinematic properties of the corresponding quasiparticles\footnote{Different authors use the word \textit{quasiparticle} with different meanings. In some cases, it is used only for those resonances that correspond to screened one particle states, in contrast to collective excitations. Some authors also use it only if the dispersion relation is approximately parabolic. Here we do not distinguish these cases and refer to any resonance that is narrow in the sense of (\ref{quasiparticle}), thus has a well-defined dispersion relation, as quasiparticle.}. They are characterised by their dispersion relation $\Omega_{\textbf{q}}$ and width $\Gamma_{\textbf{q}}$, and these are the only relevant quantities for our purpose. Note also that the dispersion relation $\Omega_{\textbf{q}}$ is not necessarily similar to that of a free particle in vacuum or even parabolic. 
A temperature dependent effective mass can be defined via the curvature of $\Omega_{\textbf{q}}$ at its minimum
as a function of $|\textbf{q}|$\footnote{Alternatively an effective mass can be defined as the momentum independent piece of ${\rm Re}\Pi^{R}$ that comes from local diagrams, the plasma frequency $\Omega_{\textbf{q}=0}$ when the quasiparticle is at rest, the minimal possible value of $\Omega_{\textbf{q}}$ as a function of $|\textbf{q}|$ or by demanding $q^{2}|_{\omega=\Omega_{\textbf{q}}}=M_{\textbf{q}}^{2}$.}
\begin{equation}\label{massendef}
M=\left(\partial_{|\textbf{q}|}^{2} {\rm min}(\Omega_{\textbf{q}})\right)^{-1}
.\end{equation}
The dispersion relation is always approximately parabolic in the vicinity of the minimum, though the minimum in some cases (e.g. fermionic holes in QED) is not at $|\textbf{q}|=0$.
For notational simplicity we denote effective masses by capital letters $M$, $M_{i}$ without making the temperature dependence explicit, while small letters $m$, $m_{i}$ refer to intrinsic (vacuum) masses.

We will in the following refer to the situation $\omega=\Omega_{\textbf{q}}$ as \textit{real} or \textit{on-shell quasiparticles}.  It is understood that the ``mass shell'' for quasiparticles defined by $\Omega_{\textbf{q}}$ can have a complicated shape. Of course these particles are not ``real'' in the sense that they could leave the plasma as their properties are determined by the interaction with the omnipresent background. Consequently we refer to contributions from energies $\omega\neq\Omega_{\textbf{q}}$ as \textit{off-shell contributions} from \textit{virtual quasiparticles}. Finally, we refer to the situation when (\ref{quasiparticle}) is fulfilled and loop integrals sufficiently strongly dominated by energies $|\omega|\simeq\Omega_{\textbf{q}}$ as \textit{quasiparticle regime}. 

It is important to notice that the above was derived without reference to any asymptotic states.
Furthermore, the nonequilibrium propagators (\ref{dmin}) and (\ref{solution}) are consistent solutions of the nonequilibrium equations of motion. They do not suffer from the known problems of approaches to nonequilibrium field theory that are based on the ad-hoc replacement of the Bose-Einstein or Fermi-Dirac distributions in the expressions for equilibrium correlation functions by some arbitrary functions. In particular, there are no ``pinching singularities'' and secular terms.

\paragraph{Correlation Functions for Quasiparticles} As $\phi$ is very weakly coupled by assumption, the spectrum of its excitations is similar to those in vacuum and can be characterised by quasiparticles corresponding to screened $\phi$ particles. Then the integrals in (\ref{dmin}), (\ref{solution}) and (\ref{fieldvalue}) are dominated by the regions near the poles, where one can approximate $\rho_{\textbf{q}}(\omega)$ as
\begin{equation}\label{BreitWigner}
\rho_{\textbf{q}}(\omega)\sim 2\frac{\Gamma_{\textbf{q}}\omega}{\left(\omega^{2}-(\Omega_{\textbf{q}})^{2}\right)+(\Gamma_{\textbf{q}}\omega)^{2}}
,\end{equation}
with $\lim_{T\rightarrow 0}\Omega_{\textbf{q}}=\omega_{\textbf{q}}$. Using Cauchy's theorem one obtains\footnote{These formulae are valid up to corrections of higher order in $\Gamma/\Omega$. The poles of $f_{B}(\omega)$ do not give sizeable contributions.}
\begin{equation}\label{exponential}
\Delta^{-}_{\q}(\tm)
\simeq \frac{\sin(\Omega_{\q}\tm)}{\Omega_{\q}}e^{-\nicefrac{\Gamma_{\q}|\tm|}{2}}
,\end{equation}
\begin{eqnarray}
\lefteqn{\Delta^{+}(\tp;\tm)\simeq\frac{\Delta^{+}_{\q,\text{in}}}{2}
\left(\cos(2\Omega_{\q}\tp)+\cos(\Omega_{\q}\tm)\right)e^{-\Gamma_{\q}\tp}+\frac{\dot{\Delta}^{+}_{\q;\text{in}}}{\Omega_{\q}}
\sin(2\Omega_{\q}\tp)e^{-\Gamma_{\q}\tp}}\nonumber\\
&-&\frac{\ddot{\Delta}^{+}_{\q,\text{in}}}{2\Omega_{\q}^{2}}
\left(\cos(2\Omega_{\q}\tp)-\cos(\Omega_{\q}\tm)\right)e^{-\Gamma_{\q}\tp}+\frac{\coth(\frac{\beta\Omega_{\textbf{q}}}{2})}{2\Omega_{\q}}\cos(\Omega_{\q}\tm)\left(e^{-\nicefrac{\Gamma_{\q}|\tm|}{2}}-e^{-\Gamma_{\q}\tp}\right)\nonumber\\\label{narrowPlus}
\end{eqnarray}
and 
\begin{equation}\label{fieldvalueapprox}
\langle\phi_{\q}(\tp)\rangle
\simeq \dot{\phi}_{\q,\text{in}}\frac{\sin(\Omega_{\q}\tp)}{\Omega_{\q}}e^{-\nicefrac{\Gamma_{\q}\tp}{2}}+\phi_{\q,\text{in}}\cos(\Omega_{\q}\tp)e^{-\nicefrac{\Gamma_{\q}\tp}{2}}
.\end{equation}
Here we have introduced centre of mass and relative time coordinates $\tp=(t_{1}+t_{2})/2$, $\tm=t_{1}-t_{2}$ and used (\ref{PiKMS}) as well as ${\rm Im}\Pi^{R}_{\textbf{q}}(\omega)=\Pi^{-}_{\textbf{q}}(\omega)/(2i)$, which follows from the definitions (\ref{PiMinusDef}) and (\ref{PiAR}). 
It is easy to see that (\ref{narrowPlus}) does not coincide with the vacuum propagator in the limit $T\rightarrow 0$. This is because $T$ is the temperature of the bath, not of $\phi$, and the system may be prepared with arbitrary correlation functions for $\phi$ even if there is no thermal bath. This case was considered in \cite{Koksma:2009wa} in the context of decoherence.
 
The solution (\ref{narrowPlus}) also differs from the free scalar vacuum propagator in the decoupling limit $\Gamma_{\textbf{q}}\rightarrow 0$, $\Omega_{\textbf{q}}\rightarrow\omega_{\textbf{q}}$. At first sight it may be worrying that in this limit the correlation functions are not time translation invariant as the $\cos(2\Omega_{\textbf{q}}\tp)$ depends on the centre of mass time. However, $\Delta^{\pm}$ are not directly observable. The dependence of the (observable) energy (\ref{Energie}) disappears in the decoupling limit, see section \ref{comparisontoBE}. 
 The reason for the dependence on centre of mass time is that $\phi$ can be prepared with arbitrary initial correlations even in a free theory. This can be seen when computing $\langle\phi\phi\rangle={\rm Tr}(\varrho\phi\phi)$ using a basis $|n\rangle$ of eigenstates of the Hamiltonian ${\rm H}$ with eigenvalues $E_{n}$,
\begin{eqnarray}
{\rm Tr}(\varrho\phi(t_{1})\phi(t_{2}))=\sum_{n}\langle n|\varrho \phi(t_{1})\phi(t_{2})|n\rangle=\sum_{n,m}\langle n|\varrho|m\rangle\langle m| \phi(t_{1})\phi(t_{2})|n\rangle.\nonumber
\end{eqnarray}
Here $\varrho$ is the initial density matrix that characterises the ensemble.
If $\varrho$ is a functional of the Hamiltonian only, $\varrho=\mathcal{P}[{\rm H}]$, then $\langle n|\varrho|m\rangle=\mathcal{P}(E_{m})\delta_{nm}$ and only terms with $n=m$ contribute. Decomposing $\phi(t_{1})\phi(t_{2})$ into creation and annihilation operators $a_{\textbf{p}}$, $a_{\textbf{p}}^{\dagger}$ it is easy to see that for $n=m$ only the terms $a_{\textbf{p}}a_{\textbf{p}}^{\dagger}$, $a_{\textbf{p}}^{\dagger}a_{\textbf{p}}$, which come with factors $\exp(\pm i(t_{1}-t_{2})E_{n})$, can contribute. Thus, the correlator depends on $t_{1}-t_{2}$ only and is time translation invariant. If, on the other hand, the ensemble contains states that are not eigenstates of the Hamiltonian, terms with $n\neq m$ appear and the combinations $a_{\textbf{p}}a_{\textbf{p}}$, $a_{\textbf{p}}^{\dagger}a_{\textbf{p}}^{\dagger}$ contribute, which come with factors that depend on $t_{1}+t_{2}$. The choice of initial correlations corresponds to a choice of quantum states to be included in the thermodynamical ensemble under consideration. 
If one imposes equilibrium initial correlations for $\phi$ with a temperature that is equal to the bath temperature, (\ref{exponential}) and (\ref{narrowPlus}) become time translation invariant
\footnote{It has been discussed in \cite{Garny:2009ni} that it is impossible to prepare a system in equilibrium (and thus construct a time translation invariant solution to the Kadanoff-Baym equations) by specifying the two-point functions only. The reason is that equilibrium is not a Gaussian state and higher $n$-point functions contain connected pieces. These generally enter the two-point function via the self energies, thus setting only the two-point functions to their equilibrium values without adjusting all other connected $n$-point functions does not give a time translation invariant solution. Here, however, self energies are computed from correlation functions of bath fields only, which are computed from an exact equilibrium density matrix. Therefore a time translation invariant equilibrium solution can be constructed by specification of the one and two point functions only as long as the bath is sufficiently large that backreaction is negligible.}.

Perturbative computations in nonequilibrium quantum field theory are sometimes performed by using the equilibrium propagators (\ref{KMSscalar}) with the Bose-Einstein distribution $f_{B}(\omega)$ replaced by some general distribution function $f(\omega_\textbf{q})\neq f_{B}(\omega_{\textbf{q}})$,
\begin{eqnarray}\label{noneqapprox}
\begin{tabular}{c c c}
$\Delta^{-}\rightarrow i\rho_{\q}(\omega)$& , &$\Delta^{+}\rightarrow \left(\frac{1}{2}+f(\omega_\textbf{q})\right)\rho_{\q}(\omega)$\\
$\Delta^{<}\rightarrow f(\omega_\textbf{q})\rho_{\q}(\omega)$& , &$\Delta^{>}\rightarrow(1+f(\omega_\textbf{q}))\rho_{\q}(\omega)$.
\end{tabular}
\end{eqnarray}
It is clear that such approximation can only be valid on timescales much shorter than the relaxation time $\sim 1/\Gamma_{\textbf{q}}$ because otherwise the correlation functions \textit{have} to depend on the centre of mass time $t$. 
The fact that (\ref{narrowPlus}) is not time translation invariant in the limit $\Gamma_{\textbf{q}}\rightarrow 0$ means that even in that case (\ref{noneqapprox})  is not a consistent limit\footnote{One can of course always parametrise $\Delta^{+}\equiv(1/2+f)\rho$ etc. if one allows the function $f$ to have additional dependencies (including time and off-shell energies): one simply \textit{defines} $f_{\textbf{q}}(\omega,t)\equiv \Delta^{+}_{\textbf{q}}(\omega;t)\rho^{-1}_{\textbf{q}}(\omega;t)-1/2$.} though it may in some cases lead to approximately correct results.
In a more general nonequilibrium system (without the assumption of weak coupling to a large bath in equilibrium) the situation becomes even worse because also $\Delta^{-}(t_1,t_2)$ depends on the centre of mass time $(t_1+t_2)/2$\footnote{In \cite{Aarts:2001qa} a toy model was used to study numerically how the quasiparticle spectral function emerges dynamically in this case.}.  
 
\paragraph{The Relaxation Rate} Equations (\ref{exponential})-(\ref{fieldvalueapprox}) show that correlations of the field $\phi$ are damped exponentially with a rate $\Gamma_{\textbf{q}}$\footnote{The initial correlations in (\ref{narrowPlus}) are not damped with respect to $y$ because we are studying an initial value problem and have assumed that the interaction is switched on at the initial time, i.e. there were no correlations between $\phi$ and the bath at earlier times.}. Knowledge of the initial conditions is lost after a relaxation time $\sim\Gamma_{\textbf{q}}^{-1}$. Thus, $\Gamma_{\textbf{q}}=-{\rm Im}\Pi^{R}_{\textbf{q}}/\Omega_{\textbf{q}}$ can be interpreted as the relaxation rate in agreement with the well-known result from thermal field theory \cite{Weldon:1983jn}.
In general the relaxation rate is given by the discontinuity of the self energy, see appendix \ref{KBEessentials}, in this case 
\begin{eqnarray}
\Gamma_{\textbf{q}}=
-\frac{1}{2i\Omega_{\textbf{q}}}(\Pi_{\textbf{q}}^{>}(\Omega_{\textbf{q}})-\Pi^{<}_{\textbf{q}}(\Omega_{\textbf{q}}))=-\frac{1}{2i\Omega_{\textbf{q}}}\Pi_{\textbf{q}}^{-}(\Omega_{\textbf{q}}) \label{GammaDefPi}.
\end{eqnarray}
The identification of discontinuity and imaginary part that is required to derive (\ref{GammaApprox}) from (\ref{GammaDefPi}) is specific to the case of a real scalar with real couplings, see (\ref{ipa}).
The self energies $\Pi^{\gtrless}$ are defined analogue to $\Delta^{\gtrless}$, see appendix \ref{KBEessentials}. They allow to define the quantities
\begin{eqnarray}\label{GainAndLossRates}
\begin{tabular}{c c c}
$\Gamma^{<}_{\textbf{q}}=-\frac{1}{2i\Omega_{\textbf{q}}}\Pi^{<}_{\textbf{q}}(\Omega_{\textbf{q}})$ & & $\Gamma^{>}_{\textbf{q}}=-\frac{1}{2i\Omega_{\textbf{q}}}\Pi^{>}(\Omega_{\textbf{q}})$.
\end{tabular} 
\end{eqnarray}
$\Gamma_{\textbf{q}}^{<}$ can be interpreted as the \textit{gain rate} of thermal production of quasiparticles from a plasma and $\Gamma_{\textbf{q}}^{>}$ as the \textit{loss rate} due to the inverse processes. Their difference 
$\Gamma_{\textbf{q}}=\Gamma_{\textbf{q}}^{>}-\Gamma_{\textbf{q}}^{<}$
then gives the total relaxation rate for the mode $\textbf{q}$ through processes involving quasiparticles with energy $\omega=\Omega_{\textbf{q}}$.   
The rates $\Gamma^{\gtrless}$ fulfil the detailed balance ratio $\Gamma^{<}/\Gamma^{>}=\exp(-\beta\omega)$ as a result of the fact that the bath is in equilibrium, see (\ref{KMS1}), (\ref{PiKMS}). Note that this definition of the relaxation rate $\Gamma_{\textbf{q}}$ does not involve any asymptotic states, the definition of which is ambiguous in the omnipresent plasma.

The interpretation of $\Gamma_{\textbf{q}}$ as a relaxation rate, i.e. the rate at which the mode $\textbf{q}$ of $\phi$ exchanges energy with other modes, holds regardless of whether the initial occupation is below or above its equilibrium value. If the initial conditions in (\ref{solution}) are chosen such that occupation numbers are below their equilibrium values, thermal production of quanta from the plasma drives the relaxation while in the opposite case the excess of energy stored in $\phi$ is dissipated into the bath\footnote{The fact that $\Gamma^{>}>e^{-\beta\omega}\Gamma^{>}=\Gamma^{<}$ does not imply that $\phi$ always loses energy because $\Gamma^{\gtrless}$ are just the rates. The gain and loss of energy for each mode also depends on the occupation number, c.f. (\ref{boltz1})}. 
The relation (\ref{GammaDefPi}) is well-known from linear response theory \cite{LeB} and has also been derived from first principles for systems where the $\phi$ occupation numbers are small at all relevant times \cite{Asaka:2006rw}.
Note that in our setup this interpretation holds irrespective of the size of the deviation of $\phi$ from equilibrium as long as the bath has sufficiently many degrees of freedom. In the following we use (\ref{GammaDefPi}) as the definition of $\Gamma_{\textbf{q}}$. 

The fact that the background plasma is in equilibrium allows to use the techniques of thermal (equilibrium) field theory to compute the self energies. There exist two different formalisms in thermal field theory, known as real and imaginary time formalism \cite{LeB}. The latter, also known as Matsubara formalism \cite{Matsubara:1955ws}, is more commonly used. Here we employ the real time formalism, which is physically more transparent and can directly be integrated into the Schwinger-Keldysh formalism used to compute the nonequilibrium correlation functions for $\phi$ \cite{Chou:1984es}.
The reason is that in both formalisms correlation functions are derived from a generating functional for fields with time arguments on a contour in the complex time plane and Feynman rules for equilibrium fields are formulated analogue to those in the Keldysh formalism sketched in appendix \ref{KBEessentials}, with (\ref{KMSscalar}) inserted into (\ref{feynmanprop})-(\ref{kleinerinliste}).

\section{Relaxation of a Scalar}\label{scalarsectioN}
In this section we compute the relaxation rate for scalars with various interactions. 
We focus on the effects that medium related changes in the dispersion relations of particles and plasma waves in the bath have on $\Gamma_{\textbf{q}}$. 
We will leave the energy $\omega$ free wherever possible for the sake of generality, but we are mainly interested in the case $\omega=\Omega_{\textbf{q}}\simeq\omega_{\textbf{q}}$, where the second equality expresses that corrections to the dispersion relations of $\phi$-resonances are small due to the weak coupling (while they have to be taken account of for all other fields in the bath that is kept in equilibrium by stronger interactions).
To be specific we chose the Lagrangian 
\begin{eqnarray}\label{L}
\mathcal{L}&=& 
\frac{1}{2}\partial_{\mu}\phi\partial^{\mu}\phi-\frac{1}{2}m^{2}\phi^{2}
+ \sum_{i=1}^2 \left(\frac{1}{2}\partial_{\mu}\chi_{i}\partial^{\mu}\chi_{i}
-\frac{1}{2}m_{i}^{2}\chi_{i}^{2}+\bar{\Psi}_{i}\left(i\Slash{\partial}-\fermionmass{m}_{i}\right)\Psi_{i}\right)\nonumber\\
&&-g\phi\chi_{1}\chi_{2}-Y\phi\left(\bar{\Psi}_{1}\Psi_{2}+\bar{\Psi}_{2}\Psi_{1}\right)+\sum_{i=1}^{2}\left(-\frac{h_{i}}{4!}\phi\chi_{i}^{3}+\mathcal{L}_{\chi_{i} \mathrm{int}}+\mathcal{L}_{\Psi_{i} \mathrm{int}}\right) 
.\end{eqnarray}
Here $\phi$ is the out-of-equilibrium field. The $\chi_{i}$ are scalar and the $\Psi_{i}$ fermionic fields that are part of the thermal bath. 
The bath is thought to contain many more fields $\mathcal{X}_{i}$ to which the $\chi_{i}$ and $\Psi_{i}$ couple via interactions contained in the terms $\mathcal{L}_{\chi_{i} \mathrm{int}}$, $\mathcal{L}_{\Psi_{i} \mathrm{int}}$. Their particular form is of no relevance at this point, they are only assumed to be strong enough to restore equilibrium in the bath fast on the time scale associated to the dynamics of $\phi$. Backreaction can be neglected if the bath is sufficiently large.
The Lagrangian (\ref{L}) may be regarded as a toy model, but it allows to study all relevant kinematic effects while avoiding the complications of gauge theories at finite temperature.
\begin{figure}
  \centering
 \psfrag{a}{$a)$}
  \psfrag{b}{$b)$}
  \psfrag{c}{$c)$}
  \psfrag{d}{$d)$}
  \psfrag{e}{$e)$}
  \psfrag{f}{$f)$}
    \includegraphics[width=12cm]{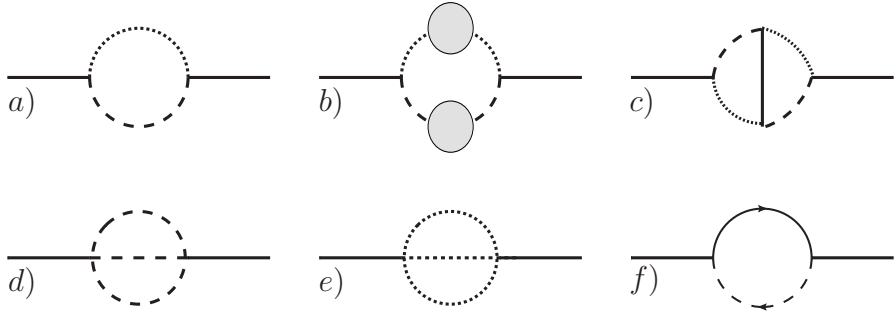}
    \caption{Feynman diagrams contributing to $\Gamma_{\textbf{q}}$, the lines represent $\phi$ (solid), $\chi_{1}$ (dashed), $\chi_{2}$ (dotted), $\Psi_{1}$ (solid with arrow) and $\Psi_{2}$ (dashed with arrow), the gray circles symbolise self energy insertions for the $\chi_{i}$ due to some interactions with themselves or other fields in the plasma.\label{diagrams}}
\end{figure}
\subsection{Trilinear Scalar Interaction}\label{trilinearsection}
We first consider the trilinear coupling $g\phi\chi_{1}\chi_{2}$ in (\ref{L}). 
The self energy is then given by the diagram shown in figure \ref{diagrams}a).
We use the fact the $\Pi^{-}$ can be expressed in terms of $\Pi^{<}$, which in the real time formalism of thermal field theory at leading order is given by a single diagram (combination of contour indices).
${\rm Im}\Pi^{R}$ to leading order is given by \cite{LeB}
\begin{eqnarray}\label{loopintegral}
{\rm Im}\Pi_{\textbf{q}}^{R}(\omega)=\frac{1}{2i}\Pi^{-}_{\textbf{q}}(\omega)=\frac{1}{2i}f_{B}^{-1}(\omega)\Pi^{<}_{\textbf{q}}(\omega)\nonumber\\
=-\frac{g^{2}}{2}f_{B}^{-1}(\omega)\int\frac{d^{4}p}{(2\pi)^{4}}\Delta^{<}_{1 \textbf{p}}(p_{0})\Delta^{ >}_{2 \textbf{p}-\textbf{q}}(p_{0}-\omega)\nonumber\\
=-\frac{g^{2}}{2}f_{B}^{-1}(\omega)\int\frac{d^{4}p}{(2\pi)^{4}}\Delta^{<}_{1 \textbf{p}}(p_{0})\Delta^{ <}_{2 \textbf{p}-\textbf{q}}(\omega-p_{0})\nonumber\\
=-\frac{g^{2}}{2}\int\frac{d^{4}p}{(2\pi)^{4}}\left(1+f(p_{0})+f(\omega-p_{0})\right)\rho_{1 \textbf{p}}(p_{0})\rho_{2 \textbf{p}-\textbf{q}}(\omega-p_{0})
.\end{eqnarray}
Here $f_{B}$ is the Bose-Einstein distribution $f_{B}(\omega)=(e^{\beta\omega}-1)^{-1}$ and $\Delta^{<}_{i \textbf{p}}$ are the spacial Fourier transforms of the connected thermal Wightman functions $\langle\chi_{i}(x_{2})\chi_{i}(x_{1})\rangle_{c}$ defined analogue to (\ref{forw}). By means of (\ref{KMSscalar}) it can be written as
\begin{equation}\label{DeltaKleiner}
\Delta_{i \textbf{p}}^{<}(p_{0})=f_{B}(p_{0})\rho_{i \textbf{p}}(p_{0})
.\end{equation}
We have also used the identity $f_{B}(a)f_{B}(b-a)(f_{B}(b))^{-1}=1+f_{B}(a)+f_{B}(b-a)$.
The spectral densities $\rho_{i \textbf{p}}(p_{0})$ for the fields $\chi_{i}$ are defined analogue to (\ref{spectralfunction}). 
In the remainder of this section quantities that carry an index $_{i}$ (such as $m_{i}$, $M_{i}$, $\rho_{i \textbf{p}}(p_{0})\ldots$) are always associated with the bath fields $\chi_{i}$ while quantities without such index ($m$, $M$, $\rho_{\textbf{q}}(\omega)\ldots$) refer to $\phi$.
Equation (\ref{loopintegral}) shows that the quantities that determine $\Gamma_{\textbf{q}}$ are the spectral densities of the bath fields $\chi_{i}$. 
\subsubsection{Relaxation Rate at leading Order}
A leading order result for $\Gamma_{\textbf{q}}$ can be obtained by inserting free spectral densities,  
\begin{equation}\label{freespectral}
\rho_{i \textbf{p}}^{{\rm free}}(p_{0})=2\pi{\rm sign}(p_{0})\delta(p^{2}-m_{i}^{2}),
\end{equation}
into (\ref{loopintegral}). Equation (\ref{freespectral}) can easily be found from (\ref{spectralfunction}) in the decoupling limit. Then (\ref{loopintegral}) reads
\begin{eqnarray}\label{DiplomFormula}
\lefteqn{{\rm Im}\Pi^{R}_{\textbf{q}}(\omega)=-\frac{g^{2}}{2}\int\frac{d^{3}p}{(2\pi)^{2}}\frac{1}{4\omega_{2}\omega_{1}}}\nonumber\\
&\times&\bigg(\Big(\big(f_{1}+1\big)\big(f_{2}+1\big)-f_{1}f_{2}\Big)\Big(\delta(\omega-\omega_1-\omega_2)-\delta(\omega+\omega_1+\omega_2)\Big)\nonumber\\
&&+\phantom{X}\Big(\big(f_1+1\big)f_2-\big(f_2+1\big)f_1\Big)\Big(\delta(\omega-\omega_1+\omega_2)-\delta(\omega+\omega_1-\omega_2\Big)\bigg)
.\end{eqnarray}
Here $\omega_{1}=(\textbf{p}^{2}+m_{1}^{2})^{1/2}$, $\omega_{2}=((\textbf{q}-\textbf{p})^{2}+m_{2}^{2})^{1/2}$ and $f_i=f_B(\omega_i)$. The well-known expression (\ref{DiplomFormula}) has a clear physical interpretation \cite{Weldon:1983jn}. 
\begin{figure}
  \centering
 \psfrag{a}{$a)$}
 \psfrag{b}{$b)$}
 \psfrag{c}{$c)$}
 \psfrag{d}{$d)$}
    \includegraphics[width=10cm]{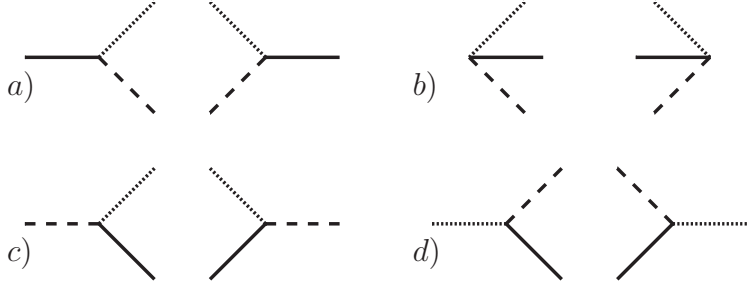}
    \caption{Diagrams obtained from cuts through the diagram in figure \ref{diagrams}a).\label{processes}}
\end{figure}
The first $\delta$-function in the first line corresponds to decays and inverse decays $\phi\leftrightarrow\chi_{1}\chi_{2}$, see figure \ref{processes}a). Looking at the statistical factor, one can easily confirm that the detailed balance relation is fulfilled, a consequence of the fact that the bath is in equilibrium. This factor leads to an amplification of the rate compared to the vacuum value due to induced transitions, typical for bosons. 
The $\delta$-function implies energy conservation\footnote{The quasiparticle momentum is conserved exactly at all orders because we restricted the analysis to systems that are invariant under spacial translations.}. The second $\delta$-function in the first line corresponds to the creation of real $\chi_{i}$ and a virtual $\phi$ quantum with negative energy from the vacuum, see figure \ref{processes}b). It comes with the same statistical factor and conserves energy as well, but never contributes in the interesting case $\omega=\Omega_{\textbf{q}}$. 
The second line corresponds to processes of the type $\chi_{i}\leftrightarrow\chi_{j}\phi$, i.e. emission or absorbtion of a $\phi$ quantum by the bath, see figure \ref{processes}c) and d). Again, detailed balance as well as energy conservation remain valid. This contribution involves interactions with real quanta from the bath and leads to a relaxation mechanism similar to Landau damping\footnote{Traditionally the term \textit{Landau damping} refers to the interaction of a particle with a wave or mean field. In the following we use this term more generally for any process in which $\phi$ interacts with the background, i.e. that has quanta of fields in the bath (other than $\phi$) in the initial state.}. In the vacuum limit $T\rightarrow 0$ the second line vanishes, and the first line reduces to the known decay rate in vacuum. 
The integral in (\ref{DiplomFormula}) can be solved analytically \cite{Boyanovsky:2004dj,Diplomarbeit} and gives 
\begin{eqnarray}\label{dampingformel}
{\rm Im}\Pi^{R}_{\textbf{q}}(\omega)=\sigma_{0}(q)+\sigma_{a}(q,T)+\sigma_{b}(q,T).
\end{eqnarray}
Here $\sigma_{0}$ is the contribution due to the decay process $\phi\rightarrow\chi_{1}\chi_{2}$, 
\begin{align}\label{sigma0}
\sigma_{0}(q) = &-\frac{g^{2}}{16\pi q^{2}}\text{sign}(\omega)
\theta(q^{2}-(m_{1}+m_{2})^{2})\nonumber\\
&\times \left((q^{2})^{2}-2q^{2}(m_{1}^{2}+m_{2}^{2})
+(m_{1}^{2}-m_{2}^{2})^{2}\right)^{\frac{1}{2}}\ ,
\end{align}
$\sigma_{a}(q,T)$ is an additional temperature dependent contribution from such processes that includes induced transitions and inverse decays,
\begin{align}\label{sigmaa}
\sigma_{a}(q,T) = &-\frac{g^{2}}{16\pi|\q|\beta}
\text{sign}(\omega) \theta(q^{2}-(m_{1}+m_{2})^{2})\nonumber\\
&\times \left(\ln\left(\frac{1-e^{-\beta\omega_{+}}}{1-e^{-\beta\omega_{-}}}
\right)+ (m_{1}\leftrightarrow m_{2})\right)\ ,
\end{align}
and $\sigma_{b}(q)$ is the contribution from Landau damping
\begin{align}\label{sigmab}
\sigma_{b}(q,T)= &-\frac{g^{2}}{16\pi|\q|\beta}
\text{sign}(\omega) \theta((m_{1}-m_{2})^{2}-q^{2})\nonumber\\
&\times\left(\ln\left(\frac{1-e^{-\beta|\omega_{-}|}}{1-e^{-\beta|\omega_{+}|}}
\right)+(m_{1}\leftrightarrow m_{2})\right)\ 
.\end{align}
We have used the abbreviations
\begin{equation}
\omega_{\pm} = \frac{|\omega|}{2q^{2}}(q^{2}+m_{1}^{2}-m_{2}^{2})
\pm\frac{|\textbf{q}|}{2|q^{2}|}\left( 
(q^{2}+m_{1}^{2}-m_{2}^{2})^{2}-4q^{2}m_{1}^{2}\right)^{\frac{1}{2}}\ ,
\end{equation}
where $q$ is the four vector $q=(\omega,\textbf{q})$.
Kinematic restrictions known from vacuum remain valid, energy and momentum are conserved in individual processes. The differences to the vacuum lie in the statistical factors and the fact that additional processes occur. The plasma can provide real $\chi_{i}$ quanta in the initial state that make inverse decays and Landau damping possible. 

\subsubsection{Resummation}\label{resummsubsub}
The energy conserving $\delta$-functions originate from the use of free spectral densities (\ref{freespectral}). They define submanifolds in the integration volume of (\ref{loopintegral}) in which the factors $\Delta^{<}_{1 \textbf{p}}(p_{0})$ and $\Delta^{ <}_{2 \textbf{p}-\textbf{q}}(\omega-p_{0})$ are non-zero, the mass shells of $\chi_{i}$ particles. Only if those intersect at some point in the integration volume, the integrand is non-zero. In vacuum this is well-known from the optical theorem and the cutting rules it implies \cite{Cutkosky:1960sp}. These imply that $\Pi^{-}$ is non-zero only if all particles in the loop integral can be on-shell somewhere in the integration volume. 
The finite temperature generalisation of these rules \cite{Bedaque:1996af}\footnote{The original formulation of the finite temperature \textit{circling rules} \cite{Kobes:1985kc} did not make this interpretation in terms of cuts and products of amplitudes obvious.} loosely state the following:
Cut the diagram in all possible ways and consider all possible amplitudes that can be built from the pieces by interpreting their external legs as initial and final state particles. 
Cut propagators are replaced by $\Delta^{>}$ or $\Delta^{<}$, depending on the momentum flow (for fermions $S^{\gtrless}$). The spectral densities appearing in $\Delta^{\gtrless}$, see (\ref{KMSscalar}), ensure that the amplitudes only contribute if all external particles can be on-shell\footnote{So called \textit{pinching singularities} in the single amplitudes, see e.g. \cite{Gelis:1997zv}, are regularised as we in the following argue that a consistent treatment requires resummation of all propagators.}, while the statistical factors lead to factors $f_{B}$ for initial and $1+f_{B}$ for final state particles.
The former ensures that in the limit $T\rightarrow 0$ contributions from processes with particles provided by the heat bath in the initial state vanish, the latter reflects the effect of induced transitions for bosons (for fermions there is a suppressing factor $1-f_{F}$, see (\ref{KMSfermi})).
Therefore, as in vacuum, only those cuts through the original diagram contribute for which all cut propagators are on-shell simultaneously somewhere in the integration volume.
The additional contributions have an intuitive physical interpretation as processes in which the external particle of the original diagram engages in reactions with real quanta from the plasma.

\begin{figure}
  \centering
 \psfrag{supset}{$\supset$}
  \psfrag{b}{$b)$}
  \psfrag{c}{$c)$}
  \psfrag{d}{$d)$}
  \psfrag{e}{$e)$}
  \psfrag{f}{$f)$}
    \includegraphics[width=12cm]{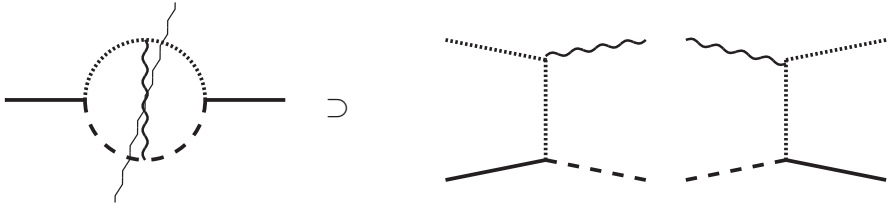}
    \caption{An example for a ladder diagram, the lines represent $\phi$ (solid), $\chi_{1}$ (dashed), $\chi_{2}$ (dotted, small spacing) and some other unspecified field $\mathcal{X}$ (wavy line) in the bath.\label{ladderdiagram}}
\end{figure}

Equation (\ref{DiplomFormula}) only accounts for the effect of the plasma on $\phi$, but not on the $\chi_{i}$. This is inconsistent as we assumed that interactions $\mathcal{L}_{\chi_{i} \mathrm{int}}$ are much stronger than the coupling of the $\chi_{i}$ to $\phi$. 
Their effect can be included by using resummed propagators in (\ref{loopintegral}), which corresponds to inserting dressed spectral functions of the form (\ref{spectralfunction})\footnote{The need to employ resummed propagators to obtain consistent results is common in finite temperature field theory\cite{Braaten:1989mz,Landau:1953um,LeB,Kraemmer:2003gd,Berges:2004yj,Anisimov:2010gy}.}. Note that this procedure, though being a consistent resummation of the propagators, does not take into account all possible contributions to the self energy. It e.g. neglects final state interactions coming from ladder diagrams as shown in figure \ref{ladderdiagram}. In the following focus on the case that diagrams of the type shown in figure \ref{diagrams}b) dominate the self energy in some controllable approximation.  
The quantitative validity of this approach depends on the details of the interactions that keep the bath in equilibrium. It is e.g. justified when there are either many more or stronger interactions giving rise to diagrams of this type. We believe that most of our conclusions qualitatively hold when this restriction is lifted\footnote{This is e.g. confirmed by the results found in \cite{Anisimov:2010gy} while this work was in progress.}, but computations become more cumbersome and a widely analytic treatment as we will perform in the following in general is not possible\footnote{If one employs $n$PI techniques as e.g. discussed in \cite{Berges:2004yj} all contributions at a given order are automatically taken into account. Here we perform the resummation ``by hand'' in order to make maximal use of the simplifications due to the weak coupling to a large thermal bath. A more general, but less transparent resummation scheme has also been proposed in \ref{Besak:2010fb}.}

\paragraph{Quasiparticle Regime} The dressed spectral densities $\rho_{i \textbf{p}}(p_{0})$ for the $\chi_{i}$ can be complicated in general as we have not specified their interactions. Let us for a moment assume that all poles fulfil (\ref{quasiparticle}). Then the $\rho_{i \textbf{p}}(p_{0})$ are characterised by a number of narrow peaks at energies $\omega=\pm\Omega_{i \textbf{p}}^{I}$, corresponding to quasiparticles with dispersion relation $\Omega_{i \textbf{p}}^{I}$ and width $\Gamma_{i \textbf{p}}^{I}$. Here the index $i$ indicates that we refer to a pole of the spectral density $\rho_{i \textbf{p}}(\omega)$ of $\chi_{i}$ and the index $I$ labels the various poles. 

Near each pole the spectral density can be approximated by a Breit Wigner function in the usual way,
\begin{displaymath}
\rho_{i \textbf{p}}(\omega)\sim \sum_{I}2\frac{\Gamma_{i \textbf{p}}^{I}\omega}{\left(\omega^{2}-(\Omega_{i \textbf{p}}^{I})^{2}\right)+(\Gamma_{i \textbf{p}}^{I}\omega)^{2}}
.\end{displaymath} 
In the limit of vanishing widths $\Gamma_{i \textbf{p}}^{I}$, this converges to a sum of $\delta$-functions of the form $\sim\delta(\omega^{2}-(\Omega_{i \textbf{p}}^{I})^{2})$. Then the previous argument can be repeated, the $\delta$-functions define submanifolds in the integration volume of (\ref{loopintegral}) on which the integrand is non-zero. Only if these submanifolds intersect somewhere in the integration volume, there is a contribution to the integral. This can be interpreted as energy and momentum conservation in scatterings and decays of real quasiparticles.
Note that, as pointed out previously, this argument is independent on whether the quasiparticles are screened particles or collective excitations. 

In the simplest case, when the only poles of $\rho_{i \textbf{p}}(p_{0})$ are $\pm(\textbf{p}^{2}+M_{i}^{2})^{1/2}$, $\Gamma_{\textbf{q}}$ is given by (\ref{dampingformel})-(\ref{sigmab}) with intrinsic masses replaced by thermal masses, $m_{i}\rightarrow M_{i}$. This allows to define two critical temperatures $T_{c}$ and $\tilde{T}_{c}$ as the solutions to
\begin{eqnarray}
0&=&\left(M-M_{1}-M_{2}\right)_{T=T_{c}}\\
0&=&\left(M-|M_{1}-M_{2}|\right)_{T=\tilde{T}_{c}}
.\end{eqnarray}
For $T<T_{c}$, (\ref{sigma0}) and (\ref{sigmaa}) contribute to $\Gamma_{\textbf{q}}$ (decays and inverse decays $\phi\leftrightarrow\chi_{1}\chi_{2}$). 
We refer to this as case (a).
For $T>\tilde{T}_{c}$, $\Gamma_{\textbf{q}}$ is given by (\ref{sigmab}) due to processes $\phi\chi_{i}\leftrightarrow\chi_{j}$, case (b). 
For $T_{c}<T<\tilde{T}_{c}$ none of these processes is allowed and $\Gamma_{\textbf{q}}$ vanishes in this approximation, case (c).
The generalisation of the definitions $T_{c}$ and $\tilde{T}_{c}$ and the cases (a), (b) and (c) to situations with more complicated dispersion relations is straightforward and we will in the following use them in this generalised sense.

When inserting the full spectral densities including continuum parts and finite widths, the integrand in (\ref{loopintegral}) is non-zero everywhere in the integration volume. However, as long as (\ref{quasiparticle}) remains fulfilled for all poles, $\Gamma_{i \textbf{p}}^{I}\ll\Omega_{i \textbf{p}}^{I}$, the integral is strongly dominated by the regions in which all quasiparticles are on-shell\footnote{Note that the loop integral in this case is \textit{not} dominated by momenta $|\textbf{p}|\sim T$, but by the on-shell regions.}. The region in the integration volume where $\rho_{i \textbf{p}}(p_{0})$ peaks extends only a small distance $\sim\Gamma_{i \textbf{p}}^{I}$ away from the quasiparticle mass shells $p_{0}^{2}=(\Omega^{I}_{i \textbf{p}})^{2}$ etc. 
Contributions from regions where one or more quasiparticles are off-shell are suppressed, the suppression is stronger in regions where more quasiparticles are off-shell.
Hence, the integral (\ref{loopintegral}) only receives a large contribution if these quasiparticle mass-shells intersect or get very close to each other. Otherwise it is non-zero, but only receives contributions from regions in the phase space where at least one of the quasiparticles in the loop is off-shell\footnote{If there are several collective resonances, the line in the loop can represent a propagator for any of these. The decay into holes has e.g. been studied in \cite{Kiessig:2010pr}.}. The suppression is analogue to the suppression of processes in vacuum that involve virtual intermediate states which are much heavier than the energy of the process. 

In the case discussed here, the relaxation of a single real scalar field, there is only one relaxation time scale $\sim 1/\Gamma_{\textbf{q}}$ per mode. In more complicated systems there can be several time scales related to kinetic equilibration of different species, effective chemical potentials or quantum coherences (i.e. correlations between different fields). In \cite{Lindner:2005kv} it was found that the on-shell approximation can lead to ``spuriously conserved quantities'', i.e. quantities that appear to be conserved though there is no corresponding symmetry in the Lagrangian. This is analogue to the apparent lack of relaxation in regime (c) observed in the on-shell approximation here. In \cite{Garbrecht:2008cb} it was pointed out that the ``spuriously conserved quantities'' disappear when the on-shell approximation is lifted. They do, however, relax on time scales much longer than the kinetic equilibration because the processes that violate the conservation involve off-shell quasiparticles. This is in complete analogy to the suppressed relaxation rate in regime (c) in our case.

The fact that the dispersion relations change with $T$ implies the phase space for various processes is dynamical and can open or close at critical temperatures. (\ref{dampingformel}) shows that this may happen rather abruptly leading to the ``thresholds'' along the temperature axis that are visible in figure \ref{ExtraBreit}. This can lead to interesting dynamics as the temperature itself can be affected by the relaxation processes (see e.g. the example discussed in section \ref{InflatonSection}).

\paragraph{Broad Resonances} If, in contrast, the widths of the $\chi_{i}$ are large, there can be a significant overlap of the spectral densities in (\ref{loopintegral}) even far away from the on-shell regions. 
Figure \ref{ExtraBreit} shows that in this case the thresholds get smeared out. $\Gamma_{\textbf{q}}$ changes rather continuously with temperature due to the smooth shape of the broad resonances. There is no critical temperature at which $\Gamma_{\textbf{q}}$ changes abruptly. 
 
A resonance that has a narrow width at temperatures well below its mass may become broad and loose its identity at hight $T$. This phenomenon, known as \textit{melting of a peak}, is theoretically well-studied and has been observed experimentally \cite{Arnaldi:2006jq} for mesons. Quantitative computations in the high temperature regime are difficult due to the poor convergence of the perturbative series even in weakly coupled theories. They usually rely on resummation methods, lattice computations or effective field theories of lower dimensionality. Recently, also the use of gravity-field theory dualities has been explored.
 
Physically the broadening of meson resonances is of course related to the approach to the QCD crossover and the dissociation of the meson states. In theories of fundamental particles, where broadening can also occur, it means that the quasiparticle cannot be viewed as an entity that is well-separated from its environment. It is taken off-shell by the statistical fluctuations of its energy due to the interactions with the dense medium. The apparent energy non-conservation in the quasiparticle decay is not surprising as only the combined energy of system and environment is conserved. From a quantum viewpoint this means that processes generally involve many quanta.
From the previous considerations and figure \ref{ExtraBreit} it is obvious that also the opposite can happen: The width of $\phi$ suddenly becomes more narrow at $T_{c}$. 

\subsubsection{Analytic Structure of the Spectral Density}\label{analsection}
\begin{figure}
  \centering
    \includegraphics[width=12cm]{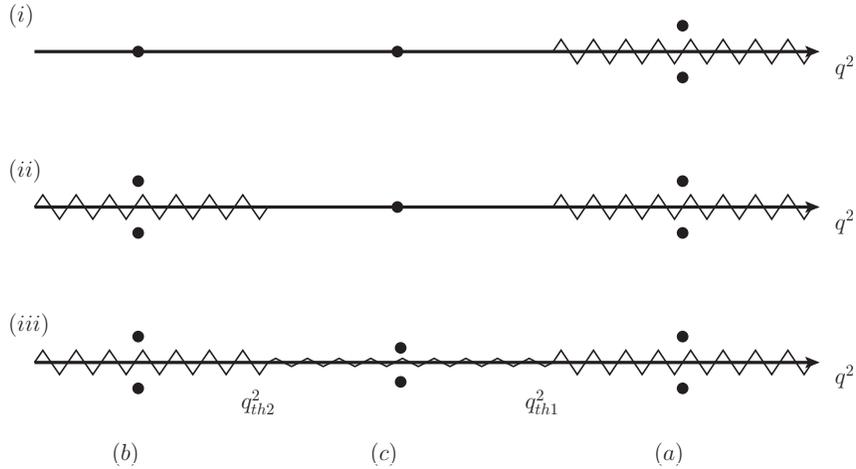}
    \caption{Analytic structure of the spectral density of $\phi$ as a function of $q^{2}$ for a self energy from trilinear interactions (i) in vacuum, (ii) at finite temperature with bare propagators and  (iii) at finite temperature after resummation. The zigzag lines represent discontinuities and the dots the positions of poles in the cases (a) $Q^{2}>q_{th1}^{2}$, (b) $Q^{2}<q_{th2}^{2}$ and (c) $q_{th2}^{2}<Q^{2}<q_{th1}^{2}$, where $Q=(\Omega_{\textbf{q}},\textbf{q})$ is the on-shell energy-momentum four vector.\label{analyticstructure}}
\end{figure}
The different kinematical constraints can be studied in terms of the analytic structure of the spectral density as a function of squared four momentum, schematically shown in figure \ref{analyticstructure}. 

\paragraph{Leading Order} In vacuum $\rho_{\textbf{q}}(\omega)$ has a continuous part above the lowest multiparticle threshold $q_{th1}^{2}=(m_{1}+m_{2})^{2}$. In this region it inherits the discontinuity of the self energy across the real axis given by (\ref{sigma0}). This discontinuity can be computed by cutting the diagram in figure \ref{diagrams}a). Poles that lie below $q_{th1}$ lead to singular, $\delta$-function shaped contributions that are interpreted as stable particles and possibly bound states. 

In a medium the diagram in figure  \ref{diagrams}a) also contains the processes $\phi\chi_{i}\leftrightarrow\chi_{j}$, leading to an additional discontinuity below a threshold $q_{th2}^{2}$. In the simplest case, when the effect of the medium on the $\chi_{i}$ dispersion relation can be parametrised by momentum independent thermal masses, it is given by $q_{th2}^{2}=(M_1-M_2)^2$, see (\ref{sigmab})\footnote{Note the difference to \cite{Weldon:1983jn}, where this discontinuity was found to vanish for $q^{2}<-|m_{1}^{2}-m_{2}^{2}|$.}.  
The discontinuity dresses any poles below $q_{th2}$ with a finite width. Defining the four vector $Q=(\Omega_{\textbf{q}},\textbf{q})$ one can distinguish the cases (a), (b) and (c). In case (a) the position of the pole $\Omega_{\textbf{q}}$ is such that $Q^{2}>q_{th1}^{2}$ and relaxation is driven by decay, see (\ref{sigma0}) and (\ref{sigmaa}). In case (b) $Q^{2}<q_{th2}^{2}$ dissipation is driven by Landau damping, see (\ref{sigmab}).  In between, there is a region $q_{th2}^{2}<Q^{2}<q_{th1}^{2}$ in which none of these processes is possible for real quasiparticles, case (c). A similar structure is found in solid states, where it determines the conductivity. A pole in region (c) leads to quasiparticles that can move freely over large distances. 
It depends on $T$ which of these cases is realised. 
The critical temperatures $T_{c}$ and $\tilde{T}_{c}$ mark the transitions from (a) to (c) and from (c) to (b), respectively.
The three temperature regimes are clearly visible in figure \ref{ExtraBreit}.

\begin{figure}
  \centering
 \psfrag{supset}{$\supset$}
  \psfrag{b}{$b)$}
  \psfrag{a}{$a)$}
    \includegraphics[width=12cm]{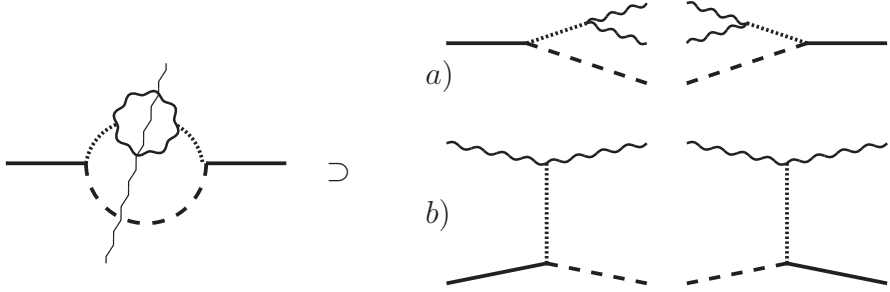}
    \caption{An example for diagrams obtained by cutting the diagram in figure \ref{diagrams}b) if $\chi_{2}$ couples to an unspecified field $\mathcal{X}$ in the bath, lines represent $\phi$ (solid), $\chi_{1}$ (dashed), $\chi_{2}$ (dotted) and $\mathcal{X}$ (wavy) propagators.\label{cutdiagram}}
\end{figure}

\paragraph{Multiparticle Final States and Scatterings} Cutting the resummed diagram in figure \ref{diagrams}b) through the $\chi_{i}$ self energy insertions leaves pieces that correspond to the decay of $\phi$ into multiparticle final states that are composed of whatever particles $\mathcal{X}_{i}$ appear in those insertions, and the inverse processes, see figure \ref{cutdiagram}a). The $\chi_{i}$ appear as intermediate state. These decays contribute if they are kinematically allowed, but they are suppressed by additional factors of the involved coupling constant. In the case (c), when the intermediate $\chi_{i}$ cannot be on-shell, they are also suppressed by the smallness of the $\chi_{i}$ width compared to its energy, which corresponds exactly to the off-shell suppression previously discussed. 

In addition, the diagram in figure \ref{diagrams}b) also contains scattering processes with $\mathcal{X}_{i}$ in the initial and final state, see figure \ref{cutdiagram}b).
In contrast to the decays, they are kinematically always possible for some momenta, thus there is always a discontinuity along the entire real axis.

We should add that the diagram in figure \ref{diagrams}c) contains scatterings of the type $\chi_{i}\phi\leftrightarrow\chi_{j}\phi$. These processes do not change the "number" of $\phi$ quasiparticles, but exchange energy between $\phi$ and the bath. We have previously omitted this diagram (and will do so in the following) as it is a ``backreaction'' term that is suppressed by two additional powers of the weak coupling $g$ and the large number of degrees of freedom in the bath.

In general there are also ladder diagrams as e.g. shown in figure \ref{ladderdiagram}, which involve other fields than $\phi$. They contain vertex corrections as well as contributions from scatterings and are more important than the diagram shown in figure \ref{diagrams}c) because the interactions in the bath are stronger than $g$ and there may be various couplings that give rise to diagrams of this type. 
Here we consider the case when contributions of the type shown in figure \ref{diagrams}b) dominate the self energy, as discussed at the end of the first part of section \ref{resummsubsub}. 
This allows to parametrise the effects of all higher order processes involving more quanta in the initial and final state in the quasiparticle widths of the $\chi_{i}$\footnote{The off-shell suppression observed in \cite{Garbrecht:2008cb} that leads to a delayed relaxation of quantities which spuriously conserved in the on-shell limit can also be interpreted in this manner: These quantities relax only via scatterings or decays with multiple quanta in the final state, which can be extracted from cuts through the dressed $\chi_{i}$ propagators.}. The fact that these processes can be included without reference to asymptotic states in the omnipresent plasma is one of the benefits of the formalism employed here.

\paragraph{Multiple Scatterings} All diagrams that contribute to scatterings contain additional vertices.
Furthermore, in case (c) at least one of the intermediate propagators must be off-shell, leading an additional suppression of $\Gamma_{\textbf{q}}$ between $T_{c}<T<\tilde{T}_{c}$.
At low temperatures scatterings are subdominant with respect to the decays. At at high temperature the small coupling suppression is compensated by the large occupation numbers\footnote{This is a physical reason for the well-known problem of poor convergence of the perturbative expansion in the coupling at high temperatures.} and the scatterings become increasingly important. 
Yet, there is another subtlety. The suppression of diagrams involving more vertices can be compensated even in region (c) if the involved quanta are collinear and almost on-shell. 
Interferences between subsequent scatterings are not negligible, similar to the Landau-Pomeranchuk-Migdal effect \cite{Landau:1953um}, see also \cite{Aurenche:2000gf,Arnold:2001ba,Gagnon:2006hi}. 
This becomes increasingly important when the temperature is larger than all involved masses and therefore affects the regions (b) and (c).  
Physically it means that the relaxation is not driven by processes involving only a handful of quasiparticles, but a large number of quanta from the bath. Intuitively it is clear that multiple scatterings should become more frequent at high temperature: even a small coupling leads to a large scattering probability if the density of scattering partners becomes sufficiently high. 
In this regime our approximation that the self energy is dominated by diagrams of the type shown in figure \ref{diagrams}b) generally becomes increasingly bad above some temperature even if it is well-justified at low $T$.
A consistent scheme to resum contributions from processes involving many soft collinear quanta has been presented in \cite{Besak:2010fb}. While this work was in progress it was applied in \cite{Anisimov:2010gy} and found that these effects can not only lead to an increase of the rate in region (b) by a factor of order one, but also almost completely overcome the suppression in region (c). This is in agreement with our results shown in figure \ref{ExtraBreit}.

The increasing importance of scatterings with temperature is of course known in the Boltzmann approach, but we would like to emphasise that the field theoretical considerations show that for a consistent treatment at temperatures larger than the masses, the use of resummed perturbation theory is necessary. This corresponds to a summation of infinitely many diagrams with large numbers of particles in the initial and final state in the Boltzmann approach \cite{FillionGourdeau:2007zz}.

It follows from the above that there can be no stable excitations in a plasma. Of course this is expected because it is clear that a particle propagating in a medium can always engage in scattering processes. If this was not the case, one could easily chose the parameters of the Lagrangian (\ref{L}) in a way that $\phi$ in some temperature regime does not exchange energy with its environment and relax despite their coupling and the possibility of scatterings. However, in case (c) the exchanged $\chi_{i}$ is off-shell and the rate for such processes suppressed in the quasiparticle regime, thus $\phi$ can be long lived and stable on the relevant timescale. In our setup, where ladder diagrams are neglected, all higher order contributions are encoded in the energy and momentum dependent finite widths of the $\chi_{i}$. Therefore the deviation from the leading order contribution from decays and inverse decays of single quasiparticles can be parameterised in terms of $\Gamma_{i}/\Omega_{i}$. 

\subsubsection{Off-Shell Contributions}\label{OffShellSection}
In the following we concretise the above considerations and derive an explicit formula for the relaxation rate $\Gamma_{\textbf{0}}$ for the zero mode of $\phi$ including off-shell contributions. For $|\textbf{q}|=0$, the only spacial momentum in (\ref{loopintegral}) is $\textbf{p}$ and we can drop the momentum index.
For simplicity we restrict the analysis to the case that the only poles of the spectral densities for the $\chi_{i}$ are those that correspond to the screened one particle states. 
We furthermore assume that the dispersion relations are in good approximation parabolic and define $\hat{\Omega}_{i}=\Omega_{i}+\frac{i}{2}\Gamma_{i}$ with $\Omega_{i}=(\textbf{p}^{2}+M_{i}^{2})^{1/2}$, where $M_{i}$ are the thermal masses for the fields $\chi_{i}$. The simplest way to realise this is to assume that the leading order contribution to the self energies $\Pi^{R}_{i}$ of the fields $\chi_{i}$ comes from local diagrams as e.g. shown in figure \ref{phi4}a) that lead to a momentum and energy independent mass shift. 
We take $\Gamma_{i}\neq 0$, but $\Gamma_i\ll\Omega_i$. 
With (\ref{loopintegral}) and (\ref{spectralfunction}) one can then write
\begin{eqnarray}
{\rm Im}\Pi_{\textbf{0}}^{R}(\omega)&=&-\frac{g^{2}}{2}\int\frac{d^{4}p}{(2\pi)^{4}}\left(1+f(p_{0})+f(\omega-p_{0})\right)\nonumber\\
&&\times{2{\rm Im}\Pi^R_{1}(p_{0})\over (p_{0}^2-\Omega_{1}^2)^2+({\rm Im}\Pi^R_{1}(p_{0}))^2}\nonumber\\
&&\times{2{\rm Im}\Pi^R_{2}(\omega-p_0)\over ((\omega-p_0)^2-\Omega_{2}^2)^2+({\rm Im}\Pi^R_{2}(\omega-p_0))^2}
\end{eqnarray}
We have dropped the $\epsilon$ prescription as we know from the previous considerations that at finite temperature ${\rm Im}\Pi^{R}_{i}$ is non-zero along the entire real energy axis.
The $p_{0}$ integral can be evaluated using Cauchy's theorem. The integral is dominated by the on-shell regions. As the smooth functions $\Pi^{R}_{i}$ do not change much within the narrow peak regions along the $p_{0}$ axis and ${\rm Im}\Pi^{R}_{i}$ in the denominator is irrelevant everywhere else, one can replace $p_{0}$ by its pole value in the argument of ${\rm Im}\Pi^{R}_{i}$. 
However, one cannot directly replace the spectral densities using (\ref{BreitWigner}) before integration as this replacement is only valid if the function that $\rho_{i}$ is multiplied with under the integral has no additional poles.  
Here one has to proceed pole by pole and ${\rm Im}\Pi^{R}$ is evaluated with different arguments.
It is straightforward to obtain 
\begin{eqnarray}
\lefteqn{{\rm Im}\Pi^{R}_{\textbf{0}}(\omega)=-\frac{g^{2}}{2}\int\frac{d^{3}\textbf{p}}{(2\pi)^{3}}\Bigg(\frac{\left(1+f_{B}(\hat{\Omega}_{1})+f_{B}(\omega-\hat{\Omega}_{1})\right){\rm Im}\Pi^{R}_{1}(\hat{\Omega}_{1}){\rm Im}\Pi^{R}_{2}(\omega-\hat{\Omega}_{1})}{\Omega_{1} \Gamma_{1} \hat{\Omega}_{1} \left(\left((\omega-\hat{\Omega}_{1})^{2}-\Omega_{2}^{2}\right)^{2}+\left({\rm Im}\Pi^{R}_{2}(\omega-\hat{\Omega}_{1})\right)^{2}\right)}}  \nonumber\\
&&+\frac{\left(f_{B}(\omega+\hat{\Omega}_{1}^{*})-f_{B}(\hat{\Omega}_{1}^{*})\right){\rm Im}\Pi^{R}_{1}(-\hat{\Omega}_{1}^{*}){\rm Im}\Pi^{R}_{2}(\omega+\hat{\Omega}_{1}^{*})}{\Omega_{1} \Gamma_{1} \hat{\Omega}_{1}^{*} \left(\left((\omega+\hat{\Omega}_{1}^{*})^{2}-\Omega_{2}^{2}\right)^{2}+\left({\rm Im}\Pi^{R}_{2}(\omega+\hat{\Omega}_{1}^{*})\right)^{2}\right)}\nonumber\\
&&+\frac{\left( f_{B}(\omega+\hat{\Omega}_{2})-f_{B}(\hat{\Omega}_{2}) \right){\rm Im}\Pi^{R}_{1}(\omega+\hat{\Omega}_{2}){\rm Im}\Pi^{R}_{2}(-\hat{\Omega}_{2})}{\Omega_{2} \Gamma_{2} \hat{\Omega}_{2} \left(\left((\omega+\hat{\Omega}_{2})^{2}-\Omega_{1}^{2}\right)^{2}+\left({\rm Im}\Pi^{R}_{1}(\omega+\hat{\Omega}_{2})\right)^{2}\right)}\nonumber\\
&&+\frac{\left(1+f_{B}(\hat{\Omega}_{2}^{*})+f_{B}(\omega-\hat{\Omega}_{2}^{*})\right){\rm Im}\Pi^{R}_{1}(\omega-\hat{\Omega}_{2}^{*}){\rm Im}\Pi^{R}_{2}(\hat{\Omega}_{2}^{*})}{\Omega_{2} \Gamma_{2} \hat{\Omega}_{2} \left(\left((\omega-\hat{\Omega}_{2}^{*})^{2}-\Omega_{1}^{2}\right)^{2}+\left({\rm Im}\Pi^{R}_{1}(\omega-\hat{\Omega}_{2}^{*})\right)^{2}\right)}\Bigg)
.\end{eqnarray}
Here we have used the identity $f_{B}(-z)=-1-f_{B}(z)$. The angular integrations can be performed straight away. We now expand the numerators in the small widths $\Gamma_{i}$. With the identity $f_{B}'(x)=-\beta f_{B}(x)(1+f_{B}(x))$ and replacements of the form ${\rm Im}\Pi^{R}_{1}(\Omega_{1})\rightarrow -\Omega_{1}\Gamma_{1}$ and ${\rm Im}\Pi^{R '}_{1}(\Omega_{1})\rightarrow -\Gamma_{1}$ one can simplify the expression to

\begin{eqnarray}
{\rm Im}\Pi^{R}_{\textbf{0}}(\omega)&=&\frac{g^{2}}{2}\int\frac{d|\textbf{p}|}{(2\pi)^{2}}\textbf{p}^{2}\Bigg(
\frac{{\rm Im}\Pi^{R}_{2}(\omega-\Omega_{1})\left(1+f_{B}(\Omega_{1})+f_{B}(\omega-\Omega_{1})\right)}{\hat{\Omega}_{1}\left(\left((\omega-\hat{\Omega}_{1})^{2}-\Omega_{2}^{2}\right)^{2}+\left({\rm Im}\Pi^{R}_{2}(\omega-\hat{\Omega}_{1})\right)^{2}\right)}\nonumber\\
&&-\frac{{\rm Im}\Pi^{R}_{2}(\omega+\Omega_{1})\left(f_{B}(\omega+\Omega_{1})-f_{B}(\Omega_{1})\right)}{\hat{\Omega}_{1}^{*}\left(\left((\omega+\hat{\Omega}_{1}^{*})^{2}-\Omega_{2}^{2}\right)^{2}+\left({\rm Im}\Pi^{R}_{2}(\omega+\hat{\Omega}_{1}^{*})\right)^{2}\right)}\nonumber\\
&&-\frac{{\rm Im}\Pi^{R}_{1}(\omega+\Omega_{2})\left(f_{B}(\omega+\Omega_{2})-f_{B}(\Omega_{2})\right)}{\hat{\Omega}_{2}\left(\left((\omega+\hat{\Omega}_{2})^{2}-\Omega_{1}^{2}\right)^{2}+\left({\rm Im}\Pi^{R}_{1}(\omega+\hat{\Omega}_{2})\right)^{2}\right)}\nonumber\\
&&+\frac{{\rm Im}\Pi^{R}_{1}(\omega-\Omega_{2})\left(1+f_{B}(\Omega_{2})+f_{B}(\omega-\Omega_{2})\right)}{\hat{\Omega}_{2}^{*}\left(\left((\omega-\hat{\Omega}_{2}^{*})^{2}-\Omega_{1}^{2}\right)^{2}+\left({\rm Im}\Pi^{R}_{1}(\omega-\hat{\Omega}_{2}^{*})\right)^{2}\right)}\Bigg)\label{OffShell}
\end{eqnarray}
Each line in (\ref{OffShell}) is proportional to a factor ${\rm Im}\Pi^{R}_{i}/\hat{\Omega}_{j}\sim\Gamma_{i}\Omega_{i}/\Omega_{j}$\footnote{For large $|\textbf{p}|$  one has $\Omega_{i}\approx\Omega_{j}$. For non-relativistic $|\textbf{p}|$ it is possible to have $\Omega_{i}\gg\Omega_{j}$. Then one of the $\chi_{i}$ quasiparticles is effectively massless in comparison to the other and there is no forbidden region (c) as $q_{th1}\approx q_{th2}$ and dissipation is always possible on-shell.} and is thus suppressed by the smallness of this ratio in comparison to the characteristic quasiparticle energy $\Omega_{i}$.
This suppression can be cancelled if the denominator is very small for some values of $\textbf{p}$. One can easily see that this happens if either $\omega\approx\pm (\Omega_{1}+\Omega_{2})$ (first and fourth line, corresponds to case (a) ) or $\omega\approx\pm(\Omega_{1}-\Omega_{2})$ (second and third line, corresponds to case (b) ). These conditions exactly require that processes $\phi\leftrightarrow\chi_{1}\chi_{2}$ (first and fourth line) and  $\chi_{i}\leftarrow\phi\chi_{j}$ (second and third line) are possible on-shell, but with particles replaced by quasiparticles. 
Figure \ref{ExtraBreit} demonstrates the efficiency of the suppression in the forbidden region corresponding to case (c) by a numerical evaluation of (\ref{OffShell}) for different $\Gamma_{i}$. For $\Gamma_i/M_{i}\sim 10^{-4}$ it is efficient and $\Gamma_{\textbf{q}}$ is indeed an order of magnitude smaller than the vacuum decay rate and two orders of magnitude smaller than at $T\lesssim T_{c}$. However, already for a moderate width $\Gamma_{i}/M_{i}\sim 10^{-2}$ the relaxation by off-shell processes can be as efficient as the vacuum decay because of the effect of induced transitions.

Using the replacement
\begin{displaymath}
\frac{{\rm Im}\Pi_{i}(x)}{(\ldots)^{2}+({\rm Im}\Pi_{i}(x))^{2}}\rightarrow -{\rm sign}(x)\pi\delta(\ldots)
\end{displaymath}
it is straightforward to verify that in the limit of vanishing ${\rm Im}\Pi_{i}^{R}$,  (\ref{OffShell}) converges to (\ref{DiplomFormula}), but with the intrinsic masses $m$, $m_{i}$ replaced by temperature dependent thermal masses $M$, $M_{i}$. 
Without the simplifying assumption of parabolic dispersion relations one would also obtain a formula containing energy conserving $\delta$-functions.
In that case these would depend on the full dispersion relations and could not be obtained by replacing intrinsic masses by thermal masses in (\ref{DiplomFormula}). 

\subsection{Yukawa Interaction}\label{ScalarwithYukawa}
The relaxation rate from the Yukawa coupling can be computed from the diagram in figure \ref{diagrams}f). Analogue to (\ref{loopintegral}) it is given by
\begin{eqnarray}\label{FermionLoopintegral}
\lefteqn{{\rm Im}\Pi^{R}_{\textbf{q}}(\omega)=\frac{Y^{2}}{2}f_{B}^{-1}(\omega)\int\frac{d^{4}p}{(2\pi)^{4}}{\rm tr}\left(S^{<}_{1 \textbf{p}}(p_{0})S^{>}_{2 \textbf{p}-\textbf{q}}(p_{0}-\omega)\right)}\nonumber\\
&=&-\frac{Y^{2}}{2}f_{B}^{-1}(\omega)\int\frac{d^{4}p}{(2\pi)^{4}}f_{F}(p_{0})(1-f_{F}(p_{0}-\omega)){\rm tr}\left(\uprho_{1 \textbf{p}}(p_{0})\uprho_{2 \textbf{p}-\textbf{q}}(p_{0}-\omega)\right),
\end{eqnarray}
where  $S^{<}_{i \textbf{p}}(p_{0})=-f_{F}(p_{0})\uprho_{i \textbf{p}}(p_{0})$ is the spacial Fourier transform of the thermal Wightman function for fermions, see (\ref{Sforw}) and (\ref{KMSfermi}).
Here $f_{F}(p_{0})=(e^{\beta p_{0}}+1)^{-1}$  is the Fermi-Dirac distribution and $\uprho_{i \textbf{p}}(p_{0})$ the fermion spectral density.

\subsubsection{Fermion Spectral Density}
Explicitly  $\uprho_{i \textbf{p}}(p_{0})$ reads
\begin{equation}\label{Sigmabeforeinversion}
\uprho_{i \textbf{p}}(p_{0})=\left(\frac{i}{\Slash{p}-\fermionmass{m}_{i}-\Sigma^{R}_{i \textbf{p}}(p_{0})+i\epsilon\gamma^{0}}-\frac{i}{\Slash{p}-\fermionmass{m}_{i}-\Sigma^{A}_{i \textbf{p}}(p_{0})-i\epsilon\gamma^{0}}\right)
.\end{equation}
The pole structure of (\ref{Sigmabeforeinversion}), which determines the spectrum of fermionic resonances, is complicated in general \cite{Klimov:1981ka,Weldon:1982bn,Weldon:1989ys,Kitazawa:2006zi}. 
Gauge and Yukawa interactions are known to give rise to additional poles that correspond to collective excitations known as holes or plasminos. We consider a fermion bath that is kept in equilibrium by an abelian gauge interaction of strength $\alpha$\footnote{The behaviour in non-abelian gauge interactions and Yukawa couplings is similar \cite{LeB,Thoma:1994yw}, the spectrum in the supersymmetric case has been studied in \cite{Czajka:2010zh}.} which, for simplicity, couples in the same way to $\Psi_{1}$ and $\Psi_{2}$. We furthermore assume that all resonances in the bath have a narrow width that can be neglected.
We consider the two limits $\fermionmass{m}_{i}\gg\alpha T$ and $\fermionmass{m}_{i}\ll\alpha T$ in which $\uprho_{\textbf{p}}(p_{0})$ is well-known.

\paragraph{$\fermionmass{m}_{i}\ll\alpha T$}: In this case the resummed spectral density in the quasiparticle regime with negligible width can be computed using the hard thermal loop approximation \cite{Braaten:1989mz}. The well-known result reads \cite{Weldon:1982bn,LeB}
\begin{equation}
\uprho_{\textbf{p}}(p_{0})=\frac{1}{2}\left((\gamma_{0}-\hat{\bf{p}}\pmb{\gamma})\uprho_{+}+(\gamma_{0}+\hat{\bf{p}}\pmb{\gamma})\uprho_{-}\right)\label{HTLrho}
\end{equation}
where $\hat{\bf{p}}\pmb{\gamma}=p_{i}\gamma_{i}/|\textbf{p}|$ is the unit vector in $\textbf{p}$-direction and 
\begin{equation}
\uprho_{\pm}(p)\simeq2\pi(\uprho_{\pm}^{\rm pole}(p)+\uprho^{\rm cont}_{\pm}(p))
\end{equation}
with
\begin{equation}
\uprho_{\pm}^{\rm pole}(p)= Z_{\pm}\delta(p_{0}-\Upomega_{\pm})+Z_{\mp}\delta(p_{0}+\Upomega_{\mp})
\end{equation}
and
\begin{eqnarray}
\lefteqn{\uprho^{\rm cont}_{\pm}(p)=\theta(1-x^{2})\frac{y^{2}}{|\textbf{p}|}(1\mp x)}\nonumber\\
&\times&\Bigg[\Bigg(1\mp x\pm y^{2}\left((1\mp x)\ln\left|\frac{x+1}{x-1}\right|\pm 2\right)\Bigg)^{2}
+\pi^{2}y^{4}(1\mp x)^{2}\Bigg]^{-1}.
\end{eqnarray}
Here $x=p_{0}/|\textbf{p}|$, $y^{2}=(\alpha T)^{2}/(4\textbf{p})^{2}$ 
and the residues are
\begin{equation}
Z_{\pm}=\frac{\Upomega_{\pm}^{2}-\textbf{p}^{2}}{4y^{2}\textbf{p}^{2}}
.\end{equation}
The dispersion relations $\Upomega_{+}$ and $-\Upomega_{-}$ are the solutions to
\begin{equation}
0=p_{0}-|\textbf{p}|\left(1+y^2\left((1-x)\ln\frac{x+1}{x-1}+2\right)\right)
.\end{equation}
There are a small continuous contribution for $p_{0}<|\textbf{p}|$ from $\uprho^{\rm cont}_{\pm}(p)$ and four poles in $\uprho^{\rm pole}_{\pm}(p)$. This complicated structure makes it obvious that simply replacing bare masses by thermal masses for fermions is not a valid procedure even in the quasiparticle regime. 
However, the poles can be interpreted as quasiparticles with dispersion relations determined by the $\textbf{p}$-dependence of their position. 
$\Upomega_{+}$ is interpreted as dressed particle and $\Upomega_{-}$ as hole or plasmino\footnote{Note that the hole has a negative chirality over helicity ratio even for fermions that are massless in vacuum.}.
The pole contributions dominate the integrand in (\ref{FermionLoopintegral}), which again is given by a product of spectral densities. Thus, the entire discussion from section \ref{trilinearsection} can be repeated. The energies of quasiparticles are given by $\Upomega_{\pm}$, and energy is approximately conserved in reactions between them. 

If $\phi$ quanta are energetic enough, they can decay or be created by inverse decays, $\phi\leftrightarrow\Psi_{i}\Psi_{j}$. If there is a combination of two fermionic quasiparticles and $\phi$ such that one can decay into $\phi$ and the other, $\Psi_{i}\leftrightarrow\phi\Psi_{j}$\footnote{$\Psi$ can symbolise any fermionic quasiparticle, screened particles or holes. The decay into holes has e.g. been studied in \cite{Kiessig:2010pr}.}, Landau damping is at work. In absence of any combination of $\phi$ and $\Psi_{1}$, $\Psi_{2}$ excitations such that one of them decay into the other on-shell, (\ref{FermionLoopintegral}) only receives contributions involving the continuous parts of the spectral functions and $\Gamma_{\textbf{q}}$ is suppressed as in case (c) in section \ref{trilinearsection}. If large widths are involved, the kinematic thresholds get smeared out as shown in figure \ref{ExtraBreit} for scalars.

The situation simplifies for large momenta $|\textbf{p}|\gg \alpha T$, meaning $y^{2}\ll 1$. Then one can approximate 
\begin{eqnarray}
\begin{tabular}{c c c}
$Z_{+}\simeq 1+y^{2}\left(1+2\ln(y)\right)$ & & $Z_{-}\simeq \alpha^{-1}y^{-2}\exp\left(-y^{-2}\right)$\label{Residuen}
\end{tabular}
\end{eqnarray}
and 
\begin{eqnarray}
\begin{tabular}{c c c}
$\Upomega_{+}\simeq |\textbf{p}|(1+2y^{2})\simeq\left(\textbf{p}^{2}+\fermionmass{M}_{f}^{2}\right)^{1/2}$ & & $\Upomega_{-}\simeq|\textbf{p}|(1+2\alpha^{-1}\exp(-y^{-2}))$\label{FermionPoles}
\end{tabular}
\end{eqnarray}
with $\fermionmass{M}_{f}=\alpha T/2$.
Thus, for $|\textbf{p}|\gg\alpha T$ the dispersion relation $\Upomega_{+}$ for the screened particle approaches that of a particle with mass $\sim\fermionmass{M}_{f}=\alpha T/2$. This thermal mass only appears in the dispersion relation and does not break the chiral symmetry \cite{Weldon:1982bn}, as one can see from (\ref{HTLrho}). $\fermionmass{M}_{f}$ is sometimes referred to as {\it asymptotic mass} and differs by a factor $\sqrt{2}$ from the plasma frequency at $|\textbf{p}|=0$.
Equation (\ref{FermionPoles}) shows that the holes effectively become massless. The contribution of the poles to the integral (\ref{FermionLoopintegral}) depends on the residues (\ref{Residuen}). As $Z_{-}$ is exponentially suppressed for large momenta, the hole effectively decouples. The physical reason is that it is a collective excitation. Such excitations appear at length scales longer than the typical inter particle distance $\sim T$.
Neglecting the continuum contribution, the spectral density can be approximated as
\begin{equation}\label{rhoFermiQED}
\uprho_{\textbf{p}}(p_{0})\simeq \pi\left((\gamma_{0}-\hat{\textbf{p}}\pmb{\gamma})\delta(p_{0}-\Upomega_{+})+(\gamma_{0}+\hat{\textbf{p}}\pmb{\gamma})\delta(p_{0}+\Upomega_{+})\right)
\end{equation}
\paragraph{$\fermionmass{m}_{i}\gg\alpha T$}: In this case medium corrections to the spectral density are negligible. Then no resummation is necessary and it is justified to use the free spectral function as a zeroth order approximation. It can easily be derived from  (\ref{Gmsimp}), 
\begin{equation}\label{RhoFermionfree}
\uprho^{{\rm free}}_{\textbf{p}}(p_{0})=2\pi(\Slash{p}+\fermionmass{m}_{i}){\rm sign}(p_{0})\delta(p^{2}-\fermionmass{m}_{i}^{2})
.\end{equation}

\subsubsection{Computation of the Relaxation Rate} 
The computation of the relaxation rate from couplings to fermions is rather involved. We restrict ourselves to the cases $\fermionmass{m}_{i}\gg\alpha T$ and $\fermionmass{m}_{i}\ll\alpha T$ previously discussed. 

\paragraph{$\fermionmass{m}_{i}\gg\alpha T$}: Insertion of (\ref{RhoFermionfree}) into (\ref{FermionLoopintegral}) in the simplest case $\fermionmass{m}_{1}=\fermionmass{m}_{2}=\fermionmass{m}<m/2$, $|\textbf{q}|=0$ yields
\begin{equation}\label{FermionGammaSimple}
\Gamma_{\textbf{0}}=\frac{Y^{2}}{8\pi}m\bigg(1-\Big(\frac{2\fermionmass{m}}{m}\Big)^{2}\bigg)^{3/2}(1-2f_{F}(m/2))\theta(m-2\fermionmass{m})
,\end{equation}
where we have used the identity $f_{B}^{-1}(\omega)f_{F}(p_{0})f_{F}(\omega-p_{0})=1-f_{F}(p_{0})-f_{F}(\omega-p_{0})$. The result (\ref{FermionGammaSimple}) can be compared to the contribution (\ref{decayrate}) from decay into scalars. The kinematical restrictions are the same, but while the decay into scalars is enhanced by induced transitions at high temperature, for fermions in the final state there is a suppression due to Pauli blocking.

\paragraph{$\fermionmass{m}_{i}\ll\alpha T$}:
Insertion of (\ref{rhoFermiQED}) into (\ref{FermionLoopintegral}) yields for $\omega>0$, $|\textbf{q}|=0$
\begin{eqnarray}\label{ImPiYukawaQED}
{\rm Im}\Pi^{R}_{\textbf{0}}(\omega)=-\frac{Y^{2}}{8\pi}\omega^{2}\left(1-\left(\frac{2\fermionmass{M}_{f}}{\omega}\right)\right)^{1/2}\left(1-2f_{F}(\omega/2)\right)\theta(\omega-2\fermionmass{M}_{f}),
\end{eqnarray}
where we approximate $\Upomega_{+}\approx(\textbf{p}^{2}+\fermionmass{M}_{f}^{2})^{1/2}$. This holds for $|\textbf{p}|\gg\fermionmass{M}_{f}$.
This leads to 
\begin{eqnarray}\label{FermionGammaSimpleII}
\Gamma_{\textbf{0}}=\frac{Y^{2}}{8\pi}m\left(1-\left(\frac{2\fermionmass{M}_{f}}{m}\right)\right)^{1/2}\left(1-2f_{F}(m/2)\right)\theta(m-2\fermionmass{M}_{f}).
\end{eqnarray}
When $\phi$ is very heavy and the $\Psi_{i}$ form a relativistic plasma, $m\gg\fermionmass{M}_{f}\gg\fermionmass{m}_{i}$, the momenta of the decay products are large, $|\textbf{p}|\sim m/2\gg\fermionmass{M}_{f}$ and (\ref{FermionGammaSimpleII}) can be used to compute the relaxation rate.
Equations (\ref{FermionGammaSimple}) and (\ref{FermionGammaSimpleII}) show that in these limits the thermal masses qualitatively act as kinetic masses and close the phase space for the decay near $T_{c}\sim m/(2\alpha)$. 
Despite of this, the correct quantitative dependence of the relaxation rate on $T$ even in this regime cannot be reproduced by replacing intrinsic masses with thermal masses, as one can easily see by comparing the exponents of the first brackets in (\ref{FermionGammaSimple}) and (\ref{FermionGammaSimpleII}). 

\begin{figure}
  \centering
 \psfrag{GoverG0}{$\Gamma(T)/\Gamma(T=0)$}
\psfrag{ToverM}{$T/m$}
 \psfrag{0.1}{$0.1$}
\psfrag{0.2}{$0.2$}
 \psfrag{0.5}{$0.5$}
\psfrag{1.0}{$1$}
\psfrag{1}{$1$}
 \psfrag{2.0}{$2$}
\psfrag{10}{$10$}
\psfrag{10.0}{$10$}
 \psfrag{0.001}{$10^{-3}$}
\psfrag{0.01}{$10^{-2}$}
\psfrag{0.1}{$0.1$}
\psfrag{5.0}{$5$}
\psfrag{-}{$\phantom{i}$}
\psfrag{212}{$2\phantom{i}10^{-12}$}
\psfrag{412}{$4\phantom{i}10^{-12}$}
\psfrag{612}{$6\phantom{i}10^{-12}$}
    \includegraphics[width=12cm]{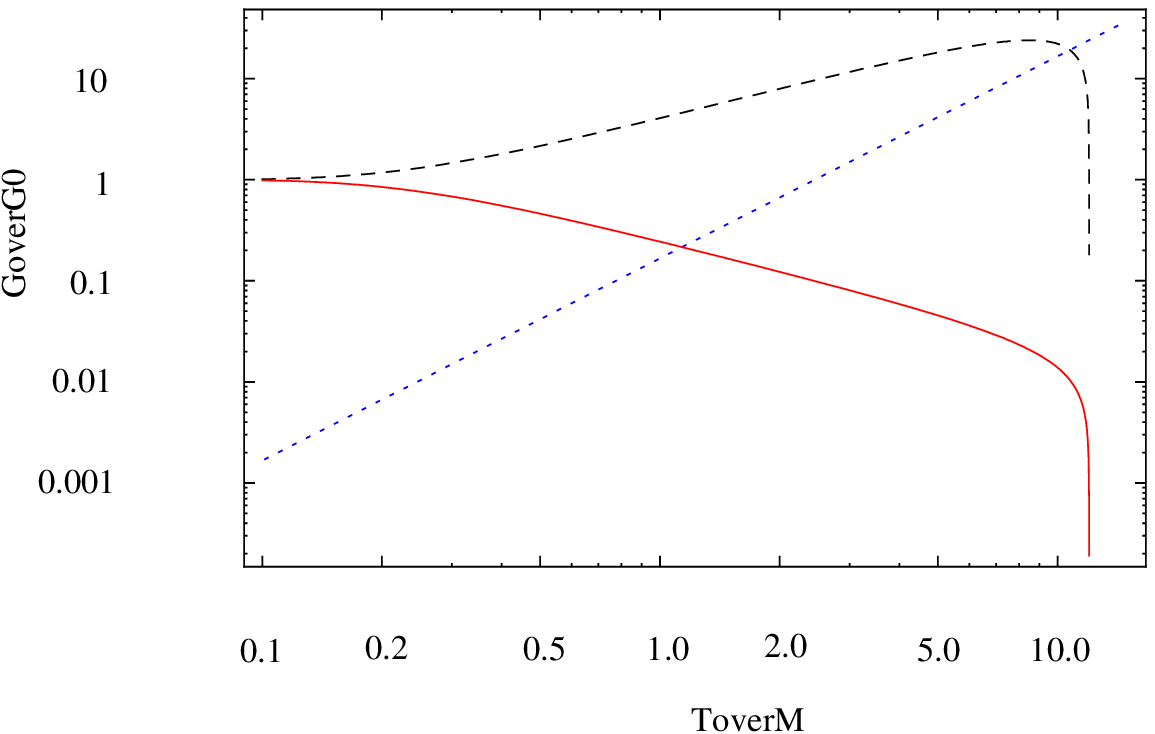}
    \caption{The different rates in comparison: trilinear (black dashed), quartic (blue dotted) and Yukawa (red solid) for $Y=h=g/(m\sqrt{2})=10^{-4}$, $\lambda=1/24$, $\alpha=1/12$. For this parameter choice the rates from Yukawa and trilinear interactions are equal at T=0. For the quartic interaction we only considered the contribution from scatterings (\ref{SimpleScatterings}).\label{gammas1}}
\end{figure}

Furthermore, the expression (\ref{FermionGammaSimpleII}) of course receives corrections near $T_{c}$, where the involved $\Psi_{i}$ quanta have momenta $|\textbf{p}|<\fermionmass{M}_{f}$ and the details of the fermionic spectrum have to be considered. However, in the regime where they would be relevant the rate is suppressed due to the effect of Pauli blocking.  The $f_{F}(m/2)$-term already leads to an effective suppression at $T\sim m\ll m/(2\alpha)$ for $\alpha\ll 1$, thus it is likely that at $T_{c}\sim m/(2/\alpha)$ contributions from processes with bosonic final state typically dominate.
This also implies that the closure of the phase space due to large thermal masses discussed for the trilinear scalar interaction is practically less important for the decay into fermions.
The effect can be seen in figure \ref{gammas1}.

Finally we would like to emphasise again that these conclusions are valid in the quasiparticle regime only. As discussed in section \ref{analsection} and shown in figure \ref{ExtraBreit} for the trilinear interaction, contributions from processes that are formally of higher order in the couplings can be of the same size as the leading order processes at high temperature. 
The suppression due to Pauli blocking that is observed in the quasiparticle regime is less efficient in this regime because diagrams of higher order also contain processes with bosons in the final state. That can be seen easily by cutting the diagram shown in figure \ref{fermiondiagram} through the scalar self-energy insertion. When there are bosons running in the loop, this cut yields contributions from scatterings with only bosons in the final state.

\subsection{Quartic Interaction}\label{quarticsection}
The leading order contribution to ${\rm Im}\Pi^{R}$ from the quartic interactions $h_{i}\phi\chi_{i}^{3}/4!$ comes from the setting sun diagrams shown in figure \ref{diagrams}d) and e).
Analogue to (\ref{loopintegral}) it can be written as  
\begin{eqnarray}\label{RisingSun}
{\rm Im}\Pi_{\textbf{q}}^{R}(\omega)=-\frac{h_{i}^{2}}{12}f_{B}^{-1}(\omega)\int\frac{d^{4}p}{(2\pi)^{4}}\frac{d^{4}k}{(2\pi)^{4}}\Delta^{<}_{i \textbf{p}}(p_{0})\Delta^{<}_{i \textbf{k}}(k_{0})\Delta^{<}_{i \textbf{q}-\textbf{p}-\textbf{k}}(\omega-p_{0}-k_{0}).
\end{eqnarray}
Again, the integrand is proportional to a product or spectral densities, in this case $\rho_{i \textbf{p}}(p_{0})\rho_{i \textbf{k}}(k_{0})\rho_{i \textbf{q}-\textbf{p}-\textbf{k}}(\omega-p_{0}-k_{0})$. 
In complete analogy to the discussion in section \ref{trilinearsection} one can use finite temperature cutting rules to interpret the discontinuities in terms of elementary processes.
\begin{figure}
  \centering
 \psfrag{a}{$a)$}
 \psfrag{b}{$b)$}
 \psfrag{c}{$c)$}
 \psfrag{d}{$d)$}
    \includegraphics[width=10cm]{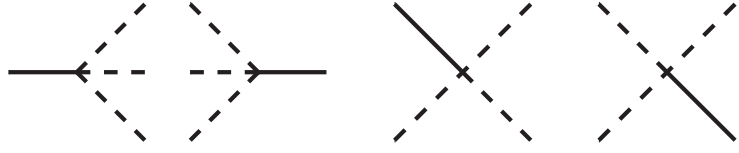}
    \caption{Processes contained in the setting sun diagram in figure \ref{diagrams}d), solid lines represent $\phi$ propagators and dashed lines $\chi_{1}$ propagators.\label{hcut}}
\end{figure}
As for the trilinear interaction, if the $\chi_{i}$ resonances have a narrow width and well-defined approximate dispersion relation, the integral (\ref{RisingSun}) only receives a significant contribution if these processes are possible for real quasiparticles in the initial and final state. However, there is a significant difference as the setting sun diagrams at finite temperature contain scattering processes $\phi\chi_{i}\leftrightarrow\chi_{i}\chi_{i}$, see figure \ref{hcut}. In contrast to decays, these processes are kinematically always possible, though at low temperatures they are suppressed by the statistical factors, reflecting the lack of scattering partners at low density. With (\ref{freespectral}) it is straightforward to derive the leading order expression 
\begin{eqnarray}\label{risingsun}
\lefteqn{{\rm Im}\Pi^{R}_{\q}(\omega)=\frac{h_{i}^{2}}{12}\int\frac{d^{3}\textbf{p}d^{3}\textbf{k}d^{3}\textbf{l}}{(2\pi)^{9}}(2\pi)^{3}\delta^{(3)}(\textbf{p}+\textbf{k}+\textbf{l}-\textbf{q})\frac{1}{8\omega_{i \textbf{p}}\omega_{i \textbf{k}}\omega_{i \textbf{l}}}}\nonumber\\
&\times&\Big(\big(\left(1+f_{1}\right)\left(1+f_{2}\right)\left(1+f_{3}\right)-f_{1}f_{2}f_{3}\big)\nonumber\\
&&\phantom{\left(1+f_{2}\right)\left(1+f_{3}\right)}\left(\delta(\omega-\omega_{i \textbf{p}}-\omega_{i \textbf{k}}-\omega_{i \textbf{l}})-\delta(\omega+\omega_{i \textbf{p}}+\omega_{i \textbf{k}}+\omega_{i \textbf{l}})\right)\nonumber\\
&&+\big(f_{1}\left(1+f_{2}\right)\left(1+f_{3}\right)-\left(1+f_{1}\right)f_{2}f_{3})\big)\nonumber\\
&&\phantom{\left(1+f_{2}\right)\left(1+f_{3}\right)}\left(\delta(\omega+\omega_{i \textbf{p}}-\omega_{i \textbf{k}}-\omega_{i \textbf{l}})-\delta(\omega-\omega_{i \textbf{p}}+\omega_{i \textbf{k}}+\omega_{i \textbf{l}})\right)\nonumber\\
&&+\big(\left(1+f_{1}\right)f_{2}\left(1+f_{3}\right)-f_{1}\left(1+f_{2}\right)f_{3})\big)\nonumber\\
&&\phantom{\left(1+f_{2}\right)\left(1+f_{3}\right)}\left(\delta(\omega-\omega_{i \textbf{p}}+\omega_{i \textbf{k}}-\omega_{i \textbf{l}})-\delta(\omega+\omega_{i \textbf{p}}-\omega_{i \textbf{k}}+\omega_{i \textbf{l}})\right)\nonumber\\
&&+\big(f_{1}f_{2}\left(1+f_{3}\right)-\left(1+f_{1}\right)\left(1+f_{2}\right)f_{3})\big)\nonumber\\
&&\phantom{\left(1+f_{2}\right)\left(1+f_{3}\right)}\left(\delta(\omega+\omega_{i \textbf{p}}+\omega_{i \textbf{k}}-\omega_{i \textbf{l}})-\delta(\omega-\omega_{i \textbf{p}}-\omega_{i \textbf{k}}+\omega_{i \textbf{l}})\right)\Big),
\end{eqnarray}
here written in a symmetric way with the notation $\omega_{1 \textbf{p}}=(\textbf{p}^{2}+m_{1}^{2})^{1/2}$, $f_{1}=f_{B}(\omega_{1 \textbf{p}})$ and so on.
Again, the $\delta$-functions ensure energy conservation and the statistical factors satisfy detailed balance.
The first line in the brackets comes from decays and inverse decays $\phi\leftrightarrow \chi_{i}\chi_{i}\chi_{i}$, which are kinematically forbidden for $m<3m_{i}$. The remaining lines correspond to scatterings $\phi\chi_{i}\leftrightarrow\chi_{i}\chi_{i}$. Kinematically they are always possible, but as they require bath quanta in the initial state, they are suppressed by the occupation numbers $f_{i}$ appearing in the statistical factors and vanish in vacuum. At high temperature they provide the dominant channel of dissipation. Using resummed perturbation theory - replacing the internal lines of the diagram in figure \ref{diagrams}d) and e) by resummed propagators as in \ref{diagrams}b) and \ref{phi4}b) - the dissipation rate at $T\gg m_{i}$ can be approximated by \cite{Parwani:1991gq}
\begin{equation}\label{SimpleScatterings}
\Gamma_{\textbf{0}}\simeq\sum_{i}\frac{h_{i}^{2}T^{2}}{768\pi m}
.\end{equation}

\section{Comparison of Relaxation Mechanisms}\label{comparisonsection}
\subsection{Decays vs Scatterings}\label{scatteringimportance}
\begin{figure}
  \centering
 \psfrag{a}{$a)$}
  \psfrag{b}{$b)$}
  \psfrag{c}{$c)$}
  \psfrag{d}{$d)$}
  \psfrag{e}{$e)$}
  \psfrag{f}{$f)$}
    \includegraphics[width=10cm]{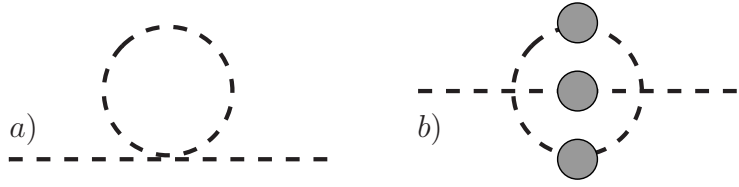}
    \caption{Feynman diagrams for the $\lambda_{i}\chi_{i}^{4}/4!$ self interaction, the gray circles indicate resummed propagators.\label{phi4}}
\end{figure}
It is instructive to compare the contributions from scatterings and decays to estimate how efficient a heavy particle $\phi$ can dissipate or be produced from the plasma around $T\approx T_{c}$. We focus on the trilinear and quartic scalar couplings because the decay into fermionic final states is suppressed by Pauli blocking.
To determine $T_{c}$, specification of $\mathcal{L}_{\chi_{i} \mathrm{int}}$ is required. For simplicity we chose $\mathcal{L}_{\chi_{i} \mathrm{int}}=\lambda_{i}\chi_{i}^{4}/4!$. The temperature dependent dispersive part of the self energy to leading order comes from the diagram shown in figure \ref{phi4}a) \footnote{The thermal mass correction here is $\sim\lambda^{1/2}T$ rather than $\sim\lambda T$ because it arises from a local (tadpole) diagram, see figure \ref{phi4}a).} 
\begin{equation}\label{thermalmass}
M_{i}^{2}=m_{i}^{2}+\lambda_{i}\int\frac{d^{3}\textbf{p}}{(2\pi)^{3}}\frac{f_{B}(\bath{\omega}_{i})}{2\bath{\omega}_{i}}\approx m_{i}^{2}+\frac{\lambda_{i}}{24}T^{2}
.\end{equation}
In the simplest case with $m_{1}=m_{2}=m_{\chi}$, $h_{1}=h_{2}=h$ and $\lambda_{1}=\lambda_{2}=\lambda$, (\ref{dampingformel}) for $|\textbf{q}|=0$ the rate (\ref{dampingformel}) reduces to
\begin{equation}\label{decayrate}
\Gamma_{{\rm decay}}=\frac{g^{2}}{16\pi m}\left(1-\left(\frac{2M_{\chi}}{m}\right)^{2}\right)^{1/2}(1+2f_{B}(M/2))\theta(\omega-2M_{\chi}),
\end{equation} 
where we have replaced the intrinsic mass $m_{\chi}$ by the thermal mass $M_{\chi}$, but neglected the thermal correction to the $\phi$ mass $m$ because of the weak coupling. The scattering rate is computed from the diagrams in figure \ref{diagrams}d) and e) with resummed $\chi_{i}$ propagators and can be found from (\ref{SimpleScatterings}). At $T_{c}\approx(6(m^{2}-4m_{\chi}^{2})/\lambda)^{1/2}$  takes the value
\begin{equation}
\Gamma_{{\rm scatter}}(T_{c})\approx2\frac{h^{2}m}{128\pi \lambda}\bigg(1-\Big(\frac{2m_{\chi}}{m}\Big)^{2}\bigg)
.\end{equation}
This can be compared to the decay rate (\ref{decayrate}) at $T=0$, $\Gamma_{\rm decay}(T=0)$ and the maximal decay rate. The temperature $T_{max}$ at which (\ref{decayrate}) is maximal fulfils the equation
\begin{equation}\label{TmaxCond}
m\left(6((2m_{\chi})^{2}-m^{2})+T_{max}^{2}\lambda\right)+T_{max}^{3}\lambda{\rm sinh}(m/T_{max})=0
\end{equation}
as one can easily see by demanding $\partial\Gamma_{\rm decay}/\partial T=0$ at $T=T_{max}$. If $T_{max}\gg m$, the ${\rm sinh}$ can be Taylor expanded and 
\begin{equation}
T_{max}\approx \left(\frac{3}{\lambda}\left(m^{2}-(2m_{\chi})^{2}\right)\right)^{1/2}
.\end{equation} 
With $M_{\chi}^{2}(T_{max})\approx(m_{\chi}^{2}+(m/2)^{2})/2$ one can estimate
\begin{equation}
\Gamma_{\rm decay}(T_{max})\approx \frac{g^{2}}{8\pi m}\sqrt{\frac{6}{\lambda}}\bigg(1-\Big(\frac{2m_{\chi}}{m}\Big)^{2}\bigg)
,\end{equation} 
where we have expanded $f_{B}(m)$ as $T_{max}\gg m$. This allows to compare
\begin{eqnarray}
\frac{\Gamma_{\rm decay}(T_{max})}{\Gamma_{\rm decay}(T=0)}&=&\sqrt{\frac{24}{\lambda}}\bigg(1-\Big(\frac{2m_{\chi}}{m}\Big)^{2}\bigg)^{1/2}\\
\frac{\Gamma_{\rm scatter}(T_{c})}{\Gamma_{\rm decay}(T=0)}&=&\left(\frac{h}{2}\frac{m}{g}\right)^{2}\frac{1}{\lambda}\bigg(1-\Big(\frac{2m_{\chi}}{m}\Big)^{2}\bigg)^{1/2}\\
\frac{\Gamma_{\rm scatter}(T_{c})}{\Gamma_{\rm decay}(T_{max})}&=&\left(\frac{h}{2}\frac{m}{g}\right)^{2}\sqrt{\frac{1}{24\lambda}}=\sqrt{\frac{\lambda}{24}}\frac{\Gamma_{\rm scatter}(T_{c})}{\Gamma_{\rm decay}(T=0)}\Big|_{m_{\chi}=0}\label{ScatterOverDecaymax}
.\end{eqnarray}
Remarkably (\ref{ScatterOverDecaymax}) does not depend on $m_{\chi}$. In the interesting case $m_{\chi}\ll m$ the ratios are very simple. The parameter that governs the maximal amount of amplification of the decay rate by induced transitions is $\sim(24/\lambda)^{1/2}$. The parameter that determines whether relaxation via scatterings is efficient at $T_{c}$ is $(h^{2}/\lambda)(m/g)^{2}$. For $g/m\sim h$ the rate $\Gamma_{\rm scatter}(T_{c})$ can easily be bigger than $\Gamma_{\rm decay}(T=0)$ or even $\Gamma_{\rm decay}(T_{max})$ if $\lambda$ is sufficiently small, where one of course has to keep $\lambda\gg h$ to be consistent with the initial assumption that the interactions that thermalise the bath are stronger than those that couple it to $\phi$.
\begin{figure}[t]
  \centering
\psfrag{a}{$(a)$}
  \psfrag{b}{$(b)$}
  \psfrag{c}{$(c)$}
  \psfrag{A}{$ $}
  \psfrag{B}{$ 0$}
  \psfrag{C}{$ $}
  \psfrag{D}{$ $}
  \psfrag{X}{$\scriptsize{T/m}$}
  \psfrag{Y}{$\scriptsize{\Gamma_{\textbf{0}}(T)/\Gamma_{\textbf{0}}(T=0)}$}
\begin{tabular}{c c}
    \includegraphics[width=7cm]{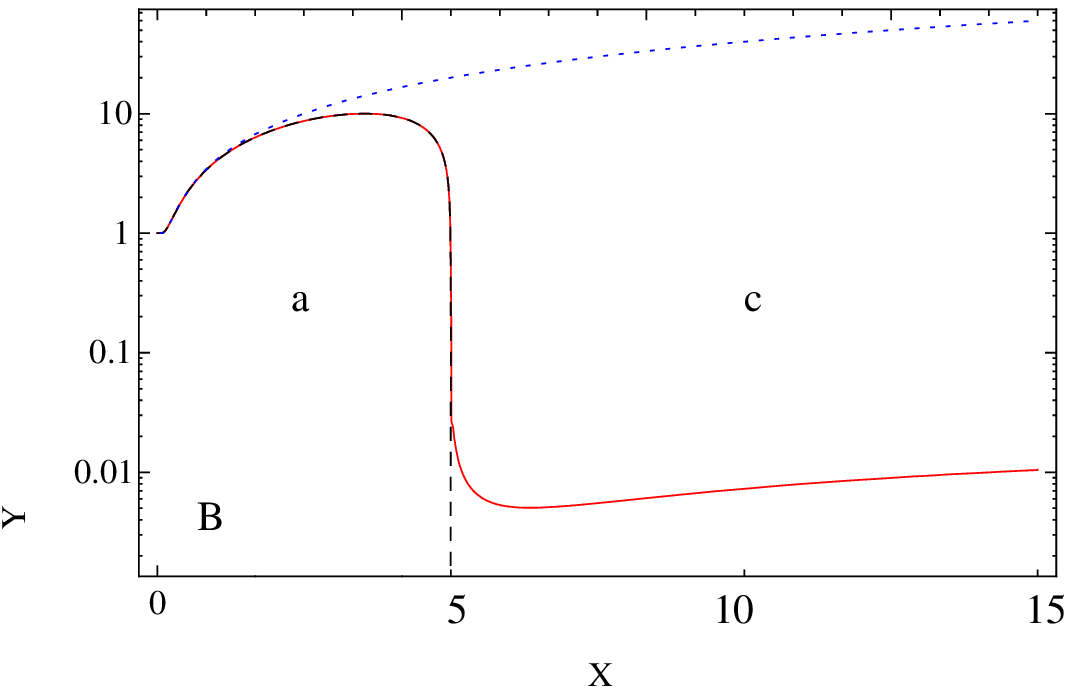} &
\includegraphics[width=7cm]{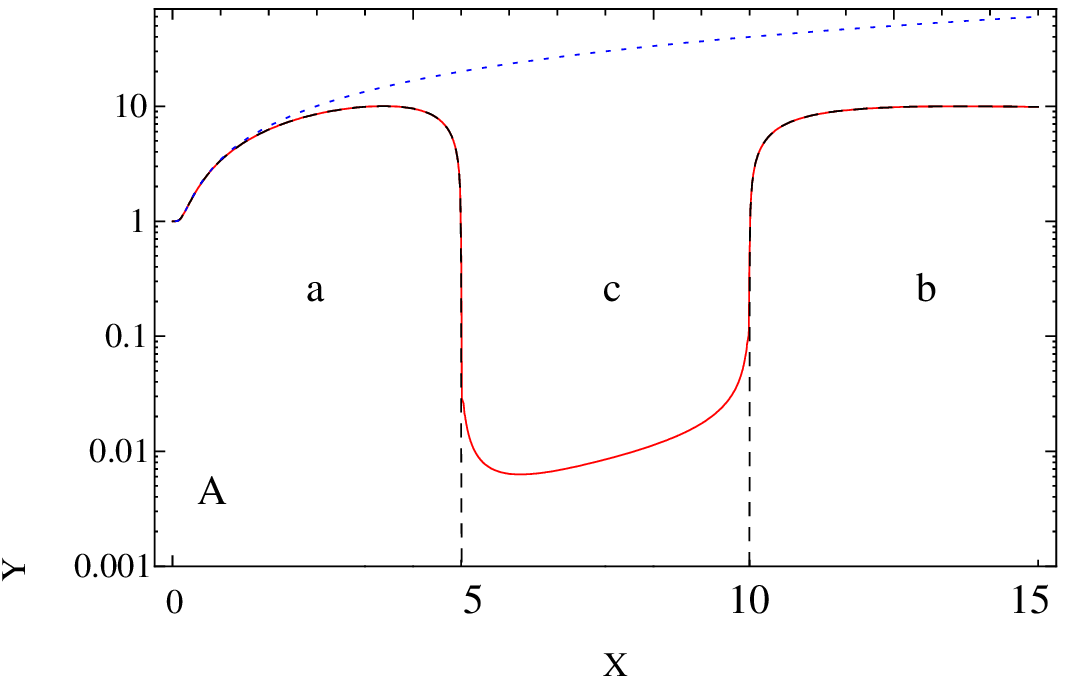}\\
    \includegraphics[width=7cm]{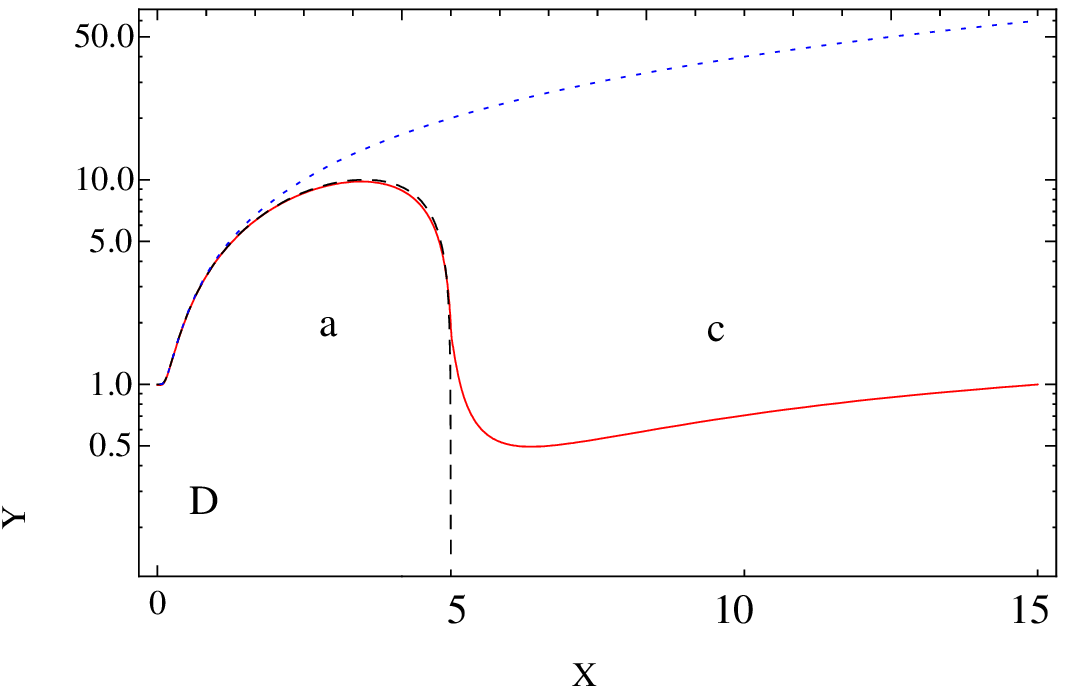} &
\includegraphics[width=7cm]{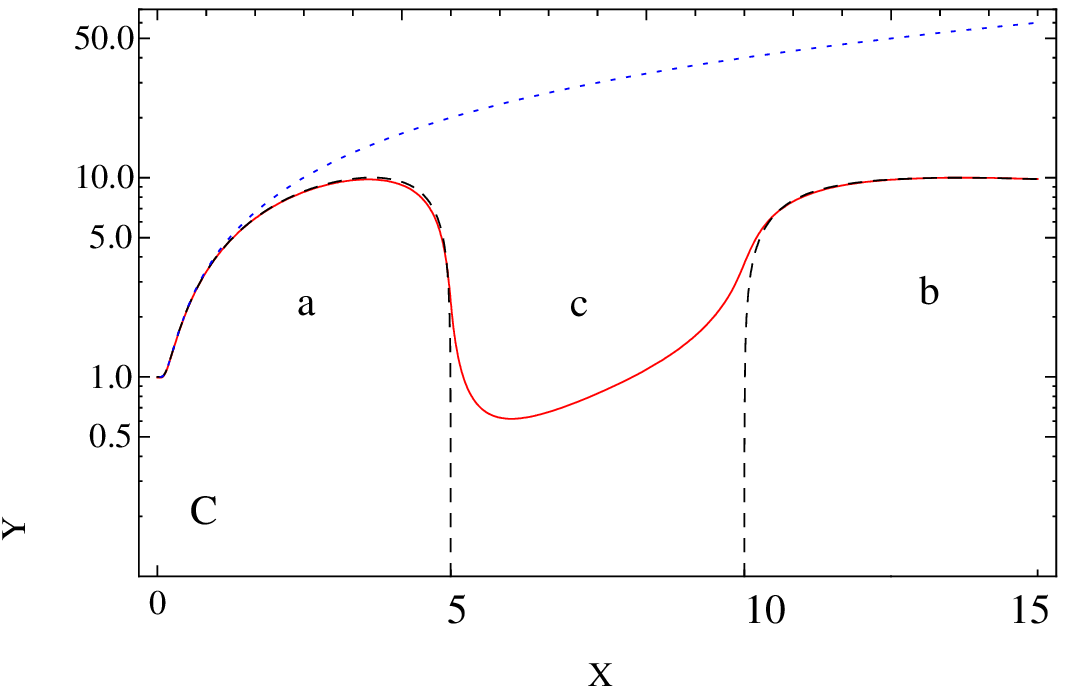}
\end{tabular}
  \caption{Dissipation rate $\Gamma_{\textbf{0}}$ for the zero mode of $\phi$ from trilinear interaction computed with bare $\chi_{i}$ propagators - dotted blue, from equation (\ref{dampingformel}) - in comparison with the result obtained using resummed $\chi_{i}$ propagators with thermal masses only (black dashed, from equation (\ref{dampingformel}) with masses replaced by thermal masses) and thermal masses as well as widths (red dotted, numerical evaluation of (\ref{OffShell}) ). In the upper row we chose $\tilde{\alpha}=0.1$ ($\Gamma_{i}/M_{i}\simeq 10^{-4}$), in the lower row $\tilde{\alpha}=1$ ($\Gamma_{i}/M_{i}\simeq 10^{-2}$). The left column corresponds to $\alpha_{1}=\alpha_{2}=0.2$ ($M_{1}=M_{2}=0.1 T$), in the right column we chose $\alpha_{1}=0.1$, $\alpha_{2}=0.3$ ($M_{1}=0.05T$, $M_{2}=0.15T$). ). Vacuum masses are small in all cases, $m_{1}=m_{2}=0.1m$. 
\label{ExtraBreit}}
\end{figure} 
\subsection{The Importance of Off-Shell Contributions}\label{OffShellDiscuss}
The importance of off-shell processes, which appear to violate energy conservation in the quasiparticle picture, can be parameterised by the widths $\Gamma_{i}$ of the bath fields. At large temperatures this effect can be stronger than naively expected. 

A quantitative evaluation of (\ref{OffShell})  requires specification of the interactions $\mathcal{L}_{\chi_{i} \mathrm{int}}$ in (\ref{L}) that keep the bath in equilibrium.
The dominant contribution to the thermal mass of scalars $\chi_{i}$ can come from local (tadpole) diagrams which give purely real, momentum independent contributions to the self energy. For a $\lambda_{i}\chi_{i}^{4}/4!$ self-interaction these come from the diagram shown in figure \ref{phi4}a). The resulting mass shift is given by (\ref{thermalmass}), $M_{i}\simeq (m_{i}^{2}+\lambda_{i}T^{2}/24)^{1/2}\sim\lambda_{i}^{1/2}T$. 
In order to study the effect of dissipation via off-shell processes, we also have to include the width of the $\chi$-resonances.
The leading order contribution to the $\chi_{i}$-width from such interaction appears only at second order, see figure \ref{phi4}b), and is much smaller, cf. (\ref{SimpleScatterings}). Thus, the leading order contribution to the width may come from another coupling. 

Therefore we in the following distinguish one coupling constant $\alpha$ that appears in the thermal mass\footnote{For the $\lambda_{i}\chi_{i}^{4}/4!$ interaction one can identify $\alpha_{i}=(\lambda_{i}/6)^{1/2}$, see (\ref{thermalmass}).} $M_{i}\simeq \alpha_{i} T/2$ and another coupling $\tilde{\alpha}_{i}$ that appears in the width.  
Equation (\ref{OffShell}) shows that the inclusion of off-shell processes requires knowledge of the functional dependence of the discontinuities on energy and momentum. Thus simply taking $\Gamma_{\chi}=\tilde{\alpha}_{i}^{2}T$ is not sufficient. This also requires specification of the interaction that gives rise to the $\chi_{i}$-widths. We chose a Yukawa coupling $\tilde{\alpha}_{i}$ to some massless fermion in the bath. That allows to use the formulae given in section \ref{ScalarwithYukawa} with $\chi_{i}$ as the external particle. This choice gives a good numerical stability when evaluating (\ref{OffShell}). Results for different parameter choices are shown in figure \ref{ExtraBreit}.

\section{Relaxation of a Fermion}\label{sectionaboutfermion}
We now proceed to the relaxation of a fermionic field $\Psi$ that is weakly coupled to a thermal bath. 
\subsection{Fermionic Nonequilibrium Correlation Functions}\label{fermioncorrelators}
As in the scalar case there are two independent two point functions $S^{>}_{\alpha\beta}(x_{1},x_{2})$ and $S^{<}_{\alpha\beta}(x_{1},x_{2})$, 
\begin{eqnarray}
S^{>}_{\alpha\beta}(x_{1},x_{2})&=&\langle \Psi_{\alpha}(x_{1})\bar{\Psi}_{\beta}(x_{2})\rangle_{c}\label{Sforw}\\
S^{<}_{\alpha\beta}(x_{1},x_{2})&=&-\langle \bar{\Psi}_{\beta}(x_{2})\Psi_{\alpha}(x_{1})\rangle_{c}\label{Sback},
\end{eqnarray}
see also (\ref{SforwA}) and (\ref{SbackA}). 
We again define a spectral function and statistical propagator
\begin{eqnarray}
S^{-}(x_{1},x_{2})&=&i\left(S^{>}(x_{1},x_{2})-S^{<}(x_{1},x_{2})\right)\\
S^{+}(x_{1},x_{2})&=&\frac{1}{2}\left(S^{>}(x_{1},x_{2})+S^{<}(x_{1},x_{2})\right).
\end{eqnarray}
Their time evolution is governed by the Kadanoff-Baym equations (\ref{Sabzug2}) and (\ref{Saddi2}). The solutions for Majorana fermions have been found in \cite{Anisimov:2010aq,ABDM3}. With the symmetry relations given in appendix \ref{KBEessentials} the generalisation to Dirac fermions is straightforward\footnote{I would like to thank Wilfried Buchm\"uller, Alexey Anisimov and Sebastian Mendizabal at this point, in collaboration with whom the propagators for Majorana fermions were first found.}, 
\begin{eqnarray}\label{Sminus}
S^{-}_{\textbf{q}}(t_{1}-t_{2})=i\int_{-\infty}^{\infty}\frac{d\omega}{2\pi} e^{-i\omega(t_{1}-t_{2})}\uprho_{\q}(\omega)
\end{eqnarray}
and
\begin{eqnarray}
\lefteqn{S^{+}_{\q}(t_{1},t_{2})=-S^{-}_{\textbf{q}}(t_{1})\gamma_{0}S^{+}_{\textbf{q}}(0,0)\gamma_{0}S^{-}_{\textbf{q}}(-t_{2})}\nonumber\\
&&+\int_{0}^{t_{1}}dt'S^{-}_{\q}(t_{1}-t')\int_{0}^{t_{2}} dt'' \Sigma^{+}_{\q}(t'-t'')S^{-}_{\q}(t''-t_{2})\label{FermiStatistical}
\end{eqnarray}
with
\begin{equation}\label{Sigmabeforeinversion2}
\uprho_{\q}(\omega)=\left(\frac{i}{\Slash{q}-\fermionmass{m}-\Sigma^{R}_{\q}(\omega)+i\epsilon\gamma_{0}}-\frac{i}{\Slash{q}-\fermionmass{m}-\Sigma^{A}_{\q}(\omega)-i\epsilon\gamma_{0}}\right)
.\end{equation}
As in the scalar case, the spectral density for the nonequilibrium field $\Psi$ coincides with the equilibrium expression (\ref{Sigmabeforeinversion}) even for large deviations from equilibrium as long as the bath is sufficiently large.
Fermion spectral functions are, as pointed out previously, in general rather complicated. Usually replacing vacuum masses by thermal masses does not even in the quasiparticle regime give a consistent approximation to the propagators. 
We are interested in the behaviour of a massive fermion that is weakly coupled to a thermal bath, thus it is consistent to consider the situation $\fermionmass{m}\gg YT$ only. Then the effect of the medium on the dispersion relation of $\Psi$ is small and can be neglected with respect to the vacuum mass $\fermionmass{m}$. The thermal correction to the width is, however, not negligible as there is no large width in vacuum in comparison to which it could be neglected.
We assume that the discontinuity of the fermion self energy is a pure vector, $\Sigma^{-}_{\textbf{p}}(\omega)=2i(a_{\q}(\omega)\Slash{p}+b_{\q}(\omega)\Slash{u})$. Here $u$ is the four velocity of the bath, which in its rest frame is simply $(1,\textbf{0})$ and $a_{\q}(\omega)$ and $b_{\q}(\omega)$ are functions of $q=(\omega,\textbf{q})$ that depend on the interactions of $\Psi$. This of course excludes a number of interesting applications such as $CP$ violating terms, but the inclusion of those is straightforward and does not affect the kinematical aspects we are interested in here. 
Then (\ref{Sigmabeforeinversion}) can be approximated by a Breit Wigner function,  
\begin{equation}
\uprho_{{\bf q}}(\omega)\simeq 2(\Slash{q}+\fermionmass{m})\frac{\omega\Upgamma_{{\bf q}}}{(\omega^{2}-\upomega_{\textbf{q}}^{2})^2+(\omega\Upgamma_{{\bf q}})^2}
\label{Gmsimp}
,\end{equation}
where $\upomega_{\textbf{p}}=(\fermionmass{m}+\textbf{q}^{2})^{1/2}$ and  
\begin{equation}\label{GammaDef}
\Upgamma_{ {\bf q}}= 2\left(a_{\q}\frac{\fermionmass{m}^{2}}{\upomega_{\bf{q}}}+b_{\q}\right)_{\omega=\upomega_{ {\bf q}}}.
\end{equation}
The $\omega$ integration in (\ref{Sminus}) can, analogue to (\ref{exponential}), be evaluated using Cauchy's theorem,
\begin{equation}\label{SMinusNarrow}
S^{-}_{\textbf{q}}(t_{1}-t_{2})\simeq e^{-\nicefrac{\Upgamma_{\q}|t_{1}-t_{2}|}{2}}\left(i\gamma_{0}\cos(\upomega_{\textbf{q}}(t_{1}-t_{2}))-\frac{\pmb{\gamma}\textbf{q}-\fermionmass{m}}{\upomega_{\textbf{q}}}\sin(\upomega_{\textbf{q}}(t_{2}-t_{2}))\right)
.\end{equation} 
Fourier transformation and another use of Cauchy's theorem allow to evaluate the memory integral in the second line of (\ref{FermiStatistical})
\begin{eqnarray}\label{Gplust1t2}
\lefteqn{S^{+}_{\q,{\rm mem}}(t_{1},t_{2})\simeq\left(1-2f_{F}(\upomega_{\textbf{q}}/2)\right)\frac{1}{2\upomega_{\textbf{q}}}\left(e^{-\nicefrac{\Upgamma_{\textbf{q}}}{2}|t_{1}-t_{2}|}-e^{-\nicefrac{\Upgamma_{\textbf{q}}}{2}(t_{1}+t_{2})}\right)}\nonumber\\
&\times&\left((\fermionmass{m}-\pmb{\gamma}\textbf{q})\cos(\upomega_{\textbf{q}}(t_{1}-t_{2}))-i\gamma_{0}\upomega_{\textbf{q}}\sin(\upomega_{\textbf{q}}(t_{1}-t_{2}))\right)
\end{eqnarray}
Together (\ref{SMinusNarrow}) and (\ref{Gplust1t2}) provide approximate analytic expressions for the correlation functions (\ref{Sminus}) and (\ref{FermiStatistical}).
We re-emphasise that it is in general not a valid procedure to obtain non-equilibrium propagators from the equilibrium correlation functions (\ref{KMSfermi}) by replacing the Fermi-Dirac distribution function $f_{F}$ with some generalised distribution\footnote{As in the scalar case on can of course \textit{define} $f(\textbf{q},\omega,t)=-S^{+}_{\textbf{q}}(t;\omega)\uprho_{\textbf{q}}^{-1}(\omega)+1/2$, but this requires a dependence of $f$ on time and off-shell energies.}, (\ref{Sminus}) and (\ref{FermiStatistical}) have been found as solutions to the full non-equilibrium equations of motion (\ref{Sabzug2}), (\ref{Saddi2}). 
Analogue in the scalar case, all correlations are damped with a rate $\Upgamma_{\textbf{q}}$, which can be interpreted as relaxation rate in accordance with \cite{Weldon:1983jn}. This interpretation holds even for large deviations of $\Psi$ from equilibrium.
 
\subsection{Relaxation via a Yukawa Coupling}\label{fermionviayukawa}
\begin{figure}
  \centering
 \psfrag{a}{$a)$}
  \psfrag{b}{$b)$}
  \psfrag{c}{$c)$}
  \psfrag{d}{$d)$}
  \psfrag{e}{$e)$}
  \psfrag{f}{$f)$}
    \includegraphics[width=5cm]{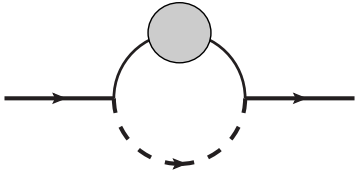}
    \caption{Fermion self energy, the lines represent $\Psi$ (solid with arrow) and the fermion (dashed with arrow) and scalar (solid) in the bath.\label{fermiondiagram}}
\end{figure}
When the dominant contribution to $\Upgamma_{\textbf{q}}$ comes from a Yukawa coupling of strength $Y$ to a Dirac fermion $\tilde{\Psi}$ and a real scalar $\tilde{\phi}$ in the bath\footnote{This means we consider a Lagrangian very similar to (\ref{L}); we could identify the out of equilibrium fermion $\Psi$ with $\Psi_{1}$ and the bath fields $\tilde{\Psi}$ and $\tilde{\phi}$ with $\Psi_{2}$ and $\phi$, respectively. We prefer to use an alternative notation here in order to avoid confusion as $\phi$ in the previous sections has always been a weakly coupled field out of equilibrium while $\tilde{\phi}$ here is a bath field and $\Psi$ is out of equilibrium.}, it can at leading order be computed from the Feynman diagram shown in figure \ref{fermiondiagram},
\begin{eqnarray}
\lefteqn{\Sigma^{-}_{\textbf{q}}(\omega)=-iY^{2}f_{F}^{-1}(\omega)\int\frac{d^{4}p}{(2\pi)^{4}}\tilde{S}^{<}_{\textbf{p}}(p_{0})\tilde{\Delta}^{>}_{\textbf{p}-\textbf{q}}(p_{0}-\omega)}\nonumber\\
&=&-iY^{2}f_{F}^{-1}(\omega)\int\frac{d^{4}p}{(2\pi)^{4}}\tilde{S}^{<}_{\textbf{p}}(p_{0})\tilde{\Delta}^{<}_{\textbf{q}-\textbf{p}}(\omega-p_{0})\nonumber\\
&=&-iY^{2}f_{F}^{-1}(\omega)\int\frac{d^{4}p}{(2\pi)^{4}}f_{F}(p_{0})f_{B}(\omega-p_{0})\tilde{\uprho}_{\textbf{p}}(p_{0})\tilde{\rho}_{\textbf{q}-\textbf{p}}(\omega-p_{0})\label{loopintfermi}
.\end{eqnarray}
Here $\tilde{S}^{<}$ and $\tilde{\Delta}^{<}$ are the thermal Wightman functions of the fermion and scalar in the bath, see (\ref{KMSfermi}) and (\ref{KMSscalar}). As in the scalar case, they have to be resummed to obtain consistent results.
The relaxation rate is again given by a product of spectral densities. 
Once again, the spectra in the bath may be highly complicated, but as long as the contributions from narrow poles dominate the integral (\ref{loopintfermi}), the discussion from section \ref{trilinearsection} can be repeated. The poles only contribute if quasiparticle decays (case (a)) or Landau damping (case (b)) are kinematically allowed on-shell. A Yukawa coupling is a three body interaction, thus a kinematically forbidden region (c) in which none of the involved quasiparticle can decay into the others is expected. In this region $\Upgamma_{\textbf{q}}$ only receives contributions from processes involving intermediate off-shell quasiparticles and is suppressed.
The phase space closure is more relevant in this case than for the decay of scalars into fermions because Pauli blocking does not suppress the efficiency of the decays near the critical temperature as one of the final state quasiparticles is a boson and the $f_{B}$-term in (\ref{loopintfermi}) can compensate the suppression by the $f_{F}$-term. 

In section \ref{ScalarwithYukawa} the two cases $\tilde{\fermionmass{m}}\gg\alpha T$ and $\tilde{\fermionmass{m}}\ll\alpha T$ were considered, where $\alpha$ is a typical coupling for the fermion in the bath. Here we focus on the case $\tilde{\fermionmass{m}}\gg\alpha T$. Then no resummation is necessary for the fermion propagator. 
For scalars replacing the vacuum masses by their thermal counterpart is a consistent approximation for the resummed spectral density if the only relevant resonances are the screened one particle states, their widths are small and the dispersion relation is approximately parabolic. This is e.g. the case if the leading order contribution to the effective mass comes from tadpole diagrams as shown in figure \ref{phi4}a) or a Yukawa coupling \cite{Thoma:1994yw}. Then we can use the expressions
\begin{eqnarray}
\tilde{\rho}_{\textbf{q}-\textbf{p}}(\omega-p_{0})&\approx& 2\pi{\rm sign}(\omega-p_{0})\delta((q-p)^{2}-\tilde{M}^{2})\label{ScalarSpectrApprox}\\
\tilde{\uprho}_{\textbf{p}}(p_{0})&\approx&2\pi{\rm sign}(p_{0})(\Slash{p}+\tilde{\fermionmass{m}})\delta(p^{2}-\tilde{\fermionmass{m}}^{2})\label{FermiSpectrApprox}
\end{eqnarray}
to evaluate (\ref{loopintfermi}), where the scalar spectral density $\tilde{\rho}$ is resummed, $\tilde{M}$ contains vacuum as well as thermal contributions while $\tilde{\fermionmass{m}}$ is a chiral vacuum  mass.
After performing the $p_{0}$ integration   reads (cf \cite{Weldon:1983jn})
\begin{eqnarray}\label{PhDFormula}
\lefteqn{\Sigma^{-}_{\textbf{q}}(\omega)=-iY^{2}\int\frac{d^{3}\textbf{p}}{(2\pi)^{2}}\frac{1}{2\tilde{\Omega}_{1}2\tilde{\Omega}_{2}}}\nonumber\\
&&\bigg((1-f_{F}(\tilde{\Omega}_{1})+f_{B}(\tilde{\Omega}_{2}))\Big((\tilde{\Omega}_{1}\gamma_{0}-\pmb{p\gamma}+\tilde{\fermionmass{m}})\delta(\omega-\tilde{\Omega}_{1}-\tilde{\Omega}_{2})\nonumber\\
&&\phantom{(1-f_{F}()+f_{B}(\tilde{\Omega}_{2}))}+(\tilde{\Omega}_{1}\gamma_{0}+\pmb{p\gamma}-\tilde{\fermionmass{m}})\delta(\omega+\tilde{\Omega}_{1}+\tilde{\Omega}_{2})\Big)\nonumber\\
&+&(f_{F}(\tilde{\Omega}_{1})+f_{B}(\tilde{\Omega}_{2}))\Big((\tilde{\Omega}_{1}\gamma_{0}-\pmb{p\gamma}+\tilde{\fermionmass{m}})\delta(\omega-\tilde{\Omega}_{1}+\tilde{\Omega}_{2})\nonumber\\
&&\phantom{(1-f_{F}()+f_{B}(\tilde{\Omega}_{2}))}+(\tilde{\Omega}_{1}\gamma_{0}+\pmb{p\gamma}-\tilde{\fermionmass{m}})\delta(\omega+\tilde{\Omega}_{1}-\tilde{\Omega}_{2})\Big)\bigg)
\end{eqnarray}
with $\tilde{\Omega}_{1}=(\textbf{p}^{2}+\tilde{\fermionmass{m}}^{2})^{\frac{1}{2}}$ and $\tilde{\Omega}_{2}=((\textbf{q}-\textbf{p})^{2}+\tilde{M}^{2})^{\frac{1}{2}}$.
The kinematic $\delta$-functions in (\ref{PhDFormula}) can be compared to those in (\ref{DiplomFormula}) and lead to similar kinematic restrictions and regions (a), (b) and (c) in which $\Psi$ exchanges energy with the bath via on-shell decays and inverse decays in case (a), on-shell Landau damping in case (b) or only off-shell processes in case (c). As in (\ref{DiplomFormula}), the integral in (\ref{PhDFormula}) is exactly zero in case (c), but would receive a  non-zero contribution if the widths of the resonances in the bath were taken into account. 

When thermal corrections to $\tilde{\uprho}_{\textbf{p}}$ are relevant, also the fermion propagator has to be resummed. The evaluation of $\Upgamma_{\textbf{q}}$ then becomes more involved, but it is obvious that, when the widths are small, (\ref{loopintfermi}) is again dominated by pole contributions and kinematic arguments based on approximate energy conservation in reactions between on-shell quasiparticles can be used.
However, as usual for fermions, this in general cannot be done by simply putting in thermal masses by hand.
The decay into fermions with resummed spectral densities (\ref{HTLrho}) has e.g. been studied in \cite{Thoma:1994yw,Kiessig:2010pr} where it was found that $\Upgamma_{\textbf{q}}$ is non-zero above the critical temperature associated with the decay into particles because phase space for the decay into scalars and on-shell holes closes at higher $T$ only. 

In the following we will simply assume that (\ref{ScalarSpectrApprox}), (\ref{FermiSpectrApprox}) hold at all temperatures of consideration. 
\begin{figure}
  \centering
 \psfrag{T}{$T/\fermionmass{m}$}
  \psfrag{Gamma}{$\Upgamma_{\textbf{0}}(T)/\Upgamma_{\textbf{0}}(T=0)$}
  \psfrag{c}{$c)$}
  \psfrag{d}{$d)$}
  \psfrag{e}{$e)$}
  \psfrag{f}{$f)$}
    \includegraphics[width=10cm]{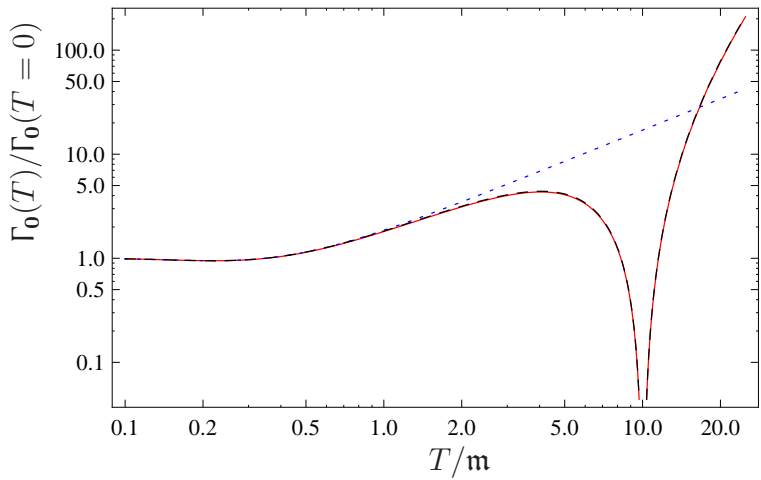}
    \caption{$\Upgamma_{\textbf{0}}$ in units of its zero temperature value as a function of $T/\fermionmass{m}$ for $|\textbf{q}|=\fermionmass{m}/5$. The analytic expression obtained in appendix \ref{YukawaSelfEnergy} without (blue dotted) and with thermal mass (red solid) for the scalar is compared to the approximation (\ref{UpgammaApprox}) with thermal mass (black dashed). For the scalar we took a vacuum mass $\tilde{m}=0.4\fermionmass{m}$ and a quartic self coupling of strength $\lambda=1/5$.  \label{YukawaRate}}
\end{figure}
If $\tilde{\fermionmass{m}}$ is negligible the integral in (\ref{PhDFormula}) can be solved analytically for arbitrary $\textbf{q}$ and $\omega$, see appendix \ref{YukawaSelfEnergy}. 
The analytic expression is rather complicated, but for nonrelativistic on-shell quasiparticles with $|\textbf{q}|<0.5\fermionmass{m}$ 
the quasiparticle width $\Upgamma_{\textbf{q}}$ 
can be well-approximated by
\begin{equation}\label{UpgammaApprox}
\Upgamma_{\textbf{q}}\simeq\frac{Y^2\fermionmass{m}_{-}^2}{16\pi\upomega_{\textbf{q}}}\left(1-f_{F}(\fermionmass{m}_{-}/2)+f_{B}(\fermionmass{m}_{+}/2)\right)
\end{equation}
with $\fermionmass{m}_{\pm}=(\fermionmass{m}^{2}\pm\tilde{M}^{2})/\fermionmass{m}$. For $|\textbf{q}|\sim\fermionmass{m}$ this approximation becomes inaccurate.  
Figure \ref{YukawaRate} compares the approximation (\ref{UpgammaApprox}) to the analytic solution of (\ref{PhDFormula}) given in appendix \ref{YukawaSelfEnergy}.
The forbidden region (c) here shrinks to a point because one of the particles in the loop is massless, thus $q_{th1}=q_{th2}$. 

Note that the requirement to keep the fermion in the loop massless is indeed quite strong. Even if the fermion is massless in the vacuum, $\tilde{\fermionmass{m}}=0$, it will in general obtain a thermal mass via radiative corrections because thermal masses are not protected by the chiral symmetry \cite{Weldon:1982bn}, see (\ref{HTLrho}).

\section{Comparison to Boltzmann Equations}\label{comparisontoBE}
Many nonequilibrium phenomena, in particular in cosmology, can be studied in terms of Boltzmann equations. They are  by construction Markovian and formulated in term of number densities or single particle phase space distributions for classical particles. There are phenomena, including coherent oscillations or memory effects, that cannot be described in this framework. 
In absence of these one should recover the Boltzmann equations as a limit of the quantum field theoretical description.

The rate $\Gamma_{\textbf{q}}$ determines at which rate the field mode $\textbf{q}$ gains or loses energy. In the quasiparticle regime, the gain and loss rates can be understood in terms of approximately energy conserving decays and scatterings of quasiparticles. This suggests that the system can be viewed as a gas of quasiparticles and one is tempted to simply replace  $m\rightarrow M$ or $\omega_{\textbf{q}}\rightarrow\Omega_{\textbf{q}}$ in the standard Boltzmann equations. In the following we show that this is not necessarily correct.

We consider the simplest case, a single real scalar field coupled to a large thermal bath.
We restrict ourselves to the quasiparticle regime as it is clear that outside this regime Boltzmann type equations have no meaning. 
For a single scalar without additional quantum numbers (such as flavour), coherent oscillations play no role. When the bath temperature is constant, the quasiparticle approximation allows to sum up all memory effects and obtain the analytic expressions (\ref{exponential}) and (\ref{narrowPlus}) for the spectral function and statistical propagator. We also assume that the dispersion relations are in good approximation parabolic. 

The Boltzmann equations are formulated in terms of one-particle phase space distribution functions. As the definition of a number density in the field theoretical approach is ambiguous, these are not well-suited for a comparison between the two. We will therefore perform the comparison in terms of the energy stored in each mode, which is a well-defined observable in both approaches.

\paragraph{Boltzmann Equations}
Let us consider the Boltzmann equation for particles of momentum $\q$
and energy $E_{\q}$. The competition between a gain and a loss term, $\gamma^{<}_{\q}$ and $\gamma^{>}_{\q}$ respectively, determines the change of the particle number $f_{\q}(t)$ for the mode $\q$,  
\begin{equation}\label{boltz1}
\partial_{t}f_{\q}(t)=(1+f_{\q}(t))\gamma^{<}_{\q}-f_{\q}(t)\gamma^{>}_{\q}\ .
\end{equation}
When the medium is in equilibrium, production and decay rates satisfy the detailed balance relation, 
\begin{align}
\gamma^{<}_{\q} &= e^{-\beta E_{\q}}\gamma^{>}_{\q}
\equiv f_{\text{B}}(E_{\q}) \gamma_{\q} \ ,
\end{align}
which implies $\gamma_{\textbf{q}}=\gamma^{>}_{\textbf{q}}-\gamma^{<}_{\textbf{q}}$, analogue to (\ref{GammaDefPi}).
The rate $\gamma_{\q}$ is usually computed at $T=0$ from scattering cross sections or, via the optical theorem, the imaginary part of the self energy, 
\begin{equation}
\gamma_{\q} = -\frac{{\rm Im}\Pi^R_{\q}(E_{\q})}{E_{\q}}\Big|_{T=0} 
.\end{equation}
Using these relations, the Boltzmann equation (\ref{boltz1}) can be 
written in the form\footnote{Note that the argument of $f_{\textbf{q}}(t)$ is a time while $f_{B}(E_{\textbf{q}})$ does not depend on time and we have explicitly written the particle energy as the argument.}
\begin{equation}
\partial_{t}f_{\q}(t) = -\gamma_{\q}(f_{\q}(t)-f_{\text{B}}(E_{\q}))\ ,
\end{equation}
with the solution
\begin{equation}\label{BoltzmannSol}
f_{\q}(t)=f_{\text{B}}(E_{\q})+\left(f_{\q}(0)
-f_{\text{B}}(E_{\q})\right)e^{-\gamma_{\q}t}\ 
.\end{equation}
The energy for each mode is obtained by multiplying (\ref{BoltzmannSol})
with $E_{\q}$ 
\begin{eqnarray}\label{EBoltzmannSol}
\epsilon^{\phi}_{\q,BE}(t)=E_{\q}\left( f_{B}(E_{\textbf{q}})+\left(f_{\q}(0)
-f_{\text{B}}(E_{\q})\right)e^{-\gamma_{\q}t}\right) 
.\end{eqnarray}

\paragraph{Kadanoff-Baym Equations}
In the field theoretical description the energy per $\phi$-mode is given by (\ref{Energie}),
\begin{displaymath}
\epsilon^{\phi}_{\q}(t) = \frac{1}{2}\left(\partial_{t_1}\partial_{t_2} 
+ \omega_{\q}^2 \right)\left(\Delta^+_{\q}(t_1,t_2)+\langle\phi_{\q}(t_{1})\rangle\langle\phi_{\q}(t_{2})\rangle\right)\big|_{t_1=t_2=t} 
.\end{displaymath}
The $\langle\phi\rangle\langle\phi\rangle$-part corresponds to the energy that 
is stored in the mean value $\langle\phi\rangle$ of the field while the $\Delta^{+}$-part comes from its fluctuations, to be interpreted as (quasi)particles. 
An explicit expression for $\epsilon^{\phi}_{\q}(t)$ can be found by inserting (\ref{narrowPlus}) and (\ref{fieldvalueapprox}),
\begin{eqnarray}\label{energiedichtephi}
\epsilon^{\phi}_{\q}(t)&\simeq& \frac{\dot{\phi}_{\q,\text{in}}^{2}}{2\Omega_{\q}^{2}}\left(\omega_{\q}^{2}\sin^{2}(\Omega_{\q}\tp)+\Omega_{\q}^{2}\cos^{2}(\Omega_{\q}\tp)\right)e^{-\Gamma_{\q}\tp}\nonumber\\
&+&\frac{\dot{\phi}_{\q,\text{in}}\phi_{\q,\text{in}}}{\Omega_{\q}}\sin(\Omega_{\q}\tp)\cos(\Omega_{\q}\tp)\left(\omega_{\q}^{2}-\Omega_{\q}^{2}\right)e^{-\Gamma_{\q}\tp}\nonumber\\
&+&\frac{\phi_{\q,\text{in}}^{2}}{2}\left(\omega_{\q}^{2}\cos^{2}(\Omega_{\q}\tp)+\Omega_{\q}^{2}\sin^{2}(\Omega_{\q}\tp)\right)e^{-\Gamma_{\q}\tp}\nonumber\\
&+&\frac{\Delta^{+}_{\q,\text{in}}}{2}
\left(\frac{\omega_{\q}^{2}-\Omega_{\q}^{2}}{2}\cos(2\Omega_{\q}\tp)+\frac{\omega_{q}^{2}+\Omega_{\q}^{2}}{2}\right)e^{-\Gamma_{\q}\tp}\nonumber\\
&-&\frac{\ddot{\Delta}^{+}_{\q,\text{in}}}{2\Omega_{\q}^{2}}
\left(\frac{\omega_{\q}^{2}-\Omega_{\q}^{2}}{2}\cos(2\Omega_{\q}\tp)-\frac{\omega_{q}^{2}+\Omega_{\q}^{2}}{2}\right)e^{-\Gamma_{\q}\tp}\nonumber\\
&+&\frac{\dot{\Delta}^{+}_{\q;\text{in}}}{\Omega_{\q}}
\frac{\omega_{\q}^{2}-\Omega_{\q}^{2}}{2}\sin(2\Omega_{\q}\tp)e^{-\Gamma_{\q}\tp}\nonumber\\
&+&\left(\frac{1}{2}+f_{B}(\Omega_{\q})\right)\frac{\omega_{q}^{2}+\Omega_{\q}^{2}}{2\Omega_{\q}}\left(1-e^{-\Gamma_{\q}\tp}\right)
.\end{eqnarray}
The first three terms represent the energy of the mean field $\langle\phi\rangle$. We will ignore them in the following and set $\dot{\phi}_{\q,\text{in}}=\phi_{\q,\text{in}}=0$.

\paragraph{Comparison}
Obviously neither identifying $E_{\q}$ with $\omega_{\q}$ nor with $\Omega_{\q}$ 
generally leads to equivalence between (\ref{EBoltzmannSol}) and (\ref{energiedichtephi}).
The oscillations in (\ref{energiedichtephi}), which contains the solution to a second order differential equation, remain present in the limit $\Omega_{\q}\rightarrow\omega_{\q}$, $\Gamma_{\textbf{q}}\rightarrow\gamma_{\textbf{q}}$. Therefore the system cannot even be described by Boltzmann equations if the relaxation rate $\Gamma_{\textbf{q}}$ is identified with its vacuum value and all modifications of the dispersion relation are neglected.

However, a consistent comparison between a quantum mechanical and a 
classical observable can only be done if the quantum system is set up 
in an initial state has a counterpart in the classical theory, namely one 
of definite initial (quasi)particle number. 

We first consider the simplest case and assume that $\phi$ initially is in equilibrium, but at a different temperature $\tilde{T}$ than the bath. One can then think of the system as a two heat baths at different temperatures $T$ and $\tilde{T}$ that are brought in touch at $t=0$. One bath ($\phi$ only) is much smaller than the other, neglecting backreaction the whole system is expected to equilibrate at temperature $T$ for times $t\gg 1/\Gamma_{\textbf{q}}$. The initial density matrix can be written as $\varrho_{ini}\propto e^{-\tilde{\beta} {\rm H}_{\phi}}\otimes e^{-\beta {\rm H}_{\rm bath}}$ and the initial statistical propagator for $\phi$ is given by
\begin{equation}
\Delta^{+}_{\textbf{q}}(t_{1},t_{2})=\frac{1}{\Omega_{\textbf{q}}}\left(\frac{1}{2}+\tilde{f}_{B}(\Omega_{\textbf{q}})\right)\cos((t_{1}-t_{2})\Omega_{\textbf{q}}) \phantom{XXX}{\rm at }\phantom{X}t_{i}
.\end{equation}
Here $\tilde{f}_{B}$ is the Bose-Einstein distribution with temperature $\tilde{T}$.
The energy (\ref{energiedichtephi}) then reads
\begin{eqnarray}
\lefteqn{\epsilon^{\phi}_{\q}(t)=\frac{\Omega_{\textbf{q}}^{2}+\omega_{\textbf{q}}^{2}}{2\Omega_{\textbf{q}}}\left( \left(\frac{1}{2}+\tilde{f}_{B}(\Omega_{\textbf{q}})\right)e^{-\Gamma_{\textbf{q}}t}+\left(\frac{1}{2}+f_{B}(\Omega_{\textbf{q}})
)\right)\left(1-e^{-\Gamma_{\q}t}\right)\right)}\nonumber\\
&=&\frac{\Omega_{\textbf{q}}^{2}+\omega_{\textbf{q}}^{2}}{2\Omega_{\textbf{q}}}\left( \left(\frac{1}{2}+f_{B}(\Omega_{\textbf{q}})\right)+\left(\tilde{f}_{B}(\Omega_{\textbf{q}})
-f_{\text{B}}(\Omega_{\q})\right)e^{-\Gamma_{\q}t}\right).\label{thermalinitial}
\end{eqnarray}
The solution (\ref{thermalinitial}) describes the transition from one thermal state to another. It can be compared to (\ref{EBoltzmannSol}) with $f_{\textbf{q}}(0)=\tilde{f}_{B}(E_{\textbf{q}})$. Apart from the additional term $1/2$ in (\ref{thermalinitial}), which is simply the vacuum energy of the mode $\textbf{q}$, the expression in the bracket looks exactly like the solution for a Boltzmann equation for quasiparticles with energy $\Omega_{\textbf{q}}$ and damping rate $\Gamma_{\textbf{q}}$. However, it is not multiplied by $\Omega_{\textbf{q}}$, but $(\Omega_{\textbf{q}}^{2}+\omega_{\textbf{q}}^{2})/(2\Omega_{\textbf{q}})$. If one loosely interprets the term in the brackets as quasiparticle number for mode $\textbf{q}$, it appears that it follows a Boltzmann equation while the energy does not.

The same effect can be observed when inserting the more general initial conditions
\begin{center}
\begin{tabular}{c c c}
$\Delta^{+}_{\q,\text{in}}  
= \frac{1}{\Omega_{\q}}
\left(\frac{1}{2}+f_{\q,{\rm ini}}\right)$, & $\dot{\Delta}^{+}_{\q,\text{in}}  
 = 0 $, & $\ddot{\Delta}^{+}_{\q,\text{in}}  = \Omega_{\q}
\left(\frac{1}{2}+f_{\q,{\rm ini}}\right)$,
\end{tabular}
\end{center}
 sometimes referred to as \textit{tsunami initial conditions}. Here $f_{\q,{\rm ini}}$ is a general function of $\textbf{q}$ that can loosely be interpreted as initial occupation number of that mode. The solution in this case reads
\begin{eqnarray}
\epsilon^{\phi}_{\q}(t)=\frac{\Omega_{\textbf{q}}^{2}+\omega_{\textbf{q}}^{2}}{2\Omega_{\textbf{q}}}\left( \left(\frac{1}{2}+f_{B}(\Omega_{\textbf{q}})\right)+\big(f_{\textbf{q},{\rm ini}}
-f_{\text{B}}(\Omega_{\q})\big)e^{-\Gamma_{\q}t}\right).\label{tsunami}
\end{eqnarray}
Again, the term in the brackets may be interpreted as the number of quasiparticles per mode $\q$ that follows a Boltzmann equation. However, the energy, which is the observable quantity, does not. Obviously (\ref{thermalinitial}) is a special case of (\ref{tsunami}) with $f_{\textbf{q},{\rm ini}}=f_{B}(\Omega_{\textbf{q}})$.
One way to interpret this result can be seen by rewriting it as
\begin{eqnarray}\label{QPBoltzman}
\lefteqn{\epsilon^{\phi}_{\q}(t)=\Omega_{\q}\left(f_{B}(\Omega_{\q})+\left(f_{\q,{\rm ini}}-f_{B}(\Omega_{\q})\right)e^{-\Gamma_{\q}\tp}\right)}\nonumber\\
&+&\frac{\omega_{\q}^{2}-\Omega_{\q}^{2}}{2\Omega_{\q}}f_{B}(\Omega_{\q})+\frac{\omega_{\q}^{2}-\Omega_{\q}^{2}}{2\Omega_{\q}}\left(f_{\textbf{q},{\rm ini}}-f_{B}(\Omega_{\q})\right)e^{-\Gamma_{\q}\tp}+\frac{1}{2}\frac{\omega_{\q}^{2}+\Omega_{\q}^{2}}{2\Omega_{\q}}
.\end{eqnarray}
The first line is a Boltzmann equation for quasiparticles. The second line may be interpreted as a vacuum energy. However, it cannot be removed by simply shifting the energy of the ground state by a constant because it depends on temperature and time. It appears even tough we have not considered the coupling of the particles to the time dependent background $\langle\phi\rangle$, which is of higher order in the coupling between $\phi$ and the bath.
Therefore, even though the exchange of energy between different field modes and fields can be understood in terms of elementary processes involving quasiparticles, our result suggests that the total energy of $\phi$ is not that of a quasiparticle gas. A similar observation was made in equilibrium in \cite{Anisimov:2008dz}, where it was found that due to this effects the contribution of $\phi$ to the total pressure of the system can be negative. 

 \section{Application to Reheating after Cosmic Inflation}\label{InflatonSection}
We now discuss an application of the previous results.
It has been suggested in \cite{Kolb:2003ke} that large thermal masses in the primordial plasma can impose an upper bound on the temperature in the early universe. The authors argue that, if reheating is fuelled by the decay of inflaton particles, this process would stop 
when the thermal masses of the would-be decay products become sufficiently large to make the decay kinematically impossible.

Any bound on the temperature in the early universe has far reaching cosmological consequences as it determines abundance of thermal relics. It is, for instance, of crucial importance in the context of the gravitino problem \cite{Pagels:1981ke} and thermal leptogenesis \cite{Fukugita:1986hr}. A large temperature can also be related to the fate of moduli \cite{Yokoyama:2006wt} and the possible decompaticfication of extra dimensions \cite{Buchmuller:2004xr}. 

\subsection{Cosmic Inflation and Reheating}
Cosmic inflation \cite{Guth:1980zm}, the idea that the universe underwent a period of accelerated expansion during its very early history, has proven an extremely successful concept in explaining various unsolved mysteries of big bang cosmology. These include the flatness and horizon problems, the absence of topological defects and the origin of the temperature fluctuations in the cosmic microwave background, thought to be the seeds of structure formation \cite{kt90}. 
Accelerated expansion requires a negative pressure component to dominate the energy content of the universe. The simplest way to realise this is by the potential energy $V(\phi)$ of a scalar field $\phi$, the inflaton. During the inflationary era all other forms of energy are diluted and virtually all energy is stored in the zero mode of the mean field $\langle\phi\rangle$, which slowly rolls down the potential $V(\phi)$ until it starts oscillating around its minimum.
It leaves a cold and empty universe with very small entropy. The conditions at the beginning of the radiation dominated era, corresponding to a hot big bang, are then created by \textit{reheating}. During this process $\langle\phi\rangle$ performs damped oscillations around its potential minimum, dissipating its energy into all other degrees of freedom. This fills the universe with particles and heats up the primordial plasma. 

We restrict ourselves to scenarios where only one scalar inflaton field $\phi$ is relevant during reheating. This includes scenarios in which the underlying physics can be parameterised in terms of a scalar degree of freedom. We describe the interactions of $\phi$ with matter by the Lagrangian (\ref{L}), where $\phi$ is identified with the inflaton. $\chi_{i}$ ad $\Psi_{i}$ are some scalar and fermionic fields that directly couple to the inflaton and via $\mathcal{L}_{\chi_{i} \mathrm{int}}$ and $\mathcal{L}_{\Psi_{i} \mathrm{int}}$ directly or indirectly interact with all other fields in the primordial plasma  (including the SM).
This can be regarded as a toy model, but is sufficient to describe the kinematic aspects of dissipation into bosons and fermions we are interested in\footnote{It is likely that gauge fields have played an important role during reheating, but the kinematic aspects we are interested in are the same as for scalars.}.

As soon as $\phi$ has dissipated some fraction of its energy into other degrees of freedom, the presence of the plasma formed by its decay products affects the ongoing reheating process, which in consequence has to be treated by methods of nonequilibrium field theory.

The details of the reheating process are largely unknown due to our lack of knowledge about the underlying physics. It is nevertheless possible to study general features and their cosmological implications.
Here we study the effect of thermal masses in the plasma on the time evolution of the temperature. We are in particular interested in two quantities, the maximal temperature $T_{MAX}$ that the universe was exposed to and the reheating temperature $T_{R}$ at onset of the radiation dominated era.

\subsubsection{Effective Masses}\label{effectivemassessubsub}
While the part of the inflaton potential that is relevant during inflation can be probed by cosmological data \cite{Finelli:2009bs}, very little is known about the region around the minimum and the inflaton mass $M$\footnote{Here we use a capital letter to denote the mass of $\phi$ at $T=0$ because it should be regarded as an effective mass that might depend on the values of other fields.} during reheating. Even without thermal effects $M$ should be regarded as an effective mass that may depend on the values of other fields. 
We make two assumptions about $M$. First, it is much larger than the vacuum masses $m_{i}$, $\fermionmass{m}_{i}$  of the other fields. Second, it does not depend on time during reheating. This is consistent with the assumptions that $\phi$ is the only relevant out-of-equilibrium field and weakly coupled, making thermal corrections to its dispersion relation negligible. Then we can treat the $\phi$-mass as a constant parameter $M$.

\begin{figure}
  \centering
 \psfrag{a}{$a)$}
  \psfrag{b}{$b)$}
  \psfrag{c}{$c)$}
  \psfrag{d}{$d)$}
  \psfrag{e}{$e)$}
  \psfrag{f}{$f)$}
    \includegraphics[width=5cm]{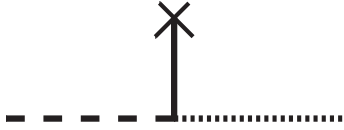}
    \caption{Contribution to the $\chi_{i}$ self energy from coupling to $\langle\phi\rangle$.\label{couplingtofield2}}
\end{figure}

Regarding the effective masses and dispersion relations of the other fields, there are three different contributions. The first is their vacuum mass.
The second comes from interactions with the plasma. If the plasma is in equilibrium it can be treated by the methods discussed in the previous sections. 
The third is due to the coupling to the mean field $\langle\phi\rangle$. For the $\chi_{i}$ it arises from diagrams as shown in figure \ref{couplingtofield2}. 

In the following we always neglect all vacuum masses except the $\phi$-mass $M$. This is justified since $T_{R}$ is likely to be far above the electroweak scale\footnote{It is clear from the previous sections that the inclusion of vacuum masses is straightforward.}.
We furthermore assume that all medium effects on dispersion relations in the plasma can be parameterised by momentum independent thermal masses. This assumption is for simplicity and due to our lack of knowledge about the exact decomposition and interactions of the primordial plasma. It can, however, also be justified when $\phi$-quanta decay into much lighter states with large momenta (cf. sections \ref{ScalarwithYukawa} and \ref{scatteringimportance}). Following the discussion in section \ref{trilinearsection}, a consistent treatment requires to use full resummed propagators in the bath. This includes not only the thermal masses, but also the widths.

For a quantitative treatment of both, the interaction terms $\mathcal{L}_{\chi_{i} \mathrm{int}}$ and $\mathcal{L}_{\Psi_{i} \mathrm{int}}$ have to be specified. 
Following the considerations in section \ref{OffShellDiscuss} we again introduce two different couplings $\alpha$ and $\tilde{\alpha}$ that govern the real and imaginary part of the self energies. 
The thermal masses can be estimated as $M_{\mathcal{X}}\simeq \alpha T/2$. 
Equation (\ref{OffShell}) shows that a consistent treatment of contributions to the relaxation rate involving virtual $\chi_{i}$, $\Psi_{i}$ quanta requires specification of the functional dependence of the discontinuities on energy and momentum, thus taking $\Gamma_{i}\sim\tilde{\alpha}^{2}T$ is not sufficient. As in section \ref{OffShellDiscuss}, we again assume that the main contributions to the $\Gamma_{i}$ comes from Yukawa couplings to some fermions in the bath. This allows to use the result from section \ref{ScalarwithYukawa}.

Finally, there is the contribution to the effective masses from diagrams as shown in figure \ref{couplingtofield2}. These terms mix $\chi_{1}$ and $\chi_{2}$.
For $m_{1},m_{2}\ll \alpha T$ the effective $\chi_{i}$-mass after diagonalisation of the mass term is 
\begin{equation}\label{chieffmass}
M_{i}^{2}\sim \alpha^{2}T^{2}+g\langle\phi\rangle 
.\end{equation}
Here we have for simplicity only considered the trilinear interaction for the $\chi_{i}$. A similar term $\sim Y\langle\phi\rangle$ arises as correction to the fermion mass.

\subsubsection{Applicability of our Results}
Denoting by $\Phi$ the amplitude of $\langle\phi\rangle$, (\ref{chieffmass}) allows to distinguish the regimes $g\Phi>\alpha^{2}T^{2}$ and $g\Phi<\alpha^{2}T^{2}$. 
The expectation value $\langle\phi\rangle=\langle\phi_{\textbf{q}}(t)\rangle$ oscillates rapidly at frequencies $\gtrsim M$, faster than the relaxation time in the plasma. Therefore the formulae used in the previous sections, valid for coupling to a large bath in equilibrium, are not applicable for $g\Phi\gtrsim\alpha^{2}T^{2}$. In this case all degrees of freedom are far from equilibrium\footnote{This also means that there is no well-defined temperature $T$ in (\ref{chieffmass}), but there is still a contribution due to the coupling to the plasma at order $\alpha^{2}$.} and there are various nontrivial collective effects such as \textit{parametric resonance}, which can not only produce particles with masses larger than $M$, but also potentially observable relics \cite{Traschen:1990sw,GarciaBellido:1999sv}.
However, we are interested in the effect of thermal masses (contributions from coupling to the background plasma). It is clear that in this regime they only give rise to a subleading correction. 

For $g\Phi\ll\alpha^{2}T^{2}$ the only relevant timescales are the relaxation times of the decay products ($\tau_{\mathcal{X}}$) and $\phi$ ($\tau_{\phi}$). The relaxation times $\tau_{\mathcal{X}}$ are usually much longer than the period of one $\phi$-oscillation, but short compared to $\tau_{\phi}$, which characterises the duration of the reheating process. On timescales $\lesssim$ $\tau_{\phi}$ the system can effectively be described as a single scalar field that is weakly coupled to a thermal bath with many degrees of freedom and the methods from the previous sections can be employed. In this regime $\phi$ dissipates via perturbative decay of single quanta. 

Since $T$ and $\Phi$ change with time, the system can pass through various stages.
In many scenarios the inflaton releases much of its energy during an initial phase of \textit{preheating} during which the time dependent mean field dominates. However, the decomposition of the primordial plasma at the onset of the radiation dominated phase may nevertheless be determined by the temperature during a later phase of reheating by perturbative decay. The reason is that the density of any particles produced during preheating is diluted by the expansion of the universe during reheating.
We denote by $\epsilon_{\phi}$ the energy density of $\phi$ while $\epsilon_{R}$ refers to the energy density of all other $g_{*}$ relativistic degrees of freedom. We refer to the period during which $\epsilon_{\phi}>\epsilon_{R}$ as \textit{reheating stage}. For $\epsilon_{\phi}<\epsilon_{R}$ the universe enters the radiation dominated era. The temperature $T_{R}$ at $\epsilon_{\phi}=\epsilon_{R}$ is the reheating temperature. It is generally lower than the highest temperature $T_{MAX}$ that the universe was exposed to, see figures \ref{alpha1isalpha2is02g5}-\ref{alpha1is001alpha2is039g4h2g}. Indeed, during most of the reheating stage the temperature actually decreases while $\epsilon_{R}/\epsilon_{\phi}$ grows. 
Since the $\phi$-oscillations are fast,  $\epsilon_{\phi}$ (averaged over a few oscillations) redshifts like nonrelativistic matter even at temperatures $T>M$. 
In contrast to that the bath quanta are relativistic and redshift like radiation as their average momenta in the plasma have values $\sim T$ while thermal masses are $\sim \alpha T$, leading to a relative dilution of particles produced at early stages. This makes $T_{R}$ a very decisive quantity.

We restrict ourselves to initial conditions corresponding to perturbative decay, including the possibility that this follows an initial stage during which other mechanisms dominated.
As $\Phi$ decreases with time the inequality $g\Phi\gtrsim \alpha^{2} T^{2}$ remains valid at all later times. The question whether large thermal masses can stop the heating of the plasma by blocking the phase space then arises once $\alpha T\sim M$.

The relevance of the perturbative decay depends on the amount of energy still stored in $\phi$ when it takes over. We demand $\epsilon_{\phi}\gg \epsilon_{R}$ while  $g\Phi\ll\alpha^{2}T^{2}$.
One can estimate $\epsilon_{\phi}\sim V(\phi)\gtrsim M^{2}\Phi^{2}$ and $\epsilon_{R}\simeq \pi^2 g_{*}T^{4}/30$. This estimate for $\epsilon_{\phi}$ can also be obtained from (\ref{energiedichtephi}) with $\Omega_{\textbf{q}}=\omega_{\textbf{q}}$ and $\ddot{\Delta}^{+}_{\q;\text{in}}=\dot{\Delta}^{+}_{\q;\text{in}}=\Delta^{+}_{\q;\text{in}}=0$, 
\begin{equation}\label{InflatonEnergie}
\epsilon^{\phi}_{\q}(t)=\frac{1}{2}\left(\dot{\phi}_{\q,\text{in}}^{2}+\omega_{\textbf{q}}^{2}\phi_{\q,\text{in}}^{2}\right)e^{-\Gamma_{\q}\tp}
,\end{equation}
which for $\dot{\phi}_{\q,\text{in}}^{2}\ll \phi_{\q,\text{in}}^{2}$ (slow roll) and $\textbf{q}=0$ reproduces the above estimate and is justified because $\phi$ is very weakly coupled and all energy initially is stored in the mean field\footnote{Note that strictly speaking $\epsilon^{\phi}_{\q}(t)$ is an energy per mode while $\epsilon_{\phi}$ is an energy density.}. The usual estimate for $\epsilon_{R}$ is derived from classical thermodynamics using single particle distribution functions, see e.g. \cite{kt90}. Our results from section \ref{comparisontoBE} suggest that it should receive corrections even in the quasiparticle regime, but we shall ignore this subtlety for the moment.
It straightforward to formulate the condition $\alpha/g\gg (\pi\sqrt{g_{*}})/(30M)$. For $g_{*}\sim 100$ this means $\alpha\gg g/M$ which is generally the case. 
\begin{figure}
  \centering
 \psfrag{T}{$T/\fermionmass{m}$}
  \psfrag{Gamma}{$\Upgamma_{\textbf{0}}(T)/\Upgamma_{\textbf{0}}(T=0)$}
  \psfrag{c}{$c)$}
  \psfrag{d}{$d)$}
  \psfrag{e}{$e)$}
  \psfrag{f}{$f)$}
    \includegraphics[width=5cm]{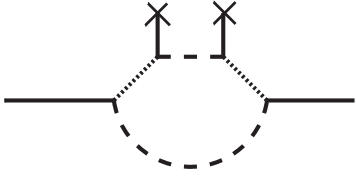}
    \caption{Example for a contribution from coupling to the mean field to the self energy $\Pi^{R}$, dashed and dotted lines denote $\chi_{1}$ and $\chi_{2}$ propagators, the crosses indicate $\langle\phi\rangle$.\label{fieldcoupling}}
\end{figure}

Equation (\ref{energiedichtephi}), or (\ref{InflatonEnergie}), shows that the dissipation rates for the oscillating field $\langle\phi\rangle$ and its fluctuations near the potential minimum are both given by $\Gamma\equiv\Gamma_{\textbf{q}=\textbf{0}}$ computed in section \ref{scalarsectioN}\footnote{Equation (\ref{fieldvalueapprox}), and thus (\ref{InflatonEnergie}), strictly holds only in the   harmonic approximation of the potential $V(\phi)$. In the late phase of reheating which determines $T_{R}$ this can be justified. Furthermore, the interpretation of $\Gamma_{\textbf{q}}$ as damping can be extended beyond the linear regime \cite{Yokoyama:2004pf}.}. The frequency of the oscillations is $\Omega_{\textbf{q}=\textbf{0}}\simeq M$.
It is not fully consistent to use (\ref{energiedichtephi}) or (\ref{InflatonEnergie}) because the solutions for the propagators $\Delta^{-}$ (\ref{dmin}), $\Delta^{+}$ (\ref{solution}) and field value $\langle\phi\rangle$ (\ref{fieldvalue}) from which they were derived were found under the assumption of time translation invariant self energies computed from couplings to a large bath of constant temperature. This condition is violated in three ways during reheating. 
On one hand, Hubble expansion cools the plasma continuously. 
On the other hand, in the case of reheating backreaction (the effect that the dissipation of $\phi$ has on the bath) is certainly not negligible by the very definition of the process. 
Finally, contributions to the $\phi$-self energy that involve $\phi$ itself (see figure \ref{fieldcoupling}) can have a sizeable effect despite the large number of degrees of freedom if $\langle\phi\rangle$ is sufficiently large.
This last point in principle imposes the strongest restrictions on the applicability of our results, but is not relevant here as we restricted ourselves to initial conditions $g\Phi_{ini}<\alpha^{2}T_{ini}^{2}$. 

Hubble expansion and backreaction can be parameterised by a time dependent temperature. As long as the relaxation time of the $\mathcal{X}_{i}$ is much shorter than the timescales associated with $\phi$-relaxation and Hubble expansion, there is a separation of timescales. 
In any case we are not interested in the details of the field oscillations.
That allows to use the expressions for $\Gamma$ obtained in section \ref{scalarsectioN} with a time dependent temperature and interpret them as the dissipation rate at each moment in time.


\subsection{Thermal History during Reheating}
\subsubsection{Boltzmann Equations}
In the following we use the results from the previous sections to study the time evolution of the temperature during reheating.
We describe the system in terms of effective Boltzmann equations for $\epsilon_{\phi}$ (averaged over a few oscillations) and $\epsilon_{R}$\footnote{This of course assumes that memory and coherence effects are not essential.},
\begin{eqnarray}
\dot{\epsilon}_{\phi}+3H\epsilon_{\phi}+\Gamma\epsilon_{\phi} &=&0\\
\dot{\epsilon}_{R}+4H\epsilon_{R}-\Gamma\epsilon_{\phi}&=&0
.\end{eqnarray}
Here $H$ is the Hubble parameter. These equations can be written more conveniently in the variables $\Phi=\epsilon_{\phi}M^{-1}a^{3}$ and $R=\epsilon_{R} a^{4}$, 
\begin{eqnarray}
\frac{d\Phi}{dx}&=&-\frac{\Gamma}{Hx}\Phi\label{Boltzmann1}\\
\frac{dR}{dx}&=&\frac{\Gamma}{H}\Phi\label{Boltzmann2}
\end{eqnarray}
with
\begin{eqnarray}
H&=&\left(\frac{8\pi}{3}\right)^{1/2}\frac{M^{2}}{M_{P}}\left(\frac{R}{x^{4}}+\frac{\Phi}{x^{3}}\right)^{1/2}\\
T&=&\frac{M}{x}\left(\frac{30}{\pi^{2}g_{*}}R\right)^{1/4}\label{TinBE}
\end{eqnarray}
where $x=aM$, $a$ is the scale factor, $M_{P}$ the Planck mass and $g_{*}$ the number of relativistic degrees of freedom in the bath.
\subsubsection{Dissipation into Bosons}
We first study the dissipation into the fields $\chi_{i}$. 
The relevant parameters are the inflaton mass $M$ and the couplings $g$, $h_{1}=h_{2}=h$. 
The fields $\chi_{1}$ and $\chi_{2}$ can have different thermal masses if their dominant couplings $\alpha_{1}$ and $\alpha_{2}$ to the rest of the bath are different. While keeping $\alpha_{1}$ and $\alpha_{2}$ independent, we assume that their widths are governed by the same parameter $\tilde{\alpha}_{1}=\tilde{\alpha}_{2}=\tilde{\alpha}$. It would be straightforward to include the possibility of different widths, but this does not lead to any new effects (while $M_{1}\neq M_{2}$ can give rise to Landau damping as a new dissipation mechanism). Without loss of generality we take $\alpha_{2}\geq\alpha_{1}$.

There are two critical temperatures $T_{c}=2M/(\alpha_{1}+\alpha_{2})$ and $\tilde{T}_{c}=2M/|\alpha_{1}-\alpha_{2}|$ associated with the trilinear coupling, defined by the conditions $M=M_{1}+M_{2}$ and $M=|M_{2}-M_{1}|$.
The initial temperature $T_{ini}$, which determines the initial value for $\epsilon_{R}=\pi^{2}g_{*}T^{4}/30$, shall be smaller than $T_{c}$, $T_{ini}\leqslant T_{c}\simeq M/\alpha$. We furthermore demand that  $\Phi_{ini}$ is large enough that the energy stored in $\phi$ can potentially heat the plasma to temperatures higher than $T_{c}$\footnote{This then also covers the case $T_{c}<T_{ini}<\tilde{T}_{c}$ because the plasma is eventually heated up to $T_{c}$. The case $T_{ini}>\tilde{T}_{c}$ is not interesting because Landau damping obviously allows heating to high temperatures.}.
We estimate the initial value for $\epsilon_{\phi}$ by $V\simeq M^{2}\Phi_{ini}^{2}$, leading to the condition
\begin{equation}\label{haveenoughenergy}
M^{2}\Phi_{ini}^{2}+\frac{\pi^{2}g_{*}}{30}T_{ini}^{4}> \frac{\pi^{2}g_{*}}{30}T_{c}^{4}
.\end{equation}
For the effective masses to be dominated by thermal contributions we demand 
\begin{equation}\label{avoidpara}
g\Phi_{ini}<\alpha^{2}T_{ini}^{2}
.\end{equation} 
If (\ref{avoidpara}) is not fulfilled, thermal masses play no essential role because the effective $\chi_{i}$-masses are dominated by coupling to the oscillating field $\langle\phi\rangle$. If (\ref{haveenoughenergy}) is not fulfilled, the plasma never reaches a temperature at which they would be relevant.

Consistent with all the above conditions we chose $T_{ini}=T_{c}=M/\alpha$, $\Phi_{ini}=0.5\alpha^{2}T_{ini}^{2}/g=M^{2}/(8g)$. We chose $\alpha=0.1$, $h_{1}=h_{2}=h$, $g_{*}=200$ and $M=10^{7}$GeV.  
Then the only parameters to vary are the ratio $g/M$ (which governs $\Gamma/M$ as $\Gamma\propto g^{2}/M$) and $\tilde{\alpha}$ (which governs $\Gamma_{\chi}/M_{\chi}$). 
These choices are for illustrative purpose and are not meant to resemble a particular inflationary model, the study of which we leave for later work.
We are mainly interested in two quantities: The maximal temperature $T_{MAX}$ achieved and the \textit{reheating temperature} $T_{R}$ at the beginning of the radiation dominated era (defined by $\epsilon_{R}=\epsilon_{\phi}$).
We consider various representative choices for the remaining parameters. 
In all cases the dissipation rate for the trilinear interaction is computed by numerical evaluation of (\ref{OffShell}) in each step in order to include off-shell corrections. For scatterings by the quartic interaction, which are always possible on-shell, we use the resummed rate (\ref{SimpleScatterings}) and neglect off-shell corrections.

\begin{figure}[t]
  \centering
\psfrag{a}{$(a)$}
  \psfrag{b}{$(b)$}
  \psfrag{c}{$(c)$}
  \psfrag{X}{$\scriptsize{a/a_{ini}}$}
  \psfrag{Y}{$\scriptsize{T/M}$}
\begin{tabular}{c c}
    \includegraphics[width=7cm]{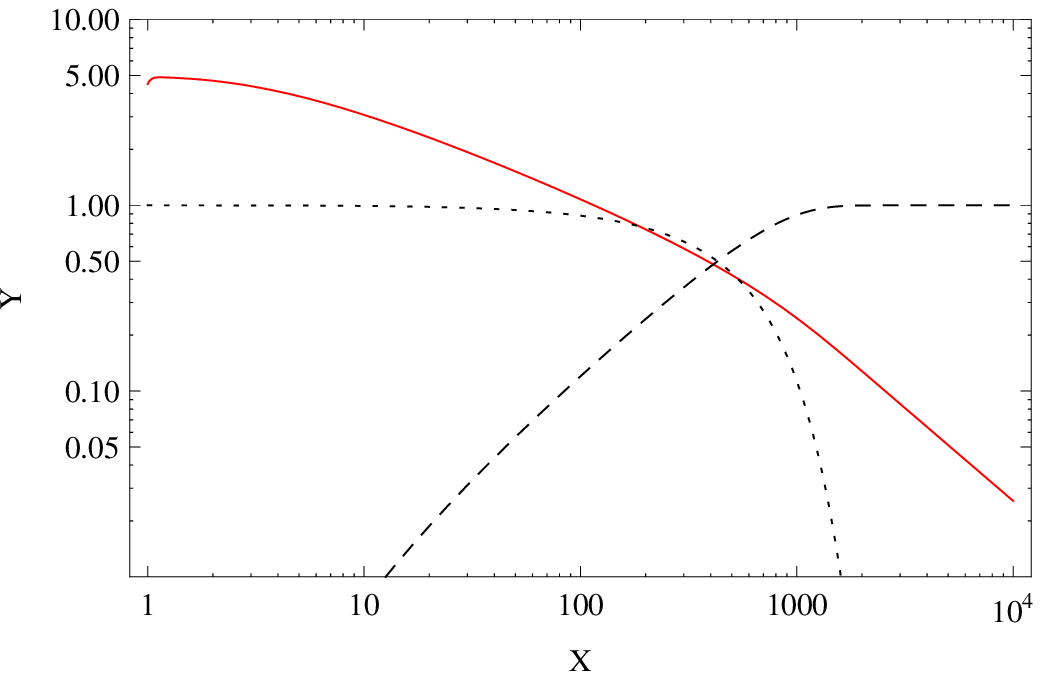} &
\includegraphics[width=7cm]{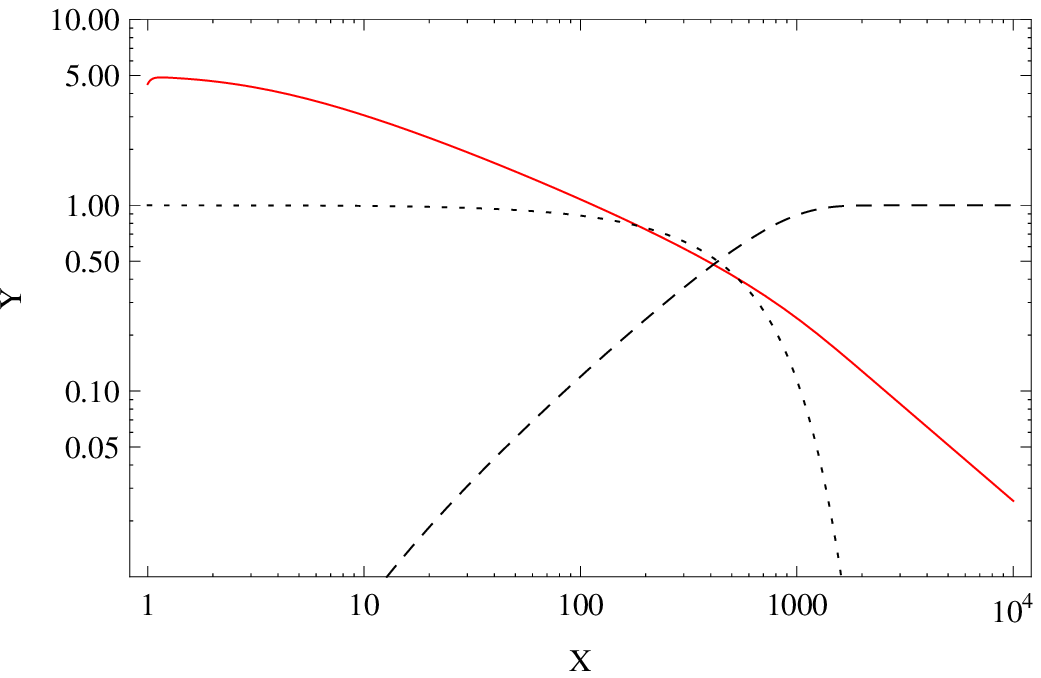}\\
    \includegraphics[width=7cm]{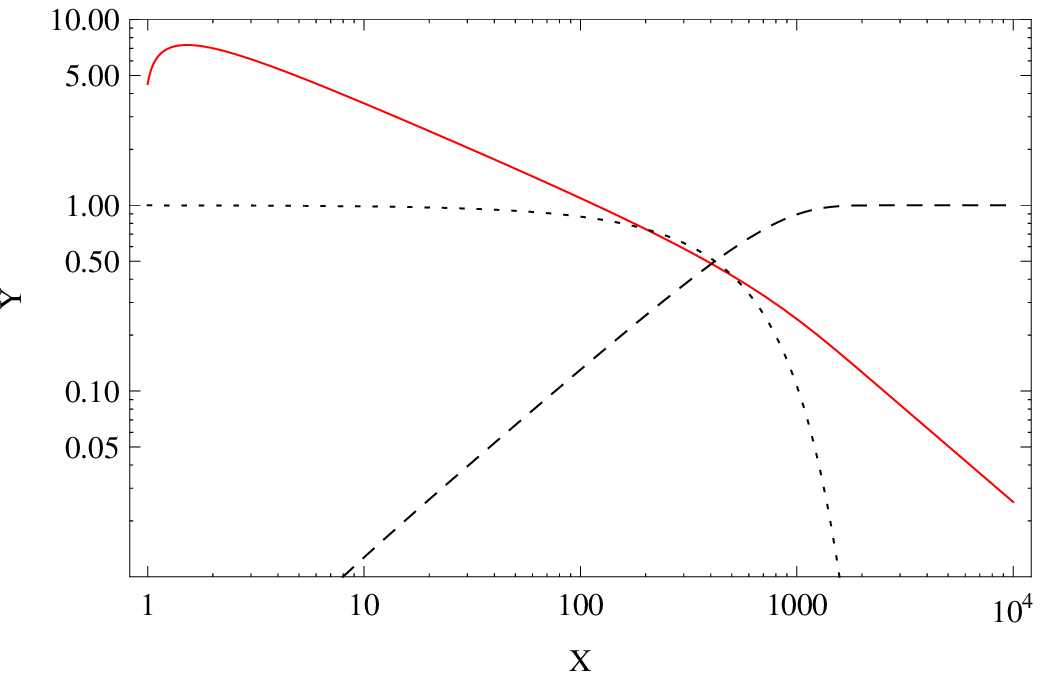} &
\includegraphics[width=7cm]{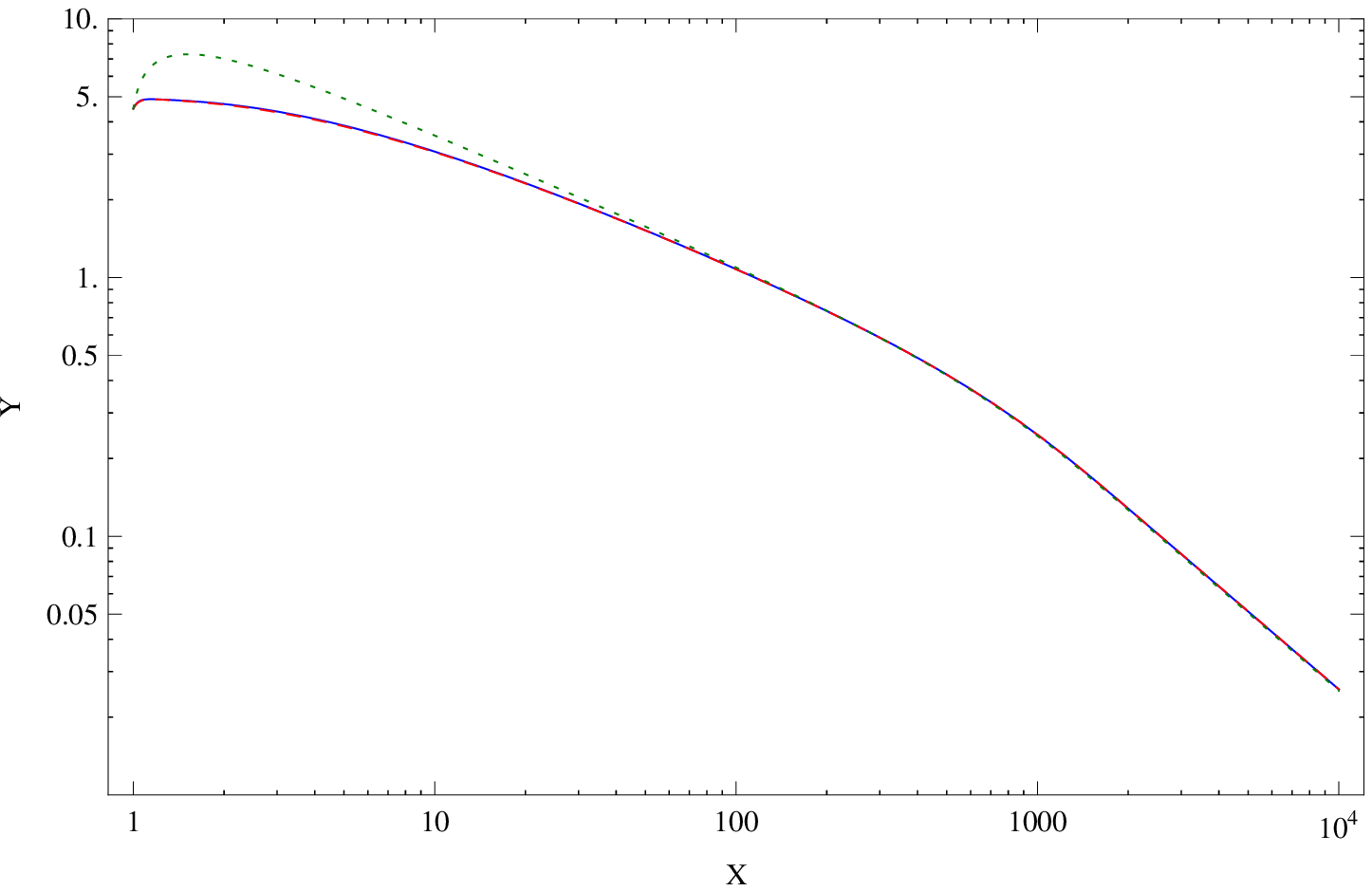}
\end{tabular}
  \caption{Temperature as a function of the scale factor for parameter choice I), $M=10^{7}$GeV, $g/M=\sqrt{2}\cdot 10^{-5}$ and $\alpha_{1}=\alpha_{2}=0.2$. The plot in the lower right corner compares the evolution of $T$ for $\tilde{\alpha}=0.1$ ($\Gamma_{i}/M_{i}\simeq 8\cdot 10^{-4}$, blue solid line), $\tilde{\alpha}=1$ ($\Gamma_{i}/M_{i}\simeq 8\cdot 10^{-2}$, red dashed line) and without resummation of the $\chi_{i}$ propagators ($\Gamma_{i}=0$, $M_{i}=m_{i}\ll M$, green dotted line). The other three plots show the temperature evolution (red line) and the relative contribution of $\epsilon_{\phi}$ (dotted line) and $\epsilon_{R}$ (dashed line) for each single case: upper left $\tilde{\alpha}=0.1$, upper right $\tilde{\alpha}=1$, lower left without resummation.\label{alpha1isalpha2is02g5}}
\end{figure}

\paragraph{I) $h,Y\ll g/M=10^{-5}$, $\alpha_{1}=\alpha_{2}=0.2$}: The results for this case are displayed in figure \ref{alpha1isalpha2is02g5}. The quartic interaction is negligible. Thermal masses close the phase space for the decay $\phi\rightarrow\chi_{1}\chi_{2}$ at $T_{c}\simeq 5M$. Due to the equal thermal masses of $\chi_{1}$ and $\chi_{2}$, $M_{1}=M_{2}=0.1 T$, Landau damping via processes $\chi_{i}\phi\leftrightarrow\chi_{j}$ is kinematically never possible, see figure \ref{ExtraBreit}. Off-shell effects are negligible for $\tilde{\alpha}=0.1$ and $\tilde{\alpha}=1$, corresponding to $\Gamma_{i}/M_{i}\simeq 8\cdot 10^{-4}$ and $\Gamma_{i}/M_{i}\simeq 8\cdot 10^{-2}$. 
Without resumming the $\chi_{i}$ propagators (i.e. neglecting thermal masses) one clearly overestimates $T_{MAX}$. However, for this choice of parameters, one would still get the correct value for $T_{R}$. 

\begin{figure}[t]
  \centering
\psfrag{a}{$(a)$}
  \psfrag{b}{$(b)$}
  \psfrag{c}{$(c)$}
  \psfrag{X}{$\scriptsize{a/a_{ini}}$}
  \psfrag{Y}{$\scriptsize{T/M}$}
\begin{tabular}{c c}
    \includegraphics[width=7cm]{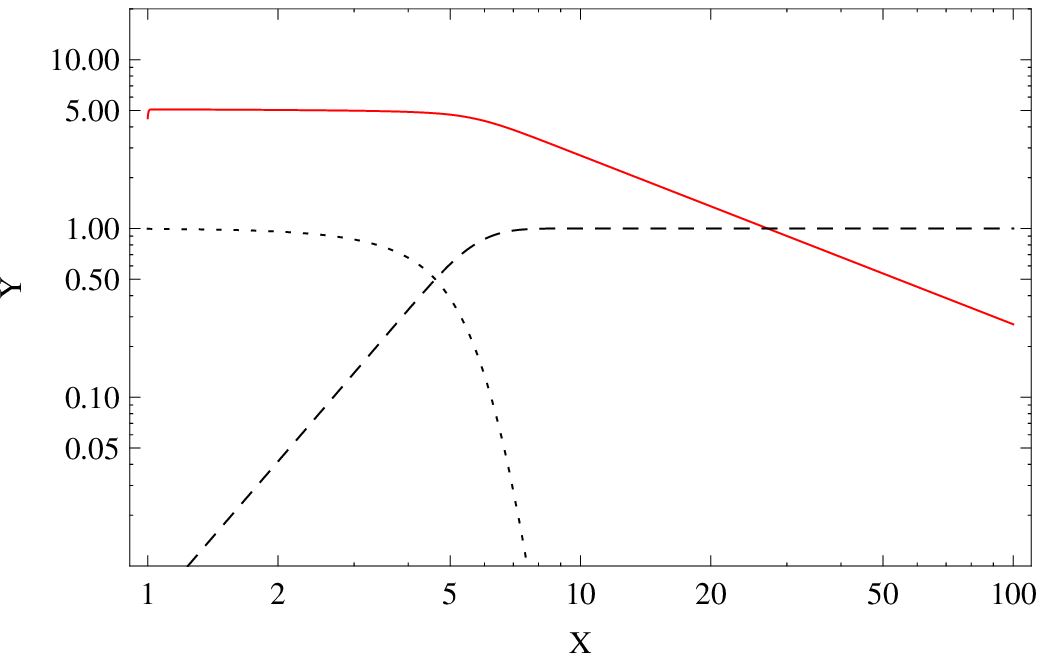} &
\includegraphics[width=7cm]{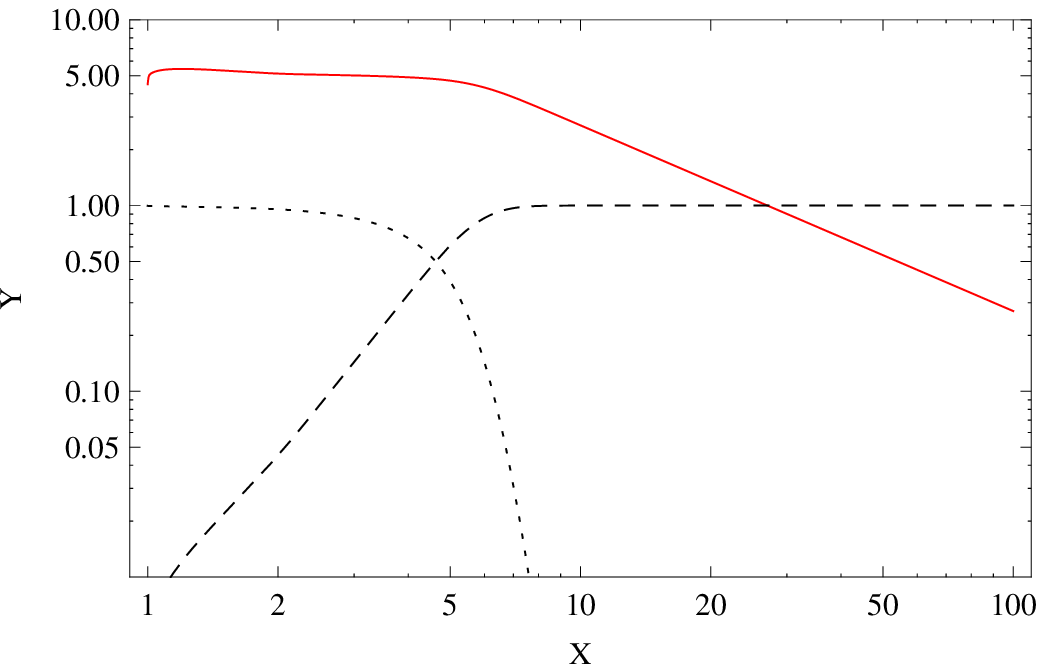}\\
    \includegraphics[width=7cm]{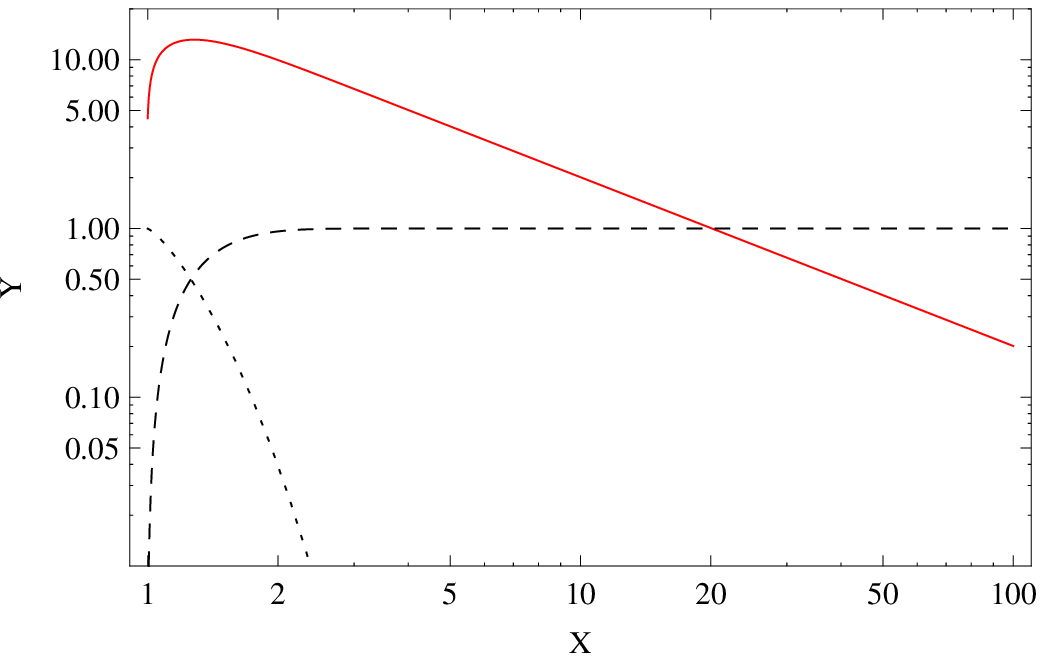} &
\includegraphics[width=7cm]{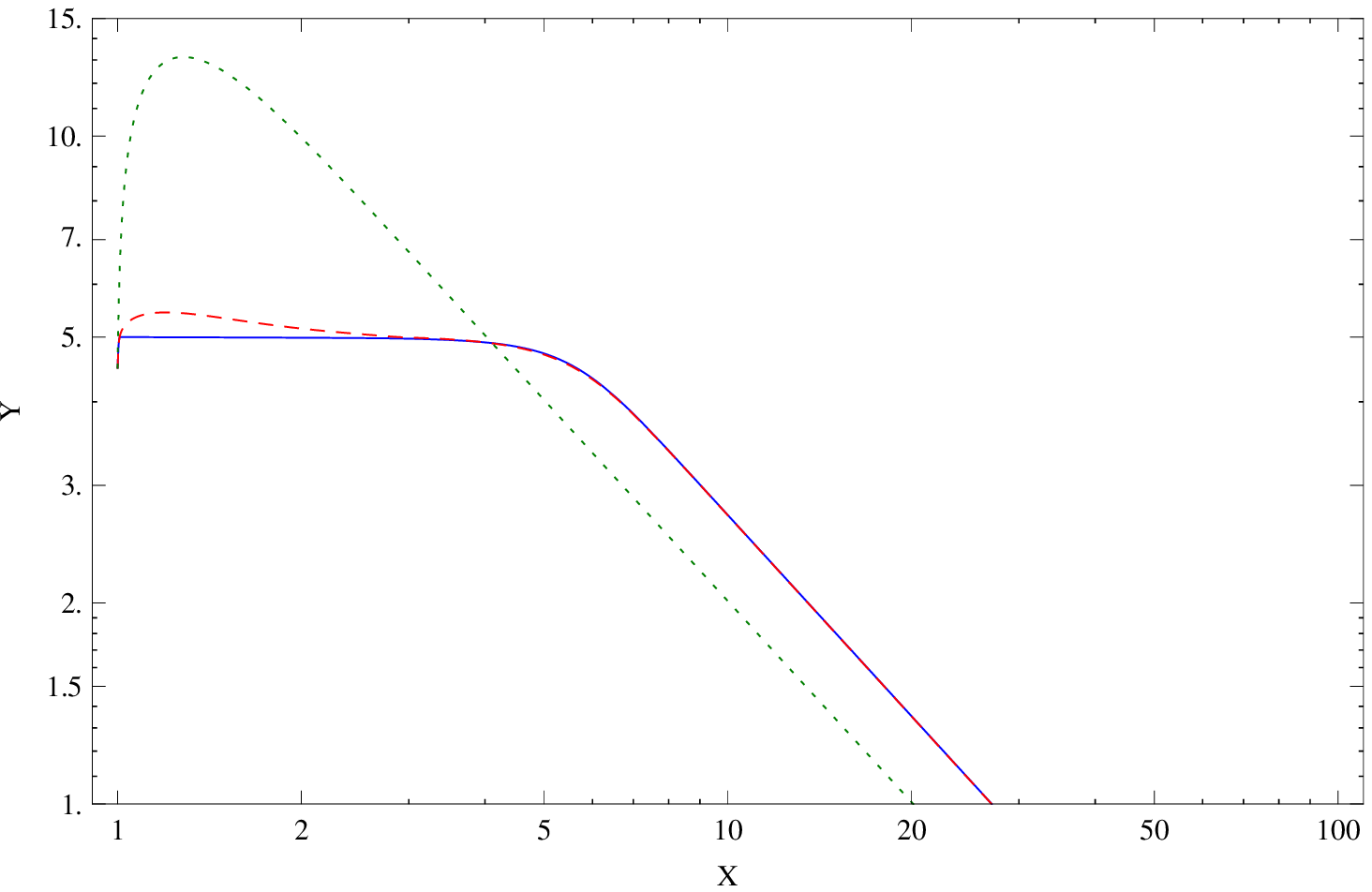}
\end{tabular}
  \caption{Same as figure \ref{alpha1isalpha2is02g5}, but for parameter choice II), $g/M=\sqrt{2}\cdot 10^{-4}$.\label{alpha1isalpha2is02g4}}
\end{figure}

\begin{figure}[t]
  \centering
\psfrag{a}{$(a)$}
  \psfrag{b}{$(b)$}
  \psfrag{c}{$(c)$}
  \psfrag{X}{$\scriptsize{a/a_{ini}}$}
  \psfrag{Y}{$\scriptsize{T/M}$}
\begin{tabular}{c c}
    \includegraphics[width=7cm]{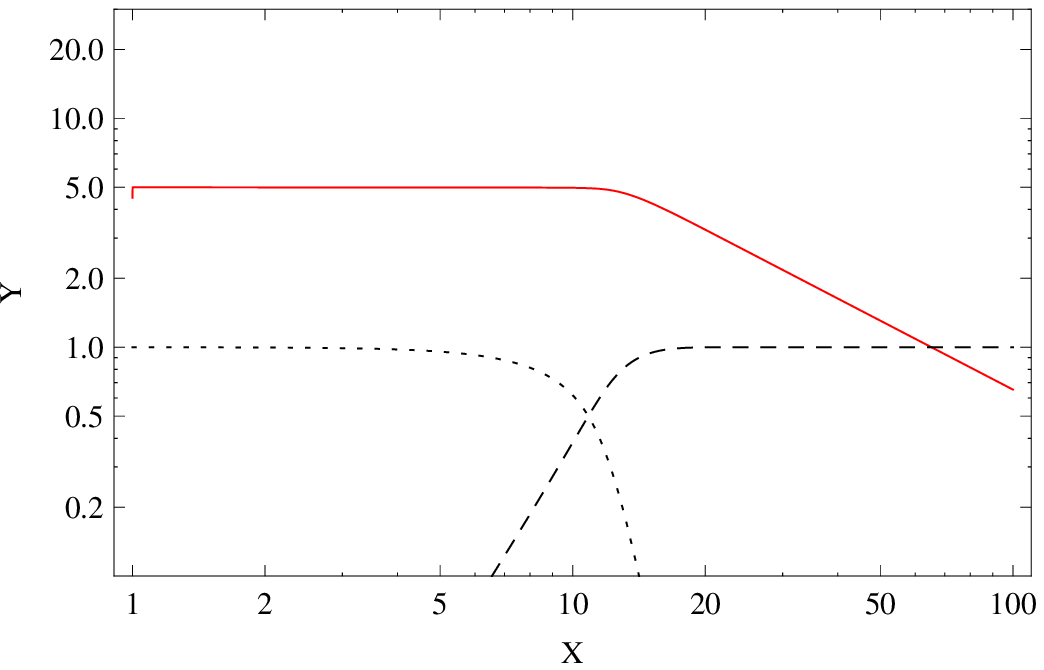} &
\includegraphics[width=7cm]{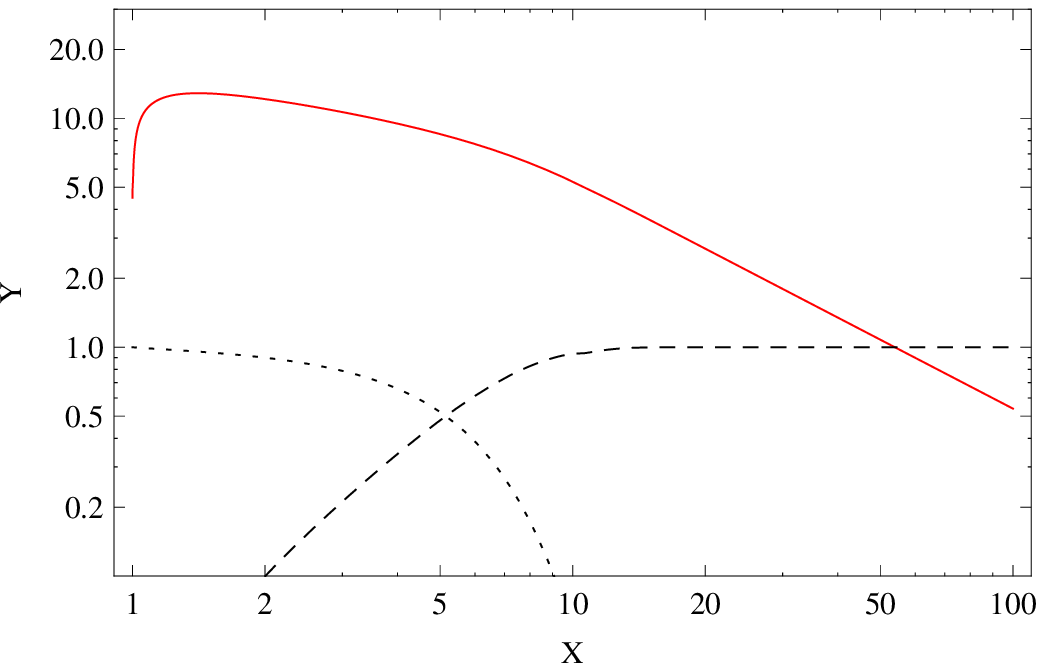}\\
    \includegraphics[width=7cm]{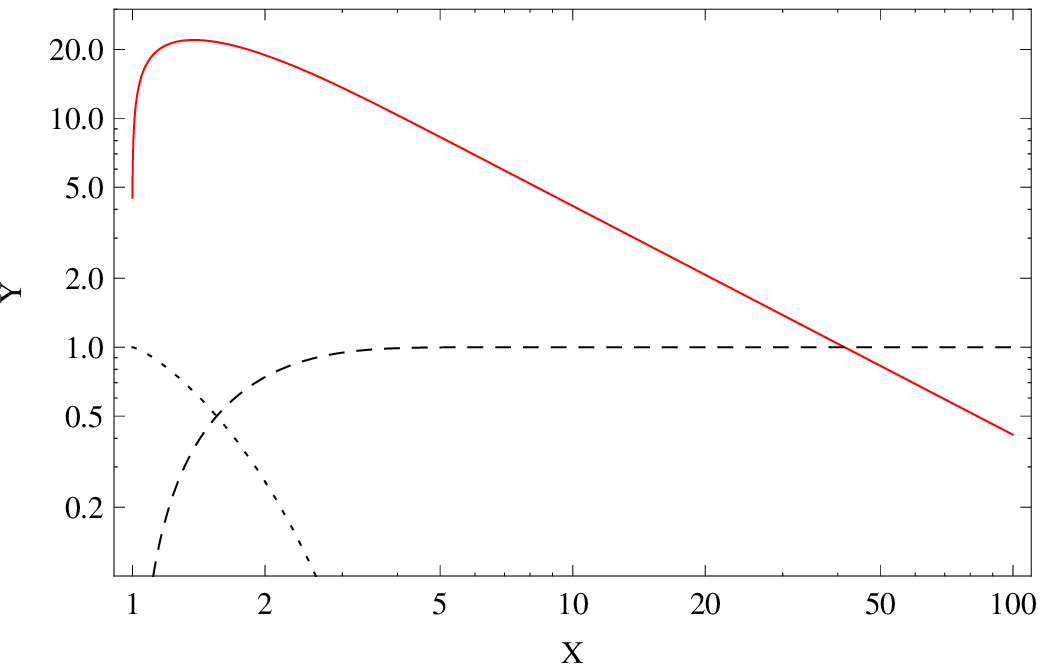} &
\includegraphics[width=7cm]{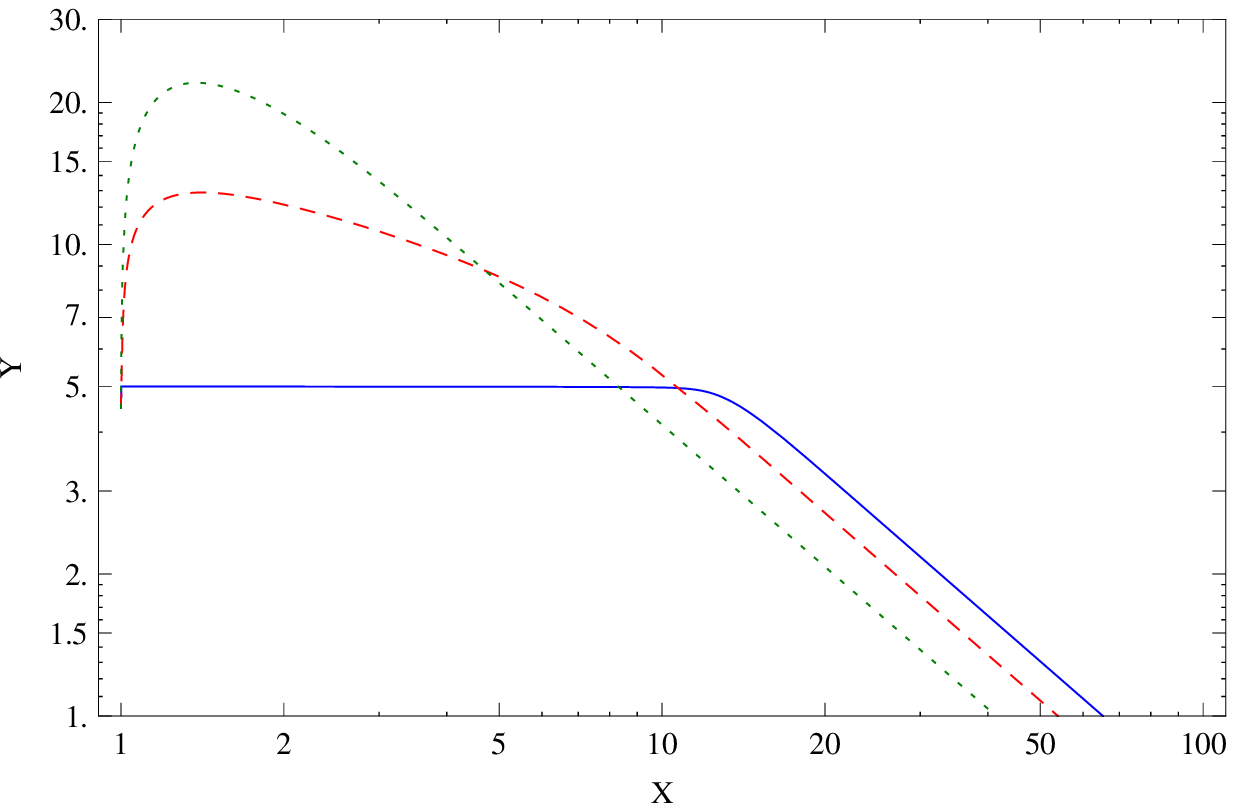}
\end{tabular}
  \caption{Same as figure \ref{alpha1isalpha2is02g5}, but for parameter choice III) with $g/M=\sqrt{2}\cdot 10^{-4}$ and $\alpha_{1}=0.01$, $\alpha_{2}=0.39$.\label{alpha1is001alpha2is039g4}}
\end{figure}

\begin{figure}
  \centering
 \psfrag{X}{$T/M$}
  \psfrag{Y}{$\Gamma_{\textbf{0}}(T)/\Gamma_{\textbf{0}}(T=0)$}
  \psfrag{2.0}{$2$}
  \psfrag{4.0}{$4$}
  \psfrag{6.0}{$6$}
  \psfrag{8.0}{$8$}
 \psfrag{1.0}{$10$}
  \psfrag{1.2}{$12$}
  \psfrag{1.2}{$14$}
  \psfrag{1}{$1$}
 \psfrag{3.0}{$3$}
  \psfrag{5.0}{$5$}
  \psfrag{20.0}{$20$}
  \psfrag{10.0}{$10$} 
    \includegraphics[width=12cm]{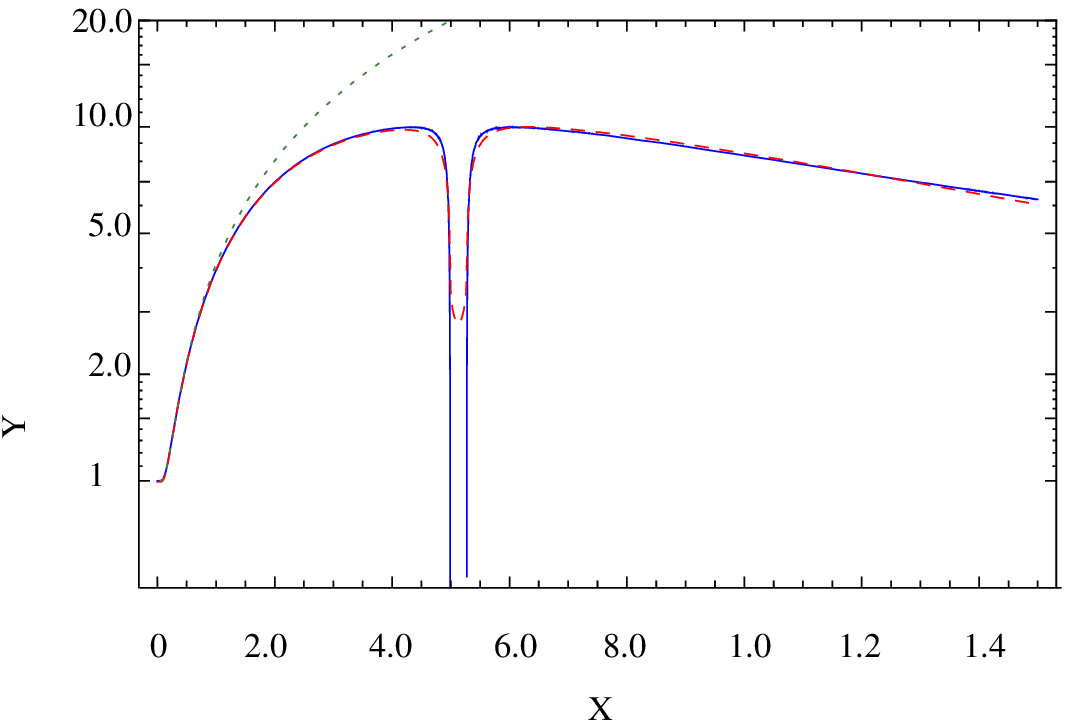}
    \caption{$\Gamma_{\textbf{q}}$ from the trilinear interaction for $\alpha_{1}=0.01$, $\alpha_{2}=0.39$ and $\tilde{\alpha}=1$ without resummation (green dotted), with thermal masses (blue solid) and with thermal masses and widths (red dashed).\label{narrowc}}
\end{figure}

\paragraph{II) $h,Y \ll g/M=10^{-4}$, $\alpha_{1}=\alpha_{2}=0.2$}: The results for this case are displayed in figure \ref{alpha1isalpha2is02g4}. The situation is similar to case I), but $\Gamma/H$ is larger and the plasma heated more efficiently.
The widths $\Gamma_{i}$ have negligible effect for $\tilde{\alpha}=0.1$ ($\Gamma_{i}/M_{i}\simeq 8\cdot 10^{-4}$), but allow a temperature $T_{MAX}$ slightly larger than $T_{c}$ for $\tilde{\alpha}=1$ ($\Gamma_{i}/M_{i}\simeq 8\cdot 10^{-2}$). Though it is conceptually interesting to see that processes involving virtual $\chi_{i}$-quasiparticles can sustain reheating for $T>T_{c}$, this small correction is phenomenologically not relevant (especially given the uncertainty about the correct model for inflation).
However, for $g/M=10^{-4}$ the beginning of the radiation dominated era falls into the epoch during which $T$ is ``frozen'' at $T_{c}$. Therefore the thermal masses affect not only $T_{MAX}$, but also $T_{R}$ and the temperature during the radiation dominated era. 
\paragraph{III) $h,Y\ll g/M=\sqrt{2}\cdot 10^{-4}$, $\alpha_{1}=0.01$, $\alpha_{2}=0.39$}: The case that $M_{1}$ and $M_{2}$ are very different and $\tilde{\alpha}=1$ is interesting. Figure \ref{alpha1is001alpha2is039g4} shows that correct values for $T_{MAX}$ \textit{and} $T_{R}$ can only be obtained using full resummed propagators, including thermal masses as well as widths. The reason is that for $\alpha_{2}\gg\alpha_{1}$ the region (c) shrinks as $\tilde{T}_{c}-T_{c}\simeq 4M\alpha_{1}/\alpha_{2}^{2}$, see figure \ref{narrowc}. Then the dissipation by off-shell processes that only led to an insignificant increase of $T_{MAX}$ slightly beyond $T_{c}$ in figure \ref{alpha1isalpha2is02g4} can be sufficient to heat the plasma up to $\tilde{T}_{c}$. After it crossed the forbidden region, reheating can continue via Landau damping.  

\begin{figure}
  \centering
 \psfrag{X}{$a/a_{ini}$}
  \psfrag{Y}{$T/M$}
  \psfrag{2}{$2$}
  \psfrag{5}{$5$}
  \psfrag{10}{$10$}
  \psfrag{20}{$20$}
 \psfrag{50}{$50$}
  \psfrag{100}{$100$}
  \psfrag{0.2}{$0.2$}
  \psfrag{0.5}{$0.5$}
 \psfrag{1.}{$1$}
  \psfrag{2.}{$2$}
  \psfrag{5.}{$5$}
  \psfrag{10.}{$10$}
    \includegraphics[width=12cm]{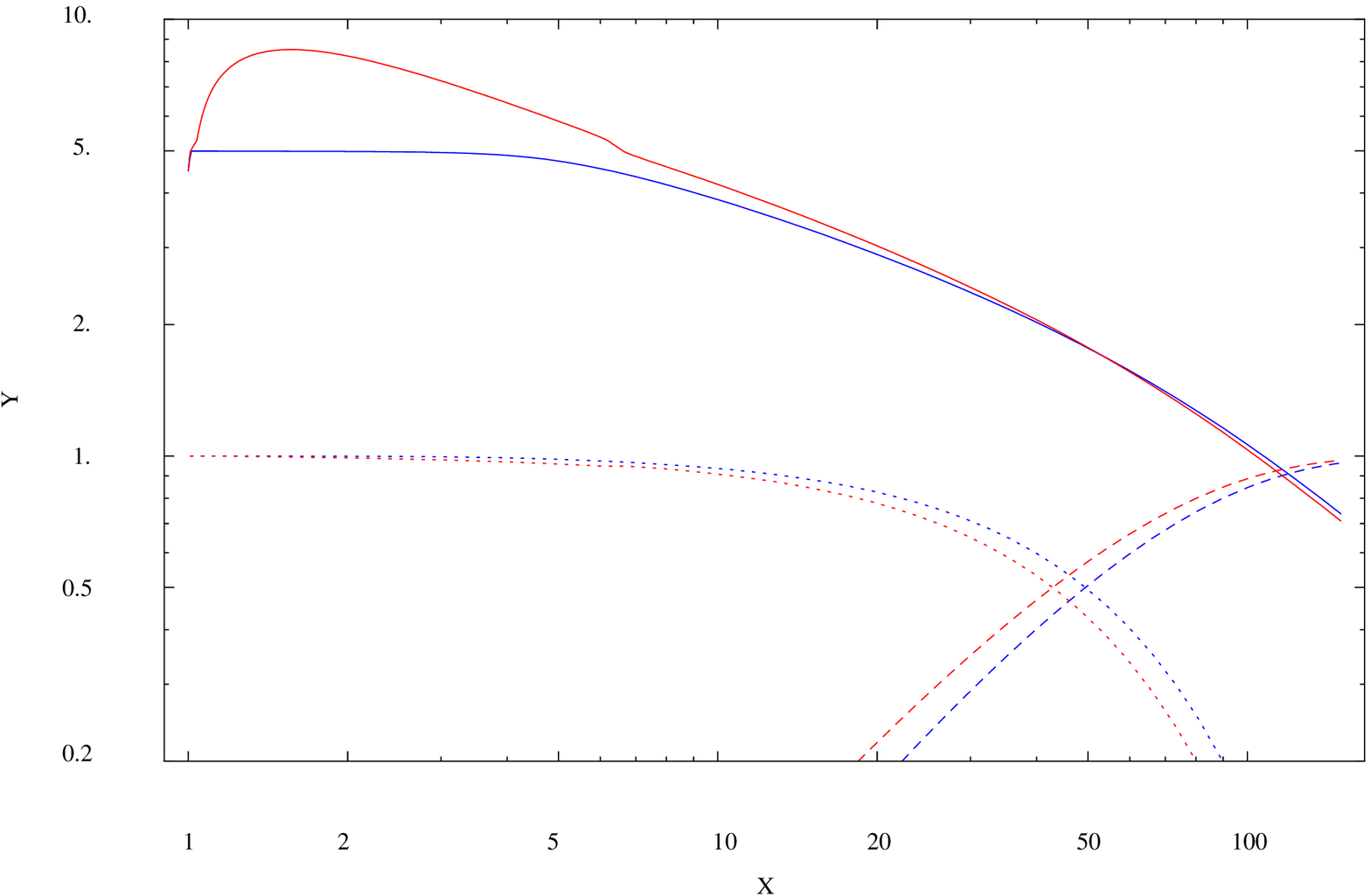}
    \caption{Evolution of $T$ (solid line) and the relative contributions of $\epsilon_{\phi}$ (dotted) and $\epsilon_{R}$ (dashed) to the energy density for parameter choice IV), $g/M=\sqrt{2}\cdot 10^{-4}$, $\alpha_{1}=0.01$, $\alpha_{2}=0.39$, $\tilde{\alpha}=0.1$. The blue lines show the result when only the trilinear interaction is taken into account, the red line shows the behaviour when quartic interactions with $h/2=g/M$ are added.  \label{alpha1is001alpha2is039g4h2g}}
\end{figure}

\paragraph{IV) $g/M=h/2=10^{-4}$, $\alpha_{1}=0.01$, $\alpha_{2}=0.39$, $\tilde{\alpha}=0.1$}
In this scenario we fix $\tilde{\alpha}=0.1$ such that the $\chi_{i}$ are in the quasiparticle regime ($\Gamma_{i}/M_{i}\simeq 8\cdot 10^{-4}$) and off-shell contributions are negligible, but add the quartic interaction. The trilinear interaction cannot heat the universe to temperatures larger than $T_{c}$. The quartic interaction mediates $\phi\chi_{i}\leftrightarrow\chi_{i}\chi_{i}$ scatterings at leading order. These are inefficient at low temperatures due to the low occupation numbers, but can become the main source of dissipation for large $T$.
Figure \ref{alpha1is001alpha2is039g4h2g} demonstrates that scatterings mediated by the quartic interaction can allow for a temperature beyond $T_{c}$.

\subsubsection{Decay into Fermions}
The contribution from processes with fermions in the final state is generally thought to be small\footnote{It has been pointed out though that fermions may play a crucial role during preheating \cite{Berges:2009bx}.}. While the dissipation rate into bosonic final states is amplified in a plasma due to induced transitions, for fermions there is a suppression due to the Pauli principle. These effects become relevant when $f_{B,F}(M/2)$ is of order one, which is the case for $T\sim M$. The effects of thermal masses kick in around $T\sim M/\alpha$. Thus, for $\alpha\ll 1$ thermal fermion masses will not affect $\Gamma_{\textbf{q}}$ significantly. In the quasiparticle regime this allows to draw two conclusions.  
First, perturbative decay into fermions is negligible at temperatures near the inflaton mass if there exists a coupling to bosons of comparable strength ($\Gamma_{\phi\rightarrow\chi\chi}\simeq\Gamma_{\phi\rightarrow\Psi\Psi}$ at $T=0$) due to the relative suppression $\Gamma_{\phi\rightarrow\Psi\Psi}/\Gamma_{\phi\rightarrow\chi\chi}\sim(1-2f_{F}(M/2))/(1+2f_{B}(M/2))=(\tanh(M\beta/4))^{2}\approx M^2(4T)^{-2}$.
Second, thermal fermion masses are generally not important because for $\alpha\ll 1$, $T\sim M$ they are much smaller than $M$ and the involved momenta.  

Nevertheless there are situations in which fermions with thermal masses can be relevant. The most obvious is $Y\gg g/M,h$. Then the inflaton effectively only couples to fermions. However, due to the Pauli blocking the maximal temperature that can be achieved by perturbative decay into fermions is usually smaller than the inflaton mass and $T_{c}$ (unless one choses a very large coupling $Y$, which is difficult to consort with a flat effective inflaton potential and the slow roll conditions), thus the effect of thermal fermion masses is again negligible.  

Another possibility is that the effective masses for bosons grow faster with temperature than those of the fermions. Then the phase space for the decay into bosons closes while that for fermions is still open. If all couplings are weak the temperature at which this happens is larger than $M$, the previous arguments apply and the decay rate into fermions is small. 
An interesting situation can occur if the thermal boson masses $M_{1}$ and $M_{2}$ are different. Then the suppressed, but nonzero decay rate into fermions can play the same role as off-shell effects in scenario III): It can allow the system to pass the temperature region $T_{c}<T<\tilde{T}_{c}$. At temperatures $T>\tilde{T}_{c}$ Landau damping via processes $\chi_{i}\phi\leftrightarrow\chi_{j}$ is efficient, see figure \ref{alpha1is001alpha2is039g4h2g}. In this case thermal fermion masses are decisive because they decide if $\Gamma$ in the forbidden region is large enough to heat the plasma up to $\tilde{T}_{c}$.
If the thermal masses of $\Psi_{1}$ and $\Psi_{2}$ are different there is a temperature regime in which the plasma can also be heated by processes $\Psi_{i}\phi\leftrightarrow\Psi_{j}$.  

A situation in which fermions with thermal masses can play an important role in perturbative reheating is $\alpha\sim 1$. In this case they are relevant at temperatures $T_{c}\sim M$. 
Then it is indeed possible that the decay into fermions gives a comparable contribution as for bosons. 
However, in this situation one would generally expect that the widths of the $\Psi_{i}$ are not negligible. Then off-shell effects are relevant, e.g. $2\rightarrow 2$ scatterings involving intermediate off-shell $\Psi_{i}$. 

The efficiency of processes involving fermions can be increased by going beyond the quasiparticle regime. It is easy to see that cuts through the self energy insertion in the diagram shown in figure \ref{fermiondiagram} can leave pieces with only bosons in the final state. Once transitions involving many quanta contribute significantly, the rate for processes involving fermions can be amplified by large occupation numbers in the bosonic sector. 

\subsection{Discussion of Thermal Masses during Reheating}
Thermal masses of the decay products can modify the rate at which the inflaton releases energy into the primordial plasma during reheating. This can impose an upper bound on the temperature if several conditions are fulfilled. First, thermal masses have to dominate the effective masses. This is only the case if the amplitude of the $\langle\phi\rangle$ oscillations is not too large (e.g. $\alpha^{2}T^{2}>g\Phi$ for the scalar trilinear interaction). The applicability of our analysis is restricted to perturbative inflaton decay.

Second, the dissipation process has to be driven by decays (or scatterings) of single (quasi)particles. At high temperature, contributions from processes involving many quanta become increasingly important. In our setup the contributions from these are parameterised by the widths of the resonances in the plasma. An upper bound on the temperature can only be imposed if all relevant fields are in the quasiparticle regime, i.e. the propagating states have a narrow width. Only then the dissipation can be pictured in terms of approximately energy conserving decays and scatterings of individual quasiparticles.

Finally, assuming the above conditions are fulfilled, a blocking of the phase space for the decay only stops reheating if no other channels of dissipation are available. 
When $\phi$ mainly couples via 3-vertices, dissipation at low temperatures is dominated by $\phi$-decay.  
If at the temperature $T_{c}$, when the phase space for the last decay channel closes, some quasiparticle in the plasma is heavy enough to decay into a final state including $\phi$, e.g. $\chi_{2}\rightarrow\phi\chi_{1}$, $\phi$ can dissipate energy into the plasma via the inverse process $\phi\chi_{1}\rightarrow\chi_{2}$. The competition between these processes leads to an overall energy transfer from $\phi$ into the plasma analogue to Landau damping as long as the $\phi$ occupation numbers are larger than their equilibrium values. 
If there is no sufficiently heavy quasiparticle in the plasma, reheating indeed becomes inefficient for $T>T_{c}$. $2\rightarrow 2$ scatterings are only possible at second order in the coupling (while the leading contribution to the decay comes at first order). Furthermore, in the regime where the decay is forbidden ($M_{1}+M_{2}>M$, $M>|M_{1}-M_{2}|$), scatterings involve a virtual intermediate quasiparticle and are suppressed by the corresponding energy scale. 
However, when $\phi$ couples to other fields via 4-vertices, $2\rightarrow 2$ scatterings between real quasiparticles are always possible at leading order. They require real bath quanta in the initial state, thus their efficiency increases with temperature (they are negligible for $T\ll M$). The same applies to effective vertices with more legs. Thus, thermal masses only imply an upper bound on the reheating temperature if they dominate the effective masses, the propagating states in the plasma are narrow resonances and scatterings and Landau damping are negligible.

These conclusions can be obtained from the results of the previous sections. We also demonstrated them numerically and computed the time evolution of the temperature for several parameter choices. We would like to emphasise again that these parameter choices are not meant to represent a particular (realistic) model of inflation, but were chosen for illustrative purpose. In models of single field inflation there exist strong bounds on the couplings of $\phi$ to other fields (and weaker bounds on the interactions of the decay products) - too strong couplings would spoil the flat effective potential necessary for slow roll inflation by radiative corrections. The field $\phi$ may also be viewed as the relevant direction in field space during reheating in multi-field models. In this situation the bounds are weaker. However, our goal here is not to impose bounds on the parameters of a particular class of models, but to perform a self-consistent study of a potentially relevant effect. We leave the application to realistic scenarios to later work.

Finally, we would like to emphasise that in reality the details of the reheating process are far more complicated than in our model. Even during perturbative decay there are various different time scales related to the equilibration of the decay products. As long as all of them are much shorter than $1/\Gamma$ we do not expect these details to affect the rate at which $\phi$ dissipates its energy considerably. We discuss these aspects at the end of section \ref{SummmaryOfResults}.

\section{Discussion and Conclusions}\label{discussion}
We studied the relaxation of scalar and fermionic fields in a large thermal bath from first principles of nonequilibrium quantum field theory. This situation is realised in various interesting physical systems, including thermal production of particles from a plasma, propagation in dense matter or cosmological freezeout processes. Depending on the initial state of the out-of-equilibrium fields, relaxation can mean dissipation of energy into the bath or thermal production of quanta. The formalism we used allows a quantum description of both processes.
 
We performed all computations in terms of correlation functions for quantum fields, avoiding the ambiguities in Boltzmann type kinetic equations related to quantum coherence, the definition of asymptotic states or particle numbers in a dense plasma and the assumption of molecular chaos. Basis of our analysis are exact expressions for nonequilibrium correlation functions for arbitrarily large deviations from thermal equilibrium that are valid for all times. They were obtained without semiclassical approximation or a gradient expansion.

The main goals of our study were to explore which effect the modified dispersion relations in the plasma have on the relaxation rate and under which circumstances a description in terms of quasiparticles is suitable for a relativistic quantum system.

\subsection{Summary of Results}\label{SummmaryOfResults}
At high temperature and density, relaxation in a plasma is not only driven by decays and scatterings of individual particles. Processes involving many quanta, that naively are of higher order in the coupling, have to be taken account. Their contribution is negligible in a dilute plasma, but grows with the occupation numbers. Therefore consistent results can only be obtained using resummed perturbation theory.
In our setup the contribution from processes involving many quanta can be parametrised by the widths of the resonances in the plasma. \\

If the widths of all resonances are small, these can be interpreted as quasiparticles with dispersion relations given by the poles of the resummed propagators. Then the exchange of energy between the modes of different fields can effectively be described in terms of decays and scatterings amongst individual quasiparticles. This is an essential condition for the validity of effective kinetic equations. However, even in the quasiparticle regime it is generally not valid to obtain such equations by simply replacing the vacuum masses by thermal masses because the dispersion relations can be more complicated. In addition, collective excitations may appear as propagating states. 

Though effective kinetic equations for quasiparticles describe the energy transfer between different degrees of freedom with good accuracy, they do not capture all properties of the system.  We explicitly showed that the energy density of the out-of-equilibrium field is not that of a quasiparticle gas and does not follow a Boltzmann equation for quasiparticles. 

The modified dispersion relations in a plasma lead to a number of interesting effects. If the out-of-equilibrium field couples via 3-vertices, the leading order contribution to the energy exchange with the bath can effectively be switched off in some temperature regime. This happens when, due to the temperature dependent dispersion relations, none of the quasiparticles connected by the vertex can kinematically decay into the other two. This can considerably delay the relaxation and have interesting cosmological implications, for instance during reheating or for the thermal production of particles. Interactions by 4-vertices always allow for relaxation at leading order as these mediate $2\rightarrow 2$ scatterings. However, since they require real quanta from the plasma in the initial state they give only small contributions at low temperature.\\

The widths of the quasiparticles parametrise the effect of processes that involve more quanta and vertices. These appear to be energy violating in the quasiparticle picture, but indeed only reflect the fact that relaxation can also happen via off-shell processes and multiple scatterings with many quanta from the bath. At high temperature there are various effects including induced transitions for bosons, the large density of scattering partners or the Landau-Pomeranchuk-Migdal effect which can make these contributions comparable to or even larger than the rate at leading order. Mathematically this manifests in the poor convergence of the perturbative series and the need for resummation. In the quasiparticle picture, this appears as an increase of the width or \textit{melting} of the resonance peaks at high temperature. Physically it means that in a dense plasma the picture of individual particles travelling freely between isolated interactions does not hold.

When the widths of the resonances are not small, the quasiparticle interpretation and description in terms of kinetic equations break down. For instance, even in a temperature region where all leading order processes involving real quasiparticles are kinematically forbidden, the relaxation rate may be as big as in the region where decays and inverse decays are allowed. Thus, naive estimates based on kinematics involving thermal masses fail to even qualitatively reproduce the behaviour.
In the quasiparticle picture this can be interpreted in terms of off-shell processes, mediated by virtual quasiparticles. At low temperatures, these effects can be included into the Boltzmann approach by adding matrix elements for higher order processes. At large temperatures, this would require the inclusion of infinitely many diagrams with large numbers of particles in the initial and final state.\\

We computed analytic expressions for the relaxation rates of scalars and fermions via a trilinear and a Yukawa interaction, respectively, for arbitrary energies and momenta in the quasiparticle regime. 
We furthermore derived a formula for the rate of relaxation via the trilinear scalar interaction, expressed as a one-dimensional integral, which includes the effects of higher order processes. Comparing to the quasiparticle result, we found that the corrections can be large even for a moderate width $\Gamma_{i}/M_{i}\sim 10^{-2}$.

We applied our results to a particular cosmological problem, namely the question whether large thermal masses of the decay products can close the phase space for perturbative inflaton decay and therefore impose an upper bound on the reheating temperature.\\

Finally, we would like to emphasise that in reality the details of the relaxation process can be far more complicated than in our model. We assumed instantaneous equilibration of the bath degrees of freedom. There may be different time scales related to this process. The word ``quasiparticles'' has been used for propagating states with definite momentum, the modes of a homogeneous field, with properties defined by averaging over a statistical ensemble. They are not localised particles (wave packets). 
When the relaxation is pictured as a statistical process, driven by the competition of decays and inverse decays (and scatterings) of quasiparticles, each individual of these processes is localised in time and space. Even though the relaxation time of $\phi$ is much larger than that of the other fields, the details of the thermalisation process and local deviations from equilibrium  do affect the decay of individual quasiparticles. However, we are not interested in the fate of individual quasiparticles here, but in the overall relaxation of the system. Therefore we believe that the thermal ensemble average is justified and gives correct results as long as the relaxation time in the plasma is much shorter than $1/\Gamma$

\subsection{Conclusions and Outlook}
Boltzmann equations and their matrix valued generalisations can accurately describe many nonequilibrium phenomena in a medium. They are known to suffer from uncertainties when the coherence lengths are large, but are usually assumed to be accurate in absence of such effects. 
On the other hand, it has been known for long that resummation is necessary in gauge theories at high temperature because processes involving many quanta that naively are of higher order in the coupling in fact contribute at leading order. 
It is important to understand in which cases and how these many-particle amplitudes can be incorporated into the framework of single particle distribution functions.

The inconsistencies are related to the assumption that the system should be described in terms of individual particles as asymptotic states. We showed that they can be removed in an intuitive way when swapping the phase space distribution functions as dynamical variables for correlation functions of quantum fields. 

The Schwinger-Keldysh formalism used in this work allows for a full quantum treatment of nonequilibrium phenomena in the framework of quantum field theory. A consistent perturbation theory can be formulated in the framework of the $n$PI effective action.
It has previously been applied to problems in which all degrees of freedom are far from equilibrium, including preheating after inflation and equilibration after heavy ion collisions. 
However, the equations of motion, coupled second order integro-differential equations, are considerably more complicated than the semiclassical Boltzmann equations. In most cases they can only be solved numerically in toy models. The use of exact equations comes at a price, the loss of the simplicity, transparency and intuitive physical interpretation that make the Boltzmann equations appealing.
 
In this work we studied systems in which only few degrees of freedom are out of equilibrium and weakly coupled to a large thermal bath. 
In this setup most of the computations can be performed analytically and the quantities in the field theoretical approach have an intuitive physical interpretation.
The analytic solutions allowed us to study in detail the effect of quantum and higher order corrections, parametrise them in quasiparticle properties and understand the limits of this approximation. The semiclassical description in terms of effective kinetic equations can be recovered in the limit of a dilute gas. 

The main simplification that allowed a widely analytical treatment lies in the negligible backreaction. We have neglected backreaction in two ways: by ignoring the effect of the out-of-equilibrium fields on the bath temperature and by dropping contributions to the self energies that have nonequilibrium propagators in the loops. This allows to use resummation techniques known from the theory in equilibrium.
There are interesting systems in which the second assumption is not justified while the first one still holds. In leptogenesis, for instance, the baryon asymmetry of the universe originates from $CP$-violating interactions of heavy right handed neutrinos that are out of equilibrium with a thermal bath of SM fields. Dropping the small Cabibbo-Kobayashi-Maskawa-phase in the SM, their Yukawa couplings are the only source of $CP$-violation. When computing the $CP$-asymmetry, the suppression of contributions from diagrams that have the out-of-equilibrium fields in the loop by the large number of degrees of freedom in the bath does not apply as none of the other interactions is $CP$-violating. Indeed, their inclusion is crucial as the generation of an asymmetry comes from a quantum interference between the two leading orders in the Yukawa couplings.

Therefore it is important to develop techniques that allow a consistent evaluation of higher order diagrams in which the nonequilibrium fields appear in the loop while making maximal use of the simplifications due to the weak coupling to a thermal bath and negligible backreaction on the temperature. 
The nonequilibrium propagators used in this work can be used for a perturbative treatment and the formulation of Feynman rules is straightforward. However, practical evaluation of higher order diagrams is technically difficult due to the dependence on centre of mass time, and it remains to be seen how to perform the resummation of all stronger (gauge) interactions in the bath.

\section*{Acknowledgements}
I would like to thank Wilfried Buchm\"uller for his advice in the initial phase of this project and Jean-Sebastien Gagnon for proof-reading and remarks. 
I would also like to thank Juan Garcia-Bellido and Fuminobu Takahashi for their comments on section \ref{InflatonSection}. 
This work was supported by the Swiss National Science Foundation.

\clearpage

\begin{appendix}
\section{Appendix}
\subsection{Nonequilibrium Quantum Field Theory}\label{KBEessentials}
Here we summarise basic concepts of nonequilibrium quantum field theory and notations used in this work. 
For detailed reviews on the topic, we refer the reader to \cite{Berges:2004yj,Chou:1984es}. Most of the notation we use here is adopted from \cite{Anisimov:2008dz}, which also contains a slightly more detailed introduction into the topic than presented here. 
 
Out of equilibrium there are two independent two point functions for each field. For a real scalar field $\phi$ these can be chosen to be the connected Wightman functions  
\begin{eqnarray}
\Delta^>(x_1,x_2) &=& \langle \phi(x_1)\phi(x_2)\rangle_{c}, \label{forwAA}\\ 
\Delta^<(x_1,x_2) &=& \langle \phi(x_2)\phi(x_1)\rangle_{c} \label{backAA}. 
\end{eqnarray}
where the $\langle\ldots\rangle$ is to be understood in the sense of (\ref{expvalue}) and includes averaging over thermal as well as quantum fluctuations.
The subscript $_{c}$ indicates that $\Delta^{\gtrless}$ are connected correlation functions.
For a Dirac fermion the corresponding definitions are
\begin{eqnarray}
S^{>}_{\alpha\beta}(x_{1},x_{2})&=&\langle \Psi_{\alpha}(x_{1})\bar{\Psi}_{\beta}(x_{2})\rangle_{c}\label{SforwA}\\
S^{<}_{\alpha\beta}(x_{1},x_{2})&=&-\langle \bar{\Psi}_{\beta}(x_{2})\Psi_{\alpha}(x_{1})\rangle_{c}\label{SbackA}
\end{eqnarray}
Here $\Psi$ is a Dirac spinor and $\alpha$ and $\beta$ are spinor indices which we will always suppress in the following.
It is convenient to define the linear combinations
\begin{eqnarray}
\Delta^{-}(x_{1},x_{2})&=&i\left(\Delta^>(x_1,x_2)-\Delta^<(x_1,x_2)\right)\label{dminus}\\
\Delta^{+}(x_{1},x_{2})&=&\frac{1}{2}\left(\Delta^>(x_1,x_2)+\Delta^<(x_1,x_2)\right)\label{dplus}\\
S^{-}(x_{1},x_{2})&=&i\left(S^{>}(x_{1},x_{2})-S^{<}(x_{1},x_{2})\right)\label{SMinus}\\
S^{+}(x_{1},x_{2})&=&\frac{1}{2}\left(S^{>}(x_{1},x_{2})+S^{<}(x_{1},x_{2})\right)\label{Splus}
.\end{eqnarray}
The functions (\ref{dminus})-(\ref{Splus}) have intuitive physical interpretations. 
The \textit{spectral functions} $\Delta^{-}$ and $S^{-}$ are the Fourier transforms of the spectral densities and, roughly speaking, characterise the spectrum of excitations in the plasma. The \textit{statistical propagators} $\Delta^{+}$ and $S^{+}$ can be viewed as a measure for the occupation numbers. 
They fulfil the symmetry relations
\begin{eqnarray}
\Delta^{-}(x_{2},x_{1})&=&-\Delta^{-}(x_{1},x_{2}),\label{dminussymm}\\
\Delta^{+}(x_{2},x_{1})&=&\Delta^{+}(x_{1},x_{2}),\label{dplussymm}\\
S^{-}(x_{2},x_{1})&=&-\gamma_{0}\left(S^{-}(x_{1},x_{2})\right)^{\dagger}\gamma_{0}\label{Sminussym}\\
S^{+}(x_{2},x_{1})&=&\gamma_{0}\left(S^{+}(x_{1},x_{2})\right)^{\dagger}\gamma_{0}\label{Splussym}
,\end{eqnarray}
which can be seen from the definitions,
and the boundary conditions
\begin{eqnarray}
\Delta^{-}(x_{1},x_{2})|_{t_{1}=t_{2}} &=& 0 , \label{con1}\\
\partial_{t_{1}}\Delta^{-}(x_{1},x_{2})|_{t_{1}=t_{2}}&=&-\partial_{t_{2}}\Delta^{-}(x_{1},x_{2})|_{t_{1}=t_{2}} = \delta^{(3)}({\bf x}_1-{\bf x}_2) , \label{con2}\\
\partial_{t_{1}}\partial_{t_{2}}\Delta^{-}(x_{1},x_{2})|_{t_{1}=t_{2}} &=& 0 ,\label{con3}\\
S^{-}(x_{1},x_{2})|_{t_{1}=t_{2}}&=&i\gamma_{0}\delta(\textbf{x}_{1}-\textbf{x}_{2})\label{DiracEqualTime}
,\end{eqnarray}
which follow from the canonical commutation and anticommutation relations. The boundary conditions for the statistical propagators are determined by the physical initial conditions in which the system is prepared. 

In thermal equilibrium\footnote{See \cite{LeB} for a detailed introduction to quantum field theory in equilibrium.} or vacuum\footnote{The correlation functions in vacuum can be obtained as the zero temperature limit of those in equilibrium and we in the following understand vacuum as a special case of thermal equilibrium.} the correlation functions only depend on the difference of coordinates $x_{1}-x_{2}$.
Furthermore they are related by the Kubo-Martin-Schwinger (KMS) relations, which are most conveniently written in terms of their Fourier transforms in $x_{1}-x_{2}$. In absence of chemical potentials they read
\begin{eqnarray}\label{KMS1}
\begin{tabular}{c c c}
$\Delta^{<}_{\textbf{q},eq}(\omega)=e^{-\beta\omega}\Delta^{>}_{\textbf{q},eq}(\omega)$& , &$S^{<}_{\textbf{q},eq}(\omega)=-e^{-\beta\omega}S^{>}_{\textbf{q},eq}(\omega)$
\end{tabular}
.\end{eqnarray}
Here $(\omega,\textbf{q})$ is a momentum space four vector and $\beta=1/T$ the inverse temperature\footnote{Here and in the following we work in the restframe of the bath, where $T$ has a physical interpretation as temperature. Due to the choice of frame the expressions are not Lorentz-invariant. This choice is for convenience, the theory can of course be formulated in a covariant manner \cite{Weldon:1982aq}.}.
The KMS relations are a consequence of the fact that the equilibrium density matrix $\varrho^{eq}=\exp(-\beta {\rm H})/{\rm Tr}\exp(-\beta {\rm H})$, where ${\rm H}$ is the Hamiltonian, acts as a time evolution operator in imaginary time and imposes boundary conditions on the $\Delta^{\gtrless}$ in the complex time plane.
They allow to express all two-point functions in terms of the spectral densities,
\begin{eqnarray}\label{KMSscalar}
\begin{tabular}{c c c}
$\Delta^{-}_{\textbf{q},eq}(\omega)=i\rho_{\q}(\omega)$& , &$\Delta^{+}_{\textbf{q},eq}(\omega)=\left(\frac{1}{2}+f_{B}(\omega)\right)\rho_{\q}(\omega)$\\
$\Delta^{<}_{\textbf{q},eq}(\omega)=f_{B}(\omega)\rho_{\q}(\omega)$& , &$\Delta^{>}_{\textbf{q},eq}(\omega)=(1+f_{B}(\omega))\rho_{\q}(\omega)$,
\end{tabular}
\end{eqnarray}
\begin{eqnarray}\label{KMSfermi}
\begin{tabular}{c c c}
$S^{-}_{\textbf{q},eq}(\omega)=i\uprho_{\q}(\omega)$& , &$S^{+}_{\textbf{q},eq}(\omega)=\left(\frac{1}{2}-f_{F}(\omega)\right)\rho_{\q}(\omega)$,\\
$S^{>}_{\textbf{q},eq}(\omega)=(1-f_{F}(\omega))\uprho_{\q}(\omega)$& , & $S^{<}_{\textbf{q},eq}(\omega)=-f_{F}(\omega)\uprho_{\q}(\omega)$.
\end{tabular}
\end{eqnarray}
Here $f_{B}=(e^{\beta\omega}-1)^{-1}$ and $f_{F}=(e^{\beta\omega}+1)^{-1}$. The Bose-Einstein and Fermi-Dirac distributions $f_{B}$ and $f_{F}$ naturally arise as a consequences of the boundary conditions for the correlation functions. The concrete shape of the spectral densities  $\rho_{\q}(\omega)$ for scalars and $\uprho_{\q}(\omega)$ for fermions will be specified below.

Out of equilibrium (\ref{forw})-(\ref{Sback}) all independently depend on both arguments $x_{1}$ and $x_{2}$.
They also cannot be related to each other by a relation analogue to (\ref{KMS1})-(\ref{KMSfermi}). The spectral function and statistical propagator both have to be found as solutions to a set of coupled integro-differential equations known as Kadanoff-Baym equations. 

\begin{figure}[t]
\centering
\includegraphics[width=12cm]{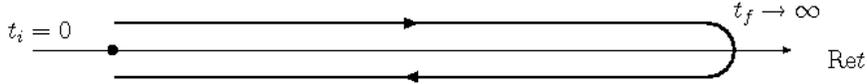}
\caption{The Keldysh contour runs from some initial time $x^{0}=t_{i}+i\epsilon$ parallel to the real axis ($x^{0}=t+i\epsilon$) up to some final time $t_{f}+i\epsilon$ and returns to $t_{i}-i\epsilon$. To compute physical correlation functions for arbitrary times $t>t_{i}$, one takes the limits $t_{f}\rightarrow\infty$ and $\epsilon \rightarrow 0$.\label{contour1}}
\end{figure}

They can be derived in the Schwinger-Keldysh formalism \cite{Schwinger:1960qe}. 
For simplicity we restrict ourselves to cases where the latter are Gaussian at initial time\footnote{For non-Gaussian initial conditions the formalism described here remains applicable, but some additional terms appear in the Kadanoff-Baym equations at initial time \cite{Garny:2009ni}.}, i.e. all $n$-point functions at initial time can be expressed by the two point function. Higher order correlations unavoidably build up at later time when system and bath are brought into touch, which is perceived as an increase of entropy by observers looking at Gaussian correlators only \cite{Koksma:2009wa}. 
Starting point is the generating functional for correlation functions of fields with time arguments on a contour in the complex time plane known as Keldysh contour, (cf. figure \ref{contour1}).
The choice of this contour (rather than the real axis) is a consequence of the fact that nonequilibrium processes are initial value problems. The system is prepared at an initial time $t_{i}$ and its state at later times is unknown. The Keldysh-contour, which starts and ends at the same time, allows to define a generating functional without knowledge of the state at asymptotic times $t=\pm\infty$. This is in contrast to the `in-out' formalism used to compute the S-matrix\footnote{Due to this fact this formalism is sometimes called `in-in' formalism.}. 
For the time ordered two-point function $\Delta_{\mathcal{C}}(x_{1},x_{2})$ for scalars on the contour $\mathcal{C}$ it yields the Dyson-Schwinger equation 
\begin{equation}\label{sde}
(\square_1 +m^2)\Delta_\C(x_{1},x_{2})+\int_{\C}d^{4}x' \Pi_{\C}(x_{1},x')
\Delta_{\C}(x',x_{2})=-i\delta_{\C}(x_{1}-x_{2})\ .
\end{equation}
Here $\square_1=\partial^{\mu}\partial_{\mu}$, where the derivatives are with respect to the components of $x_1$. $\Pi_{\mathcal{C}}(x_1,x_2)$ is the self energy and $\delta_{C}(x_{1}-x_{2})$ a $\delta$-function on the contour. Green's function and self energy can be decomposed as
\begin{eqnarray}\label{causal}
\Delta_{\mathcal{C}}(x_1,x_2)=\theta_{\mathcal{C}}(x^0_1,x^0_2)\Delta^>(x_1,x_2) + \theta_{\mathcal{C}}(x^0_2,x^0_1)\Delta^<(x_1,x_2)\\
\Pi_{\mathcal{C}}(x_1,x_2)=\theta_{\mathcal{C}}(x^0_1,x^0_2)\Pi^>(x_1,x_2) + \theta_{\mathcal{C}}(x^0_2,x^0_1)\Pi^<(x_1,x_2)
\end{eqnarray}
In (\ref{sde}) the time coordinates of $\Delta_\C$ and $\Pi_\C$ can lie on the upper or the lower branch of the contour. 
When both time arguments lie on the upper branch, $\Delta_{\C}$ can be identified with the usual Feynman propagator $\Delta_{F}$ in the limit $\epsilon\rightarrow 0$, cf. figure \ref{contour1}. For both time arguments on the lower branch this limit yields the anti-time ordered propagator $\Delta_{\bar{F}}$ due to the inverse direction of the ordering along the contour. 
In the cases that $x^0_1$ lies on the upper and $x^0_2$ on the lower branch and vice versa, $\Delta_{\C}$ can be identified with $\Delta^{<} $  and $\Delta^{>}$, respectively.
In a perturbative expansion of (\ref{sde}) in terms of Feynman diagrams, the time arguments of internal vertices can lie on either branch. Hence, the number of contributing graphs doubles with each internal vertex since all possible combinations are summed over\footnote{This fact is sometimes referred to as 
`doubling of degrees of freedom'.}. Two upper vertices are connected by $\Delta_{F}$, two lower vertices by $\Delta_{\bar{F}}$ and vertices of different type by $\Delta^{<}$ and $\Delta^{>}$. Each lower vertex leads to an additional factor $-1$. This is reminiscent of the real time formalism used in equilibrium.
All correlation functions can be obtained as linear combinations of $\Delta^{-}$ and $\Delta^{+}$ via 
\begin{eqnarray}
\Delta_{F}(x_1,x_2)   &=& \Delta^{+}(x_{1},x_{2}) 
- \frac{i}{2}{\rm sign}(x^0_1-x^0_2)\Delta^{-}(x_{1},x_{2}) \label{feynmanprop}\\
\Delta_{\bar{F}}(x_1,x_2)   &=& \Delta^{+}(x_{1},x_{2}) 
+ \frac{i}{2}{\rm sign}(x^0_1-x^0_2)\Delta^{-}(x_{1},x_{2}),\\
\Delta^{>}(x_1,x_2)   &=& \Delta^{+}(x_{1},x_{2}) 
- \frac{i}{2}\Delta^{-}(x_{1},x_{2}) \\
\Delta^{<}(x_1,x_2)   &=& \Delta^{+}(x_{1},x_{2}) 
+ \frac{i}{2}\Delta^{-}(x_{1},x_{2}).\label{kleinerinliste}
\end{eqnarray}
Using the above relations, a straightforward calculation allows to derive the equations of motion for $\Delta^{-}$ and $\Delta^{+}$ from (\ref{sde}), known as Kadanoff-Baym equations,
\begin{eqnarray}
(\square_{1}+m^{2})\Delta^{-}(x_{1},x_{2}) &=& 
-\int d^{3}\textbf{x}'\int_{t_{2}}^{t_{1}} d t' 
\Pi^{-}(x_{1},x')\Delta^{-}(x',x_{2})\;,\label{KBE1}\\
(\square_{1}+m^{2})\Delta^{+}(x_{1},x_{2}) &=&
-\int d^{3}\textbf{x}'\int_{t_{i}}^{t_{1}} dt'
\Pi^{-}(x_{1},x')\Delta^{+}(x',x_{2})\nonumber\\
&& +\int d^{3}\textbf{x}'\int_{t_{i}}^{t_{2}} dt' 
\Pi^{+}(x_{1},x')\Delta^{-}(x',x_{2})\;\label{KBE2}.
\end{eqnarray}
Here the self energies $\Pi^{\pm}$ are defined analogue to (\ref{dminus}), (\ref{dplus})\footnote{Depending on whether one defines the self energies as the sum of all 1PI-diagrams, c.f. \cite{ABDM3}, or $i$ times that sum, c.f. \cite{Anisimov:2008dz}, these definitions are either completely analogue to (\ref{dminus}), (\ref{dplus}) or vary by a factor $i$. Here chose the 'asymmetric' convention used in \cite{Anisimov:2008dz}.},
\begin{eqnarray}
\Pi^{+}(x_{1},x_{2})&=&-\frac{i}{2}\left(\Pi^{>}(x_{1},x_{2})+\Pi^{<}(x_{1},x_{2})\right)\\ 
\Pi^{-}(x_{1},x_{2})
&=&\Pi^{>}(x_{1},x_{2})-\Pi^{<}(x_{1},x_{2}),\label{PiMinusDef}
\end{eqnarray}
and can be related to the retarded and advanced self energies by
\begin{eqnarray}\label{PiAR}
\begin{tabular}{c c}
$\Pi^{R}(x_{1},x_{2})=\theta(t_{1}-t_{2})\Pi^{-}(x_{1},x_{2})$ , $\Pi^{A}(x_{1},x_{2})=-\theta(t_{2}-t_{1})\Pi^{-}(x_{1},x_{2})$.
\end{tabular}
\end{eqnarray}
All corresponding equations for the fermionic Green's functions are obtained by the replacements $(\square_1 +m^2)\rightarrow -(i\Slash{\partial}_{1}-\mathpzc{m})$, $\Delta_{\mathcal{C}}(x_{1},x_{2})\rightarrow S_{\mathcal{C}}(x_{1},x_{2})$, $\Pi_{\mathcal{C}}(x_{1},x_{2})\rightarrow \Sigma_{\mathcal{C}}(x_{1},x_{2})$ and so on.

The Kadanoff-Baym equations (\ref{KBE1}) and (\ref{KBE2}) are the exact equations of motion for the correlation functions. They do not rely on a gradient expansion or semiclassical approximation and make not restrictions on the size of the initial deviation from equilibrium.
Roughly speaking, $\Delta^{+}$ and $S^{+}$ are the field theoretical generalisations of the phase space distribution functions appearing in the semiclassical Boltzmann equations while the self energies take the role of the collision terms. This interpretation is intuitive as the total cross section is related to the discontinuity of the self energy by the optical theorem.

We are mainly interested in cosmological applications and restrict the analysis to spacially homogeneous and isotropic systems, in which the correlation functions can depend on the difference of spacial coordinates $\textbf{x}_{1}-\textbf{x}_{2}$ only. Furthermore, we are interested in the case that $\phi$ or $\Psi$  are weakly coupled to a large thermal bath. Formally this corresponds to a density matrix of the form $\varrho_{\rm sys}\otimes\varrho^{eq}_{\rm bath}$, where $\varrho^{eq}_{\rm bath}=\exp(-\beta {\rm H}_{\rm bath})/{\rm Tr}\exp(-\beta {\rm H}_{\rm bath})$ and $\varrho_{\rm sys}$ is the initial density matrix of the system. 
The interactions of the fields that make up the bath amongst each other are much stronger than the couplings to $\phi$ or $\Psi$ and bring them to equilibrium fast on the time scale associated to the dynamics of the out-of-equilibrium degrees of freedom.
Then all correlation functions of the bath fields can always be characterised by a single temperature $T$ (and possibly a handful of chemical potentials). 
We are interested in the case that the out-of-equilibrium fields mainly interact with (in particular decay into) bath fields only.
This can explicitly be realised by using coupling terms that are linear in $\phi$ and $\Psi$ and can be written in the form $\alpha\phi\mathcal{O}_{\phi}[\mathcal{X}]$ and $\tilde{\alpha}\bar{\Psi}\mathcal{O}_{\Psi}[\mathcal{X}]+h.c.$. Here $\mathcal{O}_{\phi}[\mathcal{X}]$ and $\mathcal{O}_{\Psi}[\mathcal{X}]$ are composite operators that contain bath fields only, to which we collectively refer as $\mathcal{X}$, and $\alpha$, $\tilde{\alpha}$ are coupling constants. 
The effects of backreaction on the self energies are suppressed by additional powers of the coupling constants as well as the large number of degrees of freedom in the bath and can safely be neglected.
Thus, the self energies are determined by bath properties. In particular, they depend on the relative coordinate $x_{1}-x_{2}$ only, inherit the KMS relations
\begin{equation}\label{PiKMS}
\begin{tabular}{c c c}
$\Pi^{>}_{\textbf{q}}(\omega)=e^{\beta\omega}\Pi^{<}_{\textbf{q}}(\omega)$ & & $\Sigma^{>}_{\textbf{q}}(\omega)=-e^{\beta\omega}\Sigma^{<}_{\textbf{q}}(\omega)$\\
$\Pi^{+}_{\textbf{q}}(\omega)=-i\left(\frac{1}{2}+f_{B}(\omega)\right)\Pi^{-}_{\textbf{q}}(\omega)$& , & $\Sigma^{+}_{\textbf{q}}(\omega)=-i\left(\frac{1}{2}-f_{F}(\omega)\right)\Sigma^{-}_{\textbf{q}}(\omega)$
\end{tabular}
\end{equation}
and can be computed perturbatively by methods of thermal field theory.
It can be proven that in this case also the spectral functions are time translation invariant \cite{Anisimov:2008dz}.

The definition (\ref{PiAR}) gives rise to the well-known spectral representation 
\begin{eqnarray}
\Pi^{R}(\omega)&=&i\int_{-\infty}^{\infty} \frac{dp_{0}}{2\pi}\frac{\Pi^{-}(p_{0})}{\omega-p_{0}+i\epsilon}
\end{eqnarray}
which implies
\begin{eqnarray}
\Pi^R(\omega) &=& \frac{1}{2}\Pi^-(\omega) + 
\mathcal{P}\int\frac{d\omega'}{2\pi i}
\frac{\Pi^-(\omega')}{\omega'-\omega}\ \label{pr}
\end{eqnarray}
and an analogue relation for fermions, where $\mathcal{P}$ denotes the principal value. These relate the discontinuities of the self energies to $\Pi^{-}$ and $\Sigma^{-}$. For a real scalar with real couplings (\ref{pr}) implies\footnote{A table of the analytic properties of the self energies can be found in the appendix of \cite{Anisimov:2008dz}.}
\begin{eqnarray}
-{\rm Im}\Pi^R(\omega) = \frac{i}{2}\Pi^-(\omega)\ \label{ipa}.
\end{eqnarray}
The quantities $\Pi^{\gtrless}$ individually can be related to the gain and loss terms in (\ref{GainAndLossRates}), which fulfil the detailed balance ratio due to (\ref{PiKMS}) when the bath is in equilibrium.

In the mixed time-momentum representation the Kadanoff-Baym equations can be written in the convenient form
\begin{eqnarray}
(\partial_{t_{1}}^{2}+\omega_{\textbf{q}}^{2})\Delta^{-}_{\textbf{q}}(t_{1}-t_{2}) &=& 
- \int_{t_{2}}^{t_{1}} d t' 
\Pi^{-}_{\textbf{q}}(t_{1}-t')\Delta^{-}_{\textbf{q}}(t'-t_{2})\;,\label{abzug2}\\
(\partial_{t_{1}}^{2}+\omega_{\textbf{q}}^{2})\Delta^{+}_{\textbf{q}}(t_{1},t_{2}) &=&
-\int_{t_{i}}^{t_{1}} dt'
\Pi^{-}_{\textbf{q}}(t_{1}-t')\Delta^{+}_{\textbf{q}}(t',t_{2})\nonumber\\
&& +\int_{t_{i}}^{t_{2}} dt' 
\Pi^{+}_{\textbf{q}}(t_{1}-t')\Delta^{-}_{\textbf{q}}(t'-t_{2})\;.\label{addi2}
\end{eqnarray}
\begin{eqnarray}
(i\gamma_{0}\partial_{t_{1}}-\pmb{\Slash{q}}-\fermionmass{m})S^{-}_{\textbf{q}}(t_{1}-t_{2}) &=& 
\int_{t_{2}}^{t_{1}} d t' 
\Sigma^{-}_{\textbf{q}}(t_{1}-t')S^{-}_{\textbf{q}}(t'-t_{2})\;,\label{Sabzug2}\\
(i\gamma_{0}\partial_{t_{1}}-\pmb{\Slash{q}}-\fermionmass{m})S^{+}_{\textbf{q}}(t_{1},t_{2}) &=&
\int_{t_{i}}^{t_{1}} dt'
\Sigma^{-}_{\textbf{q}}(t_{1}-t')S^{+}_{\textbf{q}}(t',t_{2})\nonumber\\
&& -\int_{t_{i}}^{t_{2}} dt' 
\Sigma^{+}_{\textbf{q}}(t_{1}-t')S^{-}_{\textbf{q}}(t'-t_{2})\;.\label{Saddi2}
\end{eqnarray}
where $\omega_{\textbf{q}}=(\textbf{q}^{2}+m^{2})^{1/2}$, $\pmb{\Slash{q}}=\gamma_{i}q_{i}$ and $t_{i}$ is the initial time at which the system is prepared, which can be set to zero without loss of generality.
 
All formulae presented here are valid in Minkowski spacetime. They are valid in good approximation if the characteristic time scale of the nonequilibrium dynamics is short in comparison to that of Hubble expansion. A generalisation to curved spacetime has e.g. been discussed in \cite{Hohenegger:2008zk}.

\subsection{The Fermion Self Energy}\label{YukawaSelfEnergy}
For vanishing $\tilde{\fermionmass{m}}$ (\ref{PhDFormula}) reads
\begin{eqnarray}\label{Weldonlike}
\lefteqn{\Sigma^{-}_{\textbf{q}}(\omega)=-iY^{2}\int\frac{d^{3}\textbf{p}}{(2\pi)^{2}}\frac{1}{2\tilde{\Omega}_{1}2\tilde{\Omega}_{2}}}\nonumber\\
&&\bigg((1-f_{F}(\tilde{\Omega}_{1})+f_{B}(\tilde{\Omega}_{2}))\Big((\tilde{\Omega}_{1}\gamma_{0}-\pmb{p\gamma})\delta(\omega-\tilde{\Omega}_{1}-\tilde{\Omega}_{2})\nonumber\\
&&\phantom{(1-f_{F}()+f_{B}(\tilde{\Omega}_{2}))}+(\tilde{\Omega}_{1}\gamma_{0}+\pmb{p\gamma})\delta(\omega+\tilde{\Omega}_{1}+\tilde{\Omega}_{2})\Big)\nonumber\\
&+&(f_{F}(\tilde{\Omega}_{1})+f_{B}(\tilde{\Omega}_{2}))\Big((\tilde{\Omega}_{1}\gamma_{0}-\pmb{p\gamma})\delta(\omega-\tilde{\Omega}_{1}+\tilde{\Omega}_{2})\nonumber\\
&&\phantom{(1-f_{F}()+f_{B}(\tilde{\Omega}_{2}))}+(\tilde{\Omega}_{1}\gamma_{0}+\pmb{p\gamma})\delta(\omega+\tilde{\Omega}_{1}-\tilde{\Omega}_{2})\Big)\bigg)
\end{eqnarray}
We take $\omega>0$, which is sufficient due to the symmetry of (\ref{loopintfermi}). 
The self energy can be decomposed as $\Sigma^{-}_{\textbf{q}}(\omega)=2i(a_{\textbf{q}}(\omega)\Slash{q}+b_{\textbf{q}}(\omega)\Slash{u})$, where $u$ is the four velocity of the bath and $a_{\textbf{q}}(\omega)$, $b_{\textbf{q}}(\omega)$ are scalar functions that depend on $\omega$ and $\textbf{q}$. In the restframe of the bath $u=(1,\textbf{0})$. 
We define the projections
\begin{eqnarray}
A_{\textbf{q}}(\omega)&=&\frac{1}{4}{\rm tr}\left(\Slash{q}\Sigma^{-}_{\textbf{q}}(\omega)\right)\label{Adef}\\
B_{\textbf{q}}(\omega)&=&\frac{1}{4}{\rm tr}\left(\Slash{u}\Sigma^{-}_{\textbf{q}}(\omega)\right)\label{Bdef} 
\end{eqnarray}
which can be related to the coefficients $a_{\textbf{q}}(\omega)$ and  $b_{\textbf{q}}(\omega)$ via
\begin{eqnarray}\label{uebersetzung}
2i a_{\textbf{q}}(\omega)&=&\frac{B_{\textbf{q}}(\omega) qu - A_{\textbf{q}}(\omega) u^{2}}{(qu)^{2}-q^{2}u^{2}}=\frac{B_{\textbf{q}}(\omega) \omega - A_{\textbf{q}}(\omega) }{\textbf{q}^{2}}\\
2i b_{\textbf{q}}(\omega)&=&\frac{- B_{\textbf{q}}(\omega) q^{2} + A_{\textbf{q}}(\omega) qu}{(qu)^{2}-q^{2}u^{2}}=\frac{\omega A_{\textbf{q}}(\omega)-\omega^{2} B_{\textbf{q}}(\omega)}{\textbf{q}^{2}}+B_{\textbf{q}}(\omega) 
\end{eqnarray}
From this one can find 
\begin{eqnarray}
A=-Y^{2}\int \frac{d^{3}\textbf{p}}{(2\pi)^{3}}\frac{2\pi}{8\tilde{\Omega}_{1}\tilde{\Omega}_{2}}&\Big(&(1-f_{1}+f_{2})\big( (\omega\tilde{\Omega}_{1}-\textbf{q}\textbf{p})\delta(\omega-\tilde{\Omega}_{1}-\tilde{\Omega}_{2})\nonumber\\
&&\phantom{(1-f_{1}+f_{2})}+(\omega\tilde{\Omega}_{1}+\textbf{q}\textbf{p})\delta(\omega+\tilde{\Omega}_{1}+\tilde{\Omega}_{2}) \big)\nonumber\\
&&+(f_{1}+f_{2})\big( (\omega\tilde{\Omega}_{1}-\textbf{q}\textbf{p})\delta(\omega-\tilde{\Omega}_{1}+\tilde{\Omega}_{2})\nonumber\\
&&\phantom{(f_{1}+f_{2})}+(\omega\tilde{\Omega}_{1}+\textbf{q}\textbf{p})\delta(\omega+\tilde{\Omega}_{1}-\tilde{\Omega}_{2}) \big) \Big)
\end{eqnarray}
with $f_{1}=f_{F}(\tilde{\Omega}_{1})$ and $f_{2}=f_{B}(\tilde{\Omega}_{2})$ during this computation.
This expression can be rewritten as 
\begin{eqnarray}
\lefteqn{A=-Y^{2}\int \frac{d^{3}\textbf{p}}{(2\pi)^{3}}\frac{2\pi}{8\tilde{\Omega}_{1}\tilde{\Omega}_{2}}}\nonumber\\
&\Big(& \omega\tilde{\Omega}_{1}\big((1-f_{1}+f_{2})(\delta_{1}+\delta_{2}) + (f_{1}+f_{2})(\delta_{3}+\delta_{4})\big)\nonumber\\
&&-\textbf{q}\textbf{p}  \big((1-f_{1}+f_{2})(\delta_{1}-\delta_{2}) + (f_{1}+f_{2})(\delta_{4}-\delta_{3})\big)\Big)
\end{eqnarray}
where
\begin{eqnarray}
\begin{tabular}{c c}
$\delta_{1}=\delta(\omega-\tilde{\Omega}_{1}-\tilde{\Omega}_{2})$, &
$\delta_{2}=\delta(\omega+\tilde{\Omega}_{1}+\tilde{\Omega}_{2})$,\\
$\delta_{3}=\delta(\omega+\tilde{\Omega}_{1}-\tilde{\Omega}_{2})$, &
$\delta_{4}=\delta(\omega-\tilde{\Omega}_{1}+\tilde{\Omega}_{2})$
\end{tabular}
\end{eqnarray}
We drop $\delta_{2}$ as it cannot contribute for $\omega>0$ and change to spherical coordinates $\varphi$, $\vartheta$, $|\textbf{p}|$. The $\varphi$ integration is trivial. As the fermion in the loop is massless  $|\textbf{p}|=\tilde{\Omega}_{1}$. Introducing $x=|\textbf{p}||\textbf{q}|\cos(\vartheta)=\textbf{p}\textbf{q}$ one can write
\begin{eqnarray}
\lefteqn{A=\frac{-Y^{2}}{16\pi|\textbf{q}|}\int_{0}^{\infty}d\tilde{\Omega}_{1}\int_{-\tilde{\Omega}_{1}|\textbf{q}|}^{\tilde{\Omega}_{1}|\textbf{q}|}dx}\nonumber\\
 &\Big(& \delta(x-x_{01})\big((\omega\tilde{\Omega}_{1}-x)(1-f_{1}+f_{2})\big)\nonumber\\
&&+\delta(x-x_{03})\big((\omega\tilde{\Omega}_{1}+x)(f_{1}+f_{2})\big)\nonumber\\
&&+\delta(x-x_{04})\big((\omega\tilde{\Omega}_{1}-x)(f_{1}+f_{2})\big)\Big) \label{Aformula}
\end{eqnarray}
where we used $\delta_{i}=\tilde{\Omega}_{2}\delta(x-x_{0i})$. The $x_{0i}$ can easily be determined as
\begin{eqnarray}
x_{01}&=&\frac{1}{2}(\textbf{q}^{2}-\omega^{2}+\tilde{M}^{2})+\tilde{\Omega}_{1}\omega\\
x_{03}&=&\frac{1}{2}(\textbf{q}^{2}-\omega^{2}+\tilde{M}^{2})-\tilde{\Omega}_{1}\omega\\
x_{04}&=&x_{01}
.\end{eqnarray}
This allows to perform the $x$ integration,
\begin{eqnarray}
A=\frac{-g^{2}}{16\pi|\textbf{q}|}\Big(\int_{1}d\tilde{\Omega}_{1}\left(\omega \mathpzc{f}_{1}+\mathpzc{g}_{1}\right)+\int_{3}d\tilde{\Omega}_{1}\left(\omega \mathpzc{f}_{3}-\mathpzc{g}_{3}\right)+\int_{4}d\tilde{\Omega}_{1}\left(\omega \mathpzc{f}_{4}+\mathpzc{g}_{4}\right) \Big)
.\end{eqnarray}
Here the subscript at the $\int_{i}$ indicates which $\delta_{i}$ determines limits for the $\tilde{\Omega}_{1}$ integration. The $\mathpzc{f}_{i}$ and $\mathpzc{g}_{i}$ are given by
\begin{eqnarray}
\mathpzc{f}_{1}&=&\tilde{\Omega}_{1}\big(1-f_{F}(\tilde{\Omega}_{1})+f_{B}(\omega-\tilde{\Omega}_{1})\big)\\
\mathpzc{f}_{3}&=&\tilde{\Omega}_{1}\big(f_{F}(\tilde{\Omega}_{1})+f_{B}(\omega+\tilde{\Omega}_{1})\big)\\
\mathpzc{f}_{4}&=&\tilde{\Omega}_{1}\big(f_{F}(\tilde{\Omega}_{1})+f_{B}(\tilde{\Omega}_{1}-\omega)\big)\\
\mathpzc{g}_{1}&=&\big(\frac{1}{2}(\omega^{2}-\textbf{q}^{2}-\tilde{M}^{2})-\tilde{\Omega}_{1}\omega\big)\big(1-f_{F}(\tilde{\Omega}_{1})+f_{B}(\omega-\tilde{\Omega}_{1})\big)\\
\mathpzc{g}_{3}&=&\big(\frac{1}{2}(\omega^{2}-\textbf{q}^{2}-\tilde{M}^{2})+\tilde{\Omega}_{1}\omega\big)\big(f_{F}(\tilde{\Omega}_{1})+f_{B}(\omega+\tilde{\Omega}_{1})\big)\\
\mathpzc{g}_{4}&=&\big(\frac{1}{2}(\omega^{2}-\textbf{q}^{2}-\tilde{M}^{2})-\tilde{\Omega}_{1}\omega\big)\big(f_{F}(\tilde{\Omega}_{1})+f_{B}(\tilde{\Omega}_{1}-\omega)\big)
\end{eqnarray}
It is easy to see from Eqs.~(\ref{Adef}), (\ref{Bdef}) and (\ref{Aformula}) that
\begin{eqnarray}
B=\frac{-g^{2}}{16\pi|\textbf{q}|}\Big(\int_{1}d\tilde{\Omega}_{1} \mathpzc{f}_{1} + \int_{3}d\tilde{\Omega}_{1} \mathpzc{f}_{3} + \int_{4}d\tilde{\Omega}_{1} \mathpzc{f}_{4} \Big)
\end{eqnarray}
$\mathpzc{f}_{1}$ and $\mathpzc{g}_{1}$ are contributions from decay and inverse decays of $\Psi$ and can lead to a zero temperature part in case (a). $\mathpzc{f}_{3}$, $\mathpzc{f}_{4}$, $\mathpzc{g}_{3}$ and $\mathpzc{g}_{4}$ are related to Landau damping. Note that
\begin{eqnarray}
\mathpzc{f}_{1}&=&-\mathpzc{f}_{4}\label{fGleichheit}\\
\mathpzc{g}_{1}&=&-\mathpzc{g}_{4}\label{gGleichheit}
\end{eqnarray}
due to the property $f_{B}(-z)=-1-f_{B}(z)$ despite the fact that the terms originate from different processes\footnote{Note that despite the equalities (\ref{fGleichheit}) and (\ref{gGleichheit}) the Landau damping terms $\mathpzc{f}_{4}$, $\mathpzc{g}_{4}$ never lead to a contribution to $\Sigma$ at zero temperature while the decay and inverse decay parts $\mathpzc{f}_{1}$ and $\mathpzc{g}_{1}$ can contribute as expected. The reason lies in the different integration limits, see (\ref{einseins}) and (\ref{viereins})}.
The $\mathpzc{f}_{i}$ are symmetric in $\omega$, the $\mathpzc{g}_{i}$ antisymmetric.
$B$ is symmetric in $\omega$ while $A$ is antisymmetric. As a consequence, $a$ is antisymmetric while $b$ is symmetric.
The stem functions of all $\mathpzc{f}_{i}$, $\mathpzc{g}_{i}$ are known analytically (see (\ref{erstestamm})-(\ref{letzte})) and the only remaining task is the determination of the integration limits for each $\delta_{i}$-function.
\paragraph{$\delta(\omega-\tilde{\Omega}_{1}-\tilde{\Omega}_{2})$}:
In order for $x_{01}$ to be a zero, the condition 
\begin{equation}\label{einseins}
\omega-\tilde{\Omega}_{1}>0
\end{equation}
has to be fulfilled. In any case, 
\begin{equation}\label{einszwei}
\tilde{\Omega}_{1}>\tilde{\fermionmass{m}}=0
.\end{equation}
In order for the $x$-integral to be non-zero it is required that 
\begin{equation}\label{einsdrei}
|x_{01}|<\tilde{\Omega}_{1}|\textbf{q}| 
.\end{equation}
The solutions to  $|x_{01}|=\tilde{\Omega}_{1}|\textbf{q}|$ are 
\begin{equation}
\tilde{\omega}_{\pm}=\frac{1}{2}\frac{q^{2}-\tilde{M}^{2}}{q^{2}}(\omega\pm |\textbf{q}|)
\end{equation}
with $q^{2}=\omega^{2}-\textbf{q}^{2}$. One has to distinguish three different regimes: 
For $0<\omega<|\textbf{q}|$ and $\tilde{\Omega}_{1}>0$ only $\tilde{\omega}_{+}$ is a solution, and it imposes a lower bound on $\tilde{\Omega}_{1}$ in order for the inequality (\ref{einsdrei}) to be fulfilled, leading to $\tilde{\Omega}_{1}>\tilde{\omega}_{+}$. On the other hand  the condition (\ref{einseins}) has to be fulfilled, and since for $\omega<|\textbf{q}|$ always $\tilde{\omega}_{+}>\omega$, there is no contribution to the integral from this region.
For $|\textbf{q}|<\omega<(\textbf{q}^{2}+\tilde{M}^{2})^{\frac{1}{2}}$ both $\tilde{\omega}_{\pm}<0$ and none of them makes (\ref{dreidrei}) an equality.
For $\omega>(\textbf{q}^{2}+\tilde{M}^{2})^{\frac{1}{2}}$ both $\tilde{\omega}_{\pm}$ are always smaller than $\omega$ and (\ref{dreidrei}) leads to $\tilde{\omega}_{-}<\omega<\tilde{\omega}_{+}$. Therefore
\begin{equation}
\int_{1}d\tilde{\Omega}_{1}=\theta(q^{2}-\tilde{M}^{2})\int_{\tilde{\omega}_{-}}^{\tilde{\omega}_{+}}d\tilde{\Omega}_{1}
\end{equation}
\paragraph{$\delta(\omega+\tilde{\Omega}_{1}-\tilde{\Omega}_{2})$}:
Here the three conditions
\begin{eqnarray}
\omega+\tilde{\Omega}_{1}&>&0\label{dreieins}\\
\tilde{\Omega}_{1}&>&\tilde{\fermionmass{m}}=0\label{dreizwei}\\
|x_{03}|&<&\tilde{\Omega}_{1}|\textbf{q}|\label{dreidrei}
\end{eqnarray}
have to be fulfilled.
In this case (\ref{dreidrei}) is made an equality for $\tilde{\Omega}_{1}=-\tilde{\omega}_{\pm}$. The same regimes have to be distinguished. 
For $\omega<|\textbf{q}|$ only $-\tilde{\omega}_{-}$ makes (\ref{dreidrei}) an equality while $-\tilde{\omega}_{+}$ is negative and not a solution. $-\tilde{\omega}_{-}$ is positive as required by (\ref{dreizwei}) and forms a lower bound.
For $|\textbf{q}|<\omega<(\textbf{q}^{2}+\tilde{M}^{2})^{\frac{1}{2}}$ both $-\tilde{\omega}_{\pm}$ are positive and solutions. Beyond its first order pole at $\omega=|\textbf{q}|$ the solution $-\tilde{\omega}_{+}$ is now the larger one and forms an upper limit, leading to $-\tilde{\omega}_{-}<\tilde{\Omega}_{1}<-\tilde{\omega}_{+}$.
For $\omega>(\textbf{q}^{2}+\tilde{M}^{2})^{\frac{1}{2}}$ both $-\tilde{\omega}_{\pm}$ are negative and not solutions of (\ref{dreidrei}) as an equality. Then there is no contribution to the integral from that region.
Therefore
\begin{equation}
\int_{3}d\tilde{\Omega}_{1}=\theta(-q^{2})\int_{-\tilde{\omega}_{-}}^{\infty}d\tilde{\Omega}_{1}+\theta(q^{2})\theta(\tilde{M}^{2}-q{^2})\int_{-\tilde{\omega}_{-}}^{-\tilde{\omega}_{+}}
\end{equation}
\paragraph{$\delta(\omega-\tilde{\Omega}_{1}+\tilde{\Omega}_{2})$}:
The situation here is exactly the same as for $\delta_{1}$, in particular $x_{04}=x_{01}$, except that the condition $\omega-\tilde{\Omega}_{1}>0$ has to be replaced by
\begin{equation}
\omega-\tilde{\Omega}_{1}<0\label{viereins}
,\end{equation}
enforcing $\tilde{\Omega}_{1}>\omega$. Again for $\omega<|\textbf{q}|$ only $\tilde{\omega}_{+}$ fulfils (\ref{einsdrei}), imposing a lower bound on $\tilde{\Omega}_{1}$ and for $\omega>(\textbf{q}^{2}+\tilde{M}^{2})^{\frac{1}{2}}$ both $\tilde{\omega}_{\pm}$ are solutions. $\tilde{\omega}_{+}$ is the upper and $\tilde{\omega}_{-}$ the lower bound here. For $0<q^{2}<\tilde{M}^{2}$ none of $\tilde{\omega}_{\pm}$ is a valid solution. This time the condition (\ref{viereins}) selects out the region $q^{2}<0$, hence the integral is
\begin{equation}
\int_{4}d\tilde{\Omega}_{1}=\theta(-q^{2})\int_{\tilde{\omega}_{+}}^{\infty}d\tilde{\Omega}_{1}
\end{equation}
The combined expressions are
\begin{eqnarray}
\lefteqn{A=\frac{-Y^{2}}{16\pi|\textbf{q}|}\Big(\theta(q^{2}-\tilde{M}^{2})\Big[\omega \mathpzc{F}_{1}+\mathpzc{G}_{1}\Big]_{\tilde{\omega}_{-}}^{\tilde{\omega}_{+}}}\nonumber\\
&+&\theta(-q^{2})\Big[\omega \mathpzc{F}_{3}-\mathpzc{G}_{3}\Big]_{-\tilde{\omega}_{-}}^{\infty}+\theta(q^{2})\theta(\tilde{M}^{2}-q^{2})\Big[\omega \mathpzc{F}_{3}-\mathpzc{G}_{3}\Big]_{-\tilde{\omega}_{-}}^{-\tilde{\omega}_{+}}\nonumber\\ 
&+&\theta(-q^{2})\Big[\omega \mathpzc{F}_{4}+\mathpzc{G}_{4}\Big]_{\tilde{\omega}_{+}}^{\infty} \Big)\label{AFormula}
\end{eqnarray}
and
\begin{eqnarray}
\lefteqn{B=\frac{-Y^{2}}{16\pi|\textbf{q}|}\Big(\theta(q^{2}-\tilde{M}^{2})\Big[ \mathpzc{F}_{1} \Big]_{\tilde{\omega}_{-}}^{\tilde{\omega}_{+}}}\nonumber\\
&+&\theta(-q^{2})\Big[ \mathpzc{F}_{3}\Big]_{-\tilde{\omega}_{-}}^{\infty}+\theta(q^{2})\theta(\tilde{M}^{2}-q^{2})\Big[ \mathpzc{F}_{3}\Big]_{-\tilde{\omega}_{-}}^{-\tilde{\omega}_{+}}\nonumber\\
&+&\theta(-q^{2})\Big[ \mathpzc{F}_{4}\Big]_{\tilde{\omega}_{+}}^{\infty} \Big)\label{BFormula}
\end{eqnarray}
with
\begin{eqnarray}
\mathpzc{F}_{1}&=&\frac{\tilde{\Omega}_{1}}{\beta}\Big(\ln\big(e^{\beta\tilde{\Omega}_{1}}+1\big)-\ln\big(1-e^{\beta(\tilde{\Omega}_{1}-\omega)}\big)\Big)\nonumber\\\label{erstestamm}
&&+\frac{1}{\beta^{2}}\Big({\rm Li}_{2}\big(-e^{\beta\tilde{\Omega}_{1}}\big)-{\rm Li}_{2}\big(e^{\beta(\tilde{\Omega}_{1}-\omega)}\big)\Big)\\
\mathpzc{F}_{3}&=&\frac{\tilde{\Omega}_{1}}{\beta}\Big(\ln\big(1-e^{\beta(\tilde{\Omega}_{1}+\omega)}\big)-\ln\big(e^{\beta\tilde{\Omega}_{1}}+1\big)\Big)\nonumber\\
&&+\frac{1}{\beta^{2}}\Big({\rm Li}_{2}\big(e^{\beta(\tilde{\Omega}_{1}+\omega)}\big)-{\rm Li}_{2}\big(-e^{\beta\tilde{\Omega}_{1}}\big)\Big)\\
\mathpzc{F}_{4}&=&-\mathpzc{F}_{1}
\end{eqnarray}
and
\begin{eqnarray}
\lefteqn{\mathpzc{G}_{1}= \frac{\tilde{M}^{2}-q^{2}}{2\beta}\Big(-\ln\big(1+e^{\beta\tilde{\Omega}_{1}}\big)+\ln\big(e^{\beta\tilde{\Omega}_{1}}-e^{\beta\omega}\big)\Big)}\nonumber\\
&+&\frac{\omega\tilde{\Omega}_{1}}{\beta}\Big(\ln\big(1-e^{\beta(\tilde{\Omega}_{1}-\omega)}\big)-\ln\big(1+e^{\beta\tilde{\Omega}_{1}}\big)\Big)\nonumber\\
&+&\frac{\omega}{\beta^{2}}\Big({\rm Li}_{2}\big(e^{\beta(\tilde{\Omega}_{1}-\omega)}\big)-{\rm Li}_{2}\big(-e^{\beta\tilde{\Omega}_{1}}\big)\Big)
\end{eqnarray}
\begin{eqnarray}
\lefteqn{\mathpzc{G}_{3}= \frac{\tilde{M}^{2}-q^{2}}{2\beta}\Big(\ln\big(1+e^{\beta\tilde{\Omega}_{1}}\big)-\ln\big(e^{\beta(\tilde{\Omega}_{1}+\omega)}-1\big)\Big)}\nonumber\\
&+&\frac{\omega\tilde{\Omega}_{1}}{\beta}\Big(\ln\big(1-e^{\beta(\tilde{\Omega}_{1}+\omega)}\big)-\ln\big(1+e^{\beta\tilde{\Omega}_{1}}\big)\Big)\nonumber\\
&+&\frac{\omega}{\beta^{2}}\Big({\rm Li}_{2}\big(e^{\beta(\tilde{\Omega}_{1}+\omega)}\big)-{\rm Li}_{2}\big(-e^{\beta\tilde{\Omega}_{1}}\big)\Big)
\end{eqnarray}
\begin{eqnarray}
\mathpzc{G}_{4}=-\mathpzc{G}_{1}\label{letzte}
.\end{eqnarray}
${\rm Li}_{2}$ is the dilogarithm function. 
The $\mathpzc{F}_{i}$ as displayed here are not real in general due to the choice of different branches of the (di)logarithms, but the imaginary terms always cancel since the choice of branch is always the same at both integration limits.
This analytic result for $\Sigma^{-}$ is in agreement with numerical plots shown in \cite{Kitazawa:2006zi}.
The $\theta$-functions are, as in (\ref{dampingformel}), a consequence of energy conservation. 
\end{appendix}

\end{document}